\documentclass[%
 reprint,
superscriptaddress,
 amsmath,amssymb,
 aps,
]{revtex4-2}

\usepackage{graphicx}
\usepackage{dcolumn}
\usepackage{bm}
\usepackage{hyperref}

\usepackage{color}
\usepackage{url}
\usepackage{subcaption}
\usepackage[labelformat=parens,labelsep=quad,skip=3pt]{caption}
\usepackage[nowatermark]{fixmetodonotes}
\usepackage[margin=1.in]{geometry}
\usepackage{floatrow}
\usepackage{caption}

\begin{document}

\preprint{APS/123-QED}

\title{Neutrino-Nucleon Cross-Section Model Tuning in GENIE v3}

\author{J\'{u}lia Tena-Vidal}
\affiliation{University of Liverpool, Dept. of Physics, Liverpool L69 7ZE, UK}
\author{Costas Andreopoulos}
\affiliation{University of Liverpool, Dept. of Physics, Liverpool L69 7ZE, UK}
\affiliation{Science and Technology Facilities Council, Rutherford Appleton Laboratory, Particle Physics Dept., Oxfordshire OX11 0QX, UK}
\author{Adi Ashkenazi}
\affiliation{Massachusetts Institute of Technology, Dept. of Physics, Cambridge, MA 02139, USA}
\author{Christopher Barry}
\affiliation{University of Liverpool, Dept. of Physics, Liverpool L69 7ZE, UK}
\author{Steve Dennis}
\altaffiliation[Now at ]{University of Cambridge}
\affiliation{University of Liverpool, Dept. of Physics, Liverpool L69 7ZE, UK}
\author{Steve Dytman}
\affiliation{University of Pittsburgh, Dept. of Physics and Astronomy, Pittsburgh PA 15260, USA}
\author{Hugh Gallagher}
\affiliation{Tufts University, Dept. of Physics and Astronomy, Medford MA 02155, USA}
\author{Steven Gardiner}
\affiliation{Fermi National Accelerator Laboratory, Batavia, Illinois 60510, USA}
\author{Walter Giele}
\affiliation{Fermi National Accelerator Laboratory, Batavia, Illinois 60510, USA}
\author{Robert Hatcher}
\affiliation{Fermi National Accelerator Laboratory, Batavia, Illinois 60510, USA}
\author{Or Hen}
\affiliation{Massachusetts Institute of Technology, Dept. of Physics, Cambridge, MA 02139, USA}
\author{Libo Jiang}
\altaffiliation[Now at ]{Virginia Polytechnic Institute and State University}
\affiliation{University of Pittsburgh, Dept. of Physics and Astronomy, Pittsburgh PA 15260, USA}
\author {Igor D. Kakorin}
\affiliation{Joint Institute for Nuclear Research (JINR), Dubna, Moscow region, 141980, Russia}
\author {Konstantin S. Kuzmin}
\affiliation{Joint Institute for Nuclear Research (JINR), Dubna, Moscow region, 141980, Russia}
\author {Anselmo Meregaglia}
\affiliation{ CENBG, Universit\'e de Bordeaux, CNRS/IN2P3, 33175 Gradignan, France}
\author {Vadim A. Naumov}
\affiliation{Joint Institute for Nuclear Research (JINR), Dubna, Moscow region, 141980, Russia}
\author{Afroditi Papadopoulou}
\affiliation{Massachusetts Institute of Technology, Dept. of Physics, Cambridge, MA 02139, USA}
\author {Gabriel Perdue}
\affiliation{Fermi National Accelerator Laboratory, Batavia, Illinois 60510, USA}
\author{Marco Roda}
\affiliation{University of Liverpool, Dept. of Physics, Liverpool L69 7ZE, UK}
\author{Vladyslav Syrotenko}
\affiliation{Tufts University, Dept. of Physics and Astronomy, Medford MA 02155, USA}
\author{Jeremy Wolcott}
\affiliation{Tufts University, Dept. of Physics and Astronomy, Medford MA 02155, USA}

\collaboration{GENIE Collaboration}

\date{\today}

\begin{abstract}

We summarise the results of a study performed within the GENIE global analysis framework, revisiting the GENIE bare-nucleon cross-section tuning and, in particular, the tuning of a) the inclusive cross-section, b) the cross-section of low-multiplicity inelastic channels (single-pion and double-pion production), and c) the relative contributions of resonance and non-resonance processes to these final states. The same analysis was performed with several different comprehensive cross-section model sets available in GENIE Generator v3. 
In this work we performed a careful investigation of the observed tensions between exclusive and inclusive data, and installed analysis improvements to handle systematics in historic data.
All tuned model configurations discussed in this paper are available through public releases of the GENIE Generator. With this paper we aim to support the consumers of these physics tunes by providing comprehensive summaries of our alternate model constructions, of the relevant datasets and their systematics, and of our tuning procedure and results. 

\end{abstract}

\maketitle

\section{Introduction}

GENIE is an international collaboration of scientists working on a global analysis of neutrino scattering data and on the incorporation of modern theoretical inputs and experimental data into robust and predictive semi-empirical comprehensive neutrino interaction simulations.
GENIE develops and maintains a suite of well-known software products for the experimental neutrino community, which includes its popular Generator product~\cite{Andreopoulos:2009rq}.
With the recent release of the GENIE Generator v3, a substantial change in the way that the GENIE Collaboration approaches the process of developing, validating, characterising, tuning and releasing comprehensive neutrino interaction simulations came into sharp focus.
The focus of the GENIE Collaboration has always been the development of universal comprehensive models, handling all probes and targets and simulating all processes across the entire kinematic phase space relevant for neutrino experiments. 
Previously, the GENIE Collaboration released a single, preferred (\emph{default}) comprehensive model that reflected our current understanding on the most predictive, robust, and self-consistent model that could be built out of GENIE neutrino interaction modelling elements.
Whereas many other alternative modelling elements were made available to users, they had to be enabled by individual users through an error-prone procedure that could bring substantial physics and logical inconsistencies, invalidate procedures for addressing double counting issues, and damage the level of agreement with data, often in ways that were unsuspected by users that had a narrow focus on some particular modelling aspect and lacked the GENIE tools and procedures to fully characterise a comprehensive model. 
To address this, and in response to the community demand for \emph{alternative} models, GENIE has released a number of comprehensive model configurations (CMCs) and is in the process of constructing several more. 
All such configurations, that are easily invoked and run out of the box, combine modelling elements in a way that is as consistent as possible, and are validated, characterised and tuned as a whole. 
This important development was underpinned by a substantial upgrade of GENIE capabilities for systematic model validation, model characterization through comparisons to large collections of complementary scattering data with neutrino, charged lepton, and hadron probes, and the development of an advanced global analysis of scattering data.

The GENIE global analysis was made possible through the continued development of curated data archives, and the successful large-scale refactoring and interfacing to the Professor tool~\cite{Professor} of a very extensive set of GENIE codes that implement comparisons to data, within a framework that allows the efficient manipulation of large ensembles of simulated events  produced from a constellation of alternative models. 
The interface to the Professor tool enabled the efficient implementation of complex multi-parameter brute-force scans and removed substantial global analysis limitations by decoupling it from event reweighting procedures that, for all but the most trivial aspects of our physics domain, require substantial development time and are not exact, or even possible at all.
Professor `reduces the exponentially expensive process of brute-force tuning to a scaling closer to a power law in the number of parameters, while allowing for massive parallelisation'~\cite{ProfessorWebPage}.
The Professor package has been extensively used for the tuning of Monte Carlo generators in the collider community.

The above developments allowed the GENIE Collaboration to fulfil its \emph{dual purpose} described in its mission statement: GENIE develops a popular Monte Carlo event generation platform and implements, within its platform, universal and comprehensive physics simulations for lepton scattering, as well as simulations for several Beyond the Standard Model processes. 
But, in addition, and separately from the previous mission, GENIE develops a global analysis of scattering data for the tuning and uncertainty characterization of comprehensive neutrino interaction models.
The GENIE Generator is the main outlet for the GENIE global analysis results, and our goal is that, for each supported comprehensive model, several selected tuned versions shall be made available. 

Typically, nuclear modifications to the cross-section are computed separately, and the decomposition of the total cross-section into the possible exclusive final states proceeds via separate hadronization, intranuclear rescattering and particle decay codes.
Therefore, bare-nucleon cross-section are a crucial first modelling component to tune in the process of building a global fit of all relevant scattering data.
Tunes for several aspects of GENIE modelling, including neutrino-induced hadronization and nuclear cross-sections for low-multiplicity channels, are near completion and will be released and published in the future. 
This paper summarises the results of the first analysis performed within the GENIE global analysis framework, revisiting the GENIE  bare-nucleon cross-section tune and, in particular, the tuning of the empirical non-resonance background contribution to one and two pion final states. 
A similar, albeit much simpler, analysis underpinned the tune of the well known and widely used  comprehensive model that was included as the default model throughout the very long GENIE v2 series of releases.
At that time, not sufficiently explored and understood tensions between inclusive and exclusive data, and an executive decision to anchor the GENIE v2 model on inclusive data, led to some expected and well known discrepancies with exclusive data that were increasingly brought into focus as new experiments started performing increasingly precise measurements of low-multiplicity, exclusive final states~\cite{PRPaper}. 
Here, we perform a careful investigation of the observed tensions between exclusive and inclusive data, retune the bare-nucleon cross-section model for all GENIE comprehensive models available in GENIE v3, and provide best-fit values and correlations for several parameters influencing the GENIE bare-nucleon cross-sections.
The work presented here was based on the model implementations of GENIE v3.0.6 (released on 23 July 2019), and the results of this work will be included in the GENIE v3.2.0 release.
Preliminary versions of this work appeared in earlier releases of the GENIE v3 series (v3.0.0 - v3.0.6).

In Sec.~\ref{sec:BareNucleonXSec}, we summarise relevant aspects of the free nucleon cross-section modelling in GENIE, while in Sec.~\ref{sec:ComprehensiveConfigurations} we provide further details for the construction of comprehensive GENIE models considered in this work.
In Sec.~\ref{sec:model_uncertaintines} we provide details of the datasets, parameterisation of model and data uncertainties for this particular tune. Sec.~\ref{sec:TuningProcedure} describes the tuning procedure as well as the statistical methodology used.
Finally, our tuning results are presented in Sec.~\ref{sec:TuningResults}.

%
%

\section{Bare nucleon cross-section modelling in GENIE}
\label{sec:BareNucleonXSec}

In very simplified terms, neglecting diffractive production, as well as $|\Delta S| = 1$ and $|\Delta C| = 1 $ processes, the total inelastic differential cross section for neutrino scattering off bare nucleons, $d^{2}\sigma^{inel}/dQ^{2}dW$, is computed in GENIE as
\begin{equation}
  \frac{\displaystyle d^{2}\sigma^{\text{inel}}}{\displaystyle dQ^{2}dW} =
  \begin{cases}
    \frac{\displaystyle d^{2}\sigma^{\text{RES}}}{\displaystyle dQ^{2}dW} +
    \frac{\displaystyle d^{2}\sigma^{\text{SIS}}}{\displaystyle dQ^{2}dW} & \text{for} \enskip W < W_{\text{cut}} \\
     \\
    \frac{\displaystyle d^{2}\sigma^{\text{DIS}}}{\displaystyle dQ^{2}dW} & \text{for} \enskip W \ge W_{\text{cut}}\\
   \end{cases} \label{eq:total_inelastic_xsec}
\end{equation}
The term $d^{2}\sigma^{\text{RES}}/dQ^{2}dW$ represents the contribution from all
low multiplicity inelastic channels proceeding via resonance production
and, in present versions of GENIE, it is computed as an {\em incoherent} 
sum over several resonances. 
The resonances included in GENIE v3 are the ones specified by the Rein-Seghal paper~\cite{REIN198179}.
The 9 lightest $N^*$ and the 8 lightest $\Delta$ labeled by the PDG with 3 or 4 stars are considered. 
The following resonances are included in GENIE v3:
$N$(1440), $N$(1520), $N$(1535), $N$(1650), $N$(1675), $N$(1680), $N$(1700), $N$(1720), $N$(1710),
$\Delta$(1232), $\Delta$(1600), $\Delta$(1620), $\Delta$(1700), $\Delta$(1905), $\Delta$(1910), $\Delta$(1920) and $\Delta$(1950).
$W_{cut}$ is a free parameter that determines the end of the SIS region. The nominal value is set to $W_{cut} = 1.7 \, GeV/c^2$.

In the version of GENIE used in this work, there is the option to select one of several neutrino-induced resonance production calculations performed by Rein and Sehgal~\cite{REIN198179}, Kuzmin, Lyubushkin and Naumov~\cite{Kuzmin2006axial,Kuzmin:2003ji}, and Berger and Sehgal~\cite{0709.4378}. 
The last two models are extensions of the first one, that account for nonzero lepton masses.  
Both models are based on the same formalism and the only difference between them is that the latter includes the pion-pole contribution to the hadronic axial current.
The term $d^{2}\sigma^{\text{DIS}}/dQ^{2}dW$ represents the GENIE calculation of the deep inelastic cross-section that, in all relevant GENIE comprehensive model configurations, is carried out using an effective leading order model with the modifications suggested by Bodek and Yang~\cite{BODEK200270} to describe scattering at low momentum-transfers. This model is the foundation of both the DIS model and the SIS model in GENIE.

\begin{figure}
    \centering
    \begin{subfigure}{8cm}
        \includegraphics[width=\textwidth]{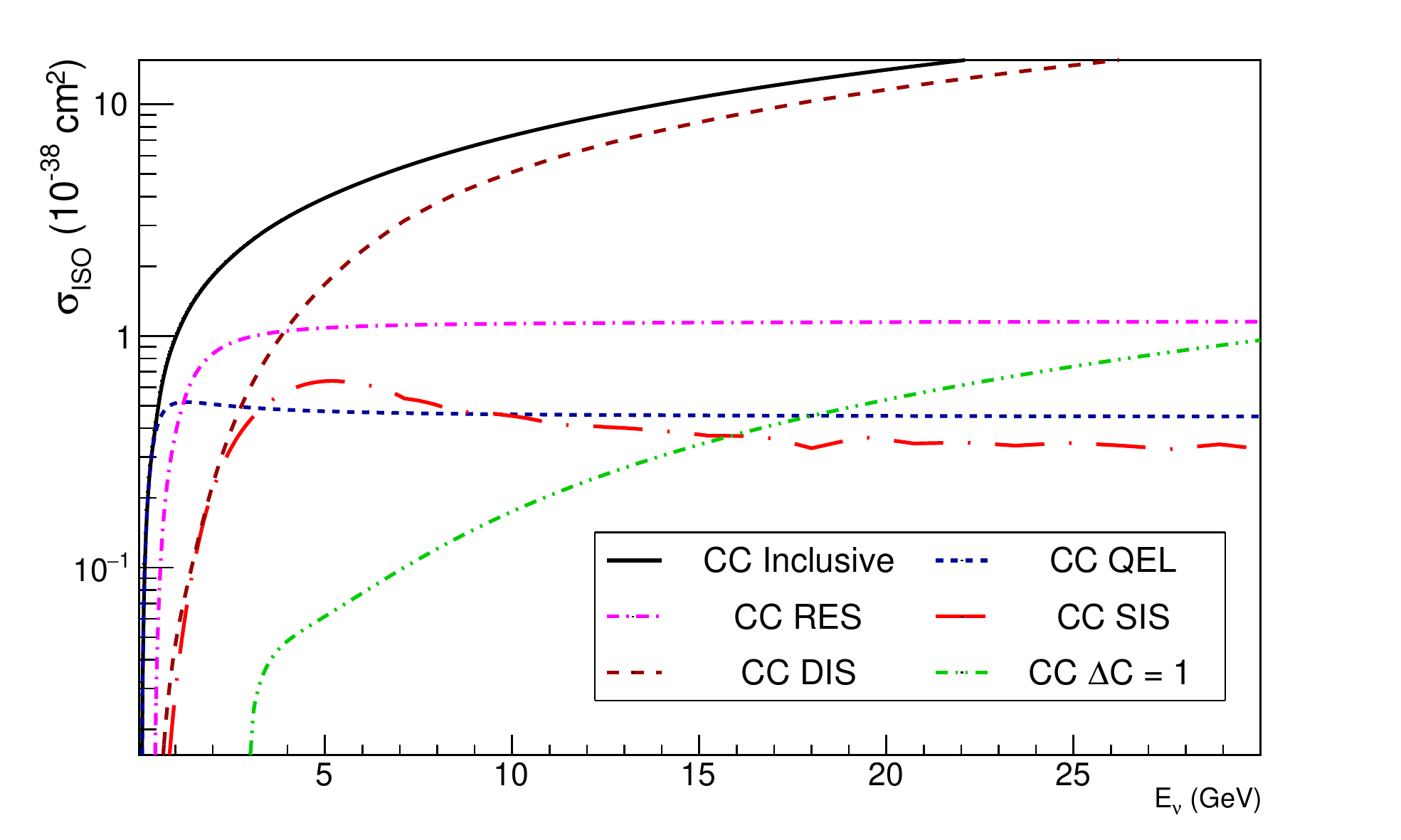}
        \caption{$\nu_{\mu}$ CC on isoscalar targets. }
    \end{subfigure} \,
    \begin{subfigure}{8cm}
        \includegraphics[width=\textwidth]{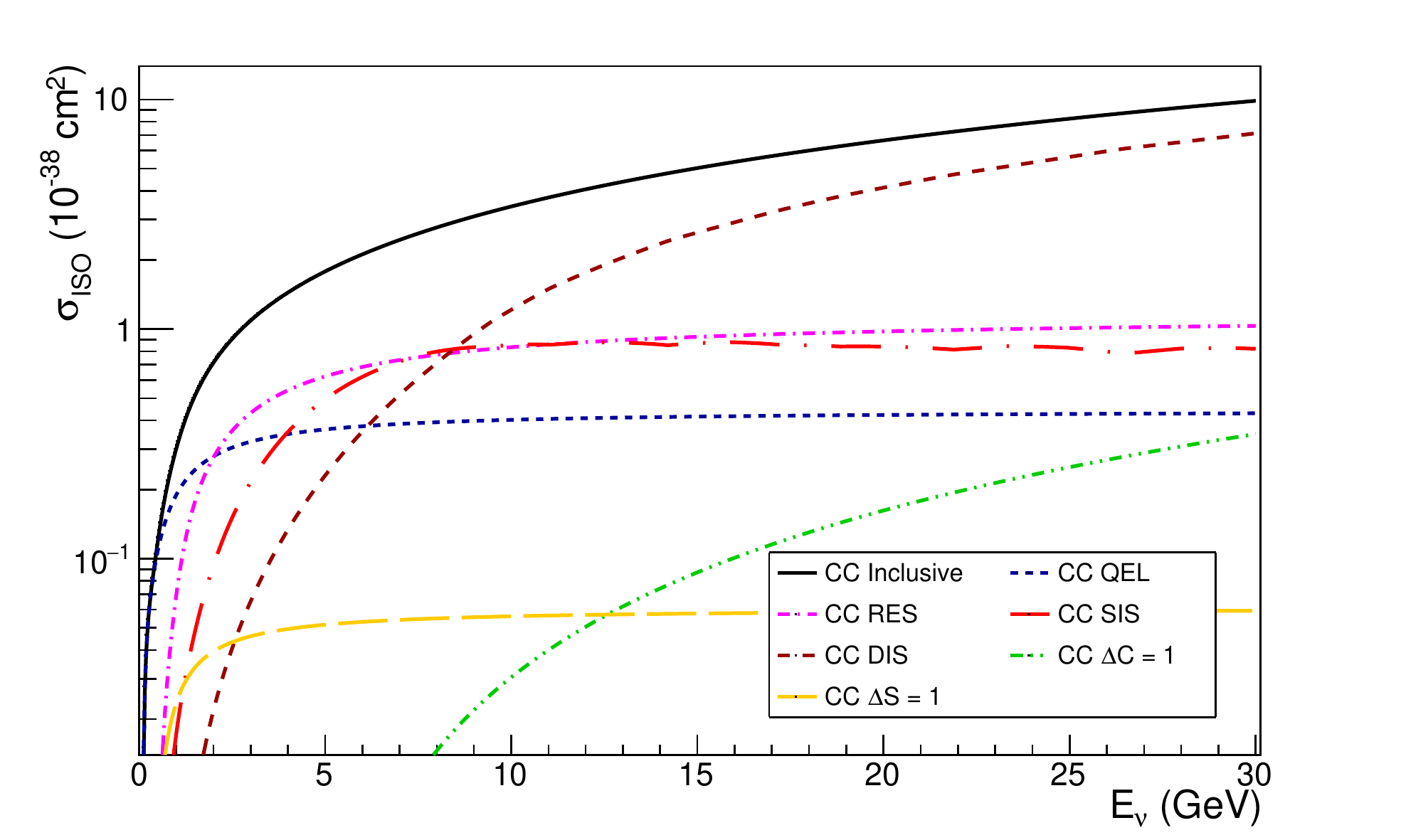}
        \caption{$\bar{\nu}_{\mu}$ CC on isoscalar targets. }
    \end{subfigure} 
    \caption{ Summary of neutrino and anti-neutrino cross sections on isoscalar targets.  } 
    \label{fig:SummaryXSec}
\end{figure}

The term $d^{2}\sigma^{\text{SIS}}/dQ^{2}dW$ requires some elaboration. 
It represents the cross-section contribution from non-resonance shallow inelastic scattering in the resonance region. 
In GENIE, this cross section is computed with an empirical model where the Bodeck and Yang inclusive deep inelastic cross section is extrapolated into the resonance region and it is decomposed, via the GENIE AGKY~\cite{Yang:2007zzt} hadronization model, into the cross sections for different hadronic multiplicity channels. 
The extrapolation of the DIS model down to the inelastic threshold, $W<W_{\text{cut}}$, includes, on average, the effect of the resonances~\cite{Mandula:1969ce}. 
Notice that, even though the Bodeck and Yang model is capable of describing the inclusive cross section at inelastic threshold, we prefer to utilize an explicit resonance model. 
The contribution for hadronic multiplicities 2 and 3, that are responsible for producing many final states similar to those produced via resonance excitation, are tuned to remove double counting. 
This tuning is the main topic of this work.

The non-resonance shallow inelastic scattering cross section
can be written as
\begin{eqnarray}
   \frac{d^{2}\sigma^{\text{SIS}}}{dQ^{2}dW} =
   \frac{d^{2}\tilde{\sigma}^{\text{DIS}}}{dQ^{2}dW} & \cdot {\Theta}(W_{\text{cut}}-W) \nonumber\\
   & \cdot {\sum\limits_m}f_{m}(Q^{2},W) 
\end{eqnarray}
where $\tilde{\sigma}^{\text{DIS}}$ represents the extrapolated deep inelastic
cross section into the resonance region, and $m$ refers to the multiplicity of the hadronic system. The factor $f_{m}$ relates the total calculated DIS cross section to the DIS contribution to this particular multiplicity channel.
These factors are computed as 
\begin{equation}
f_{m}(Q^2,W) = R_{m} P^{\text{had}}_{m}(Q^2,W) 
\label{eq:R_parameters_definition}
\end{equation}
where $R_{m}$ is an adjustable parameter and $P^{\text{had}}_{m}$ is the probability, taken from the GENIE hadronization model, that the DIS final state hadronic system multiplicity would be equal to $m$,
\begin{equation}
  P^{\text{had}}_{m}(Q^2,W) = \frac{1}{ \langle m \rangle} \psi\left(\frac{m}{ \langle m \rangle}\right) 
\end{equation}
The average hadronic multiplicity $ \langle m \rangle $ is computed, for each value of hadronic invariant mass W, by
\begin{eqnarray}
  \langle m \rangle (Q^2,W) = \alpha & + \beta \ln \left( \frac{W^2}{\text{GeV}^2\text{/c}^4} \right) \nonumber\\
                                     & + \beta'\ln \left( \frac{Q^2}{\text{GeV}^2\text{/c}^2} \right) 
\end{eqnarray}
and the function $\psi$ has the following asymptotic form 
\begin{equation}
  \psi(z) = \frac{2e^{-c}c^{cz+1}}{\Gamma(cz+1)}
\end{equation}
In the above expressions, $\alpha$, $\beta$, $\beta'$ and $c$ are adjustable parameters.
In principle, $\alpha$, $\beta$, $\beta'$ $c$ and $R_{m}$, are different for each initial state ($\nu+p$, $\nu+n$, $\bar{\nu}+p$, $\bar{\nu}+n$)
and are different for charged current and neutral current interactions.
A new tune of the neutrino-induced hadronization models in GENIE is
currently in progress and, in future, it may be possible to perform a
joint tuning of the GENIE cross section and hadronization modeling
components for bare-nucleon targets. However, at this present work, 
the parameters $\alpha$, $\beta$, $\beta'$ and $c$ were kept at the default values of the AGKY model in GENIE v3. For easy reference, the relevant values
for the channels studied in this work are included in Tab.~\ref{tab:agkyParameters}. No dependence on $Q^2$ has been observed in $\nu$ and $\bar{\nu}$ scattering data~\cite{GRASSLER1983269}, hence  $\beta'=0$ for all channels.

\begin{figure*}
    \centering
    \includegraphics[width=0.8\textwidth]{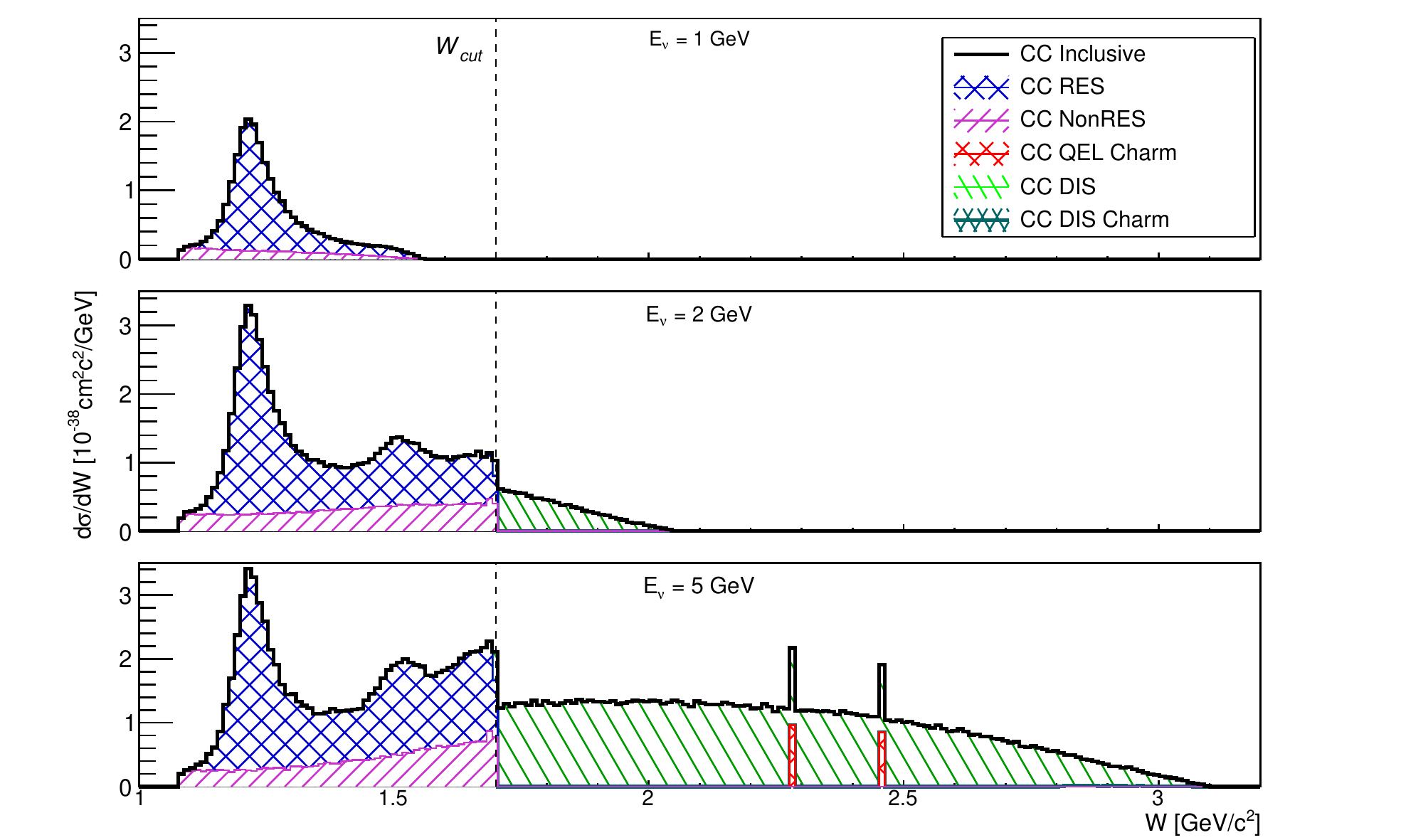}
    \caption{RES and SIS/DIS differential cross section as a function of the invariant mass for three different neutrino energies using a $1/E$-like flux. $W_{\text{cut}}$ divides the SIS and the DIS regions.}
    \label{fig:sisxsec_nenergy}
\end{figure*}

For most inelastic processes simulated in neutrino-nucleus scattering by all current GENIE comprehensive model configurations, the total inelastic differential cross section for scattering off bare nucleons takes centre stage. 
In Fig.~\ref{fig:SummaryXSec}, the contribution to the $\nu_\mu$ CC and $\bar{\nu}_\mu$ CC inclusive cross sections on isoscalar targets in GENIE is shown for the different interaction processes. 
The CC RES and SIS/DIS CC cross-section contribution for different neutrino energies is shown in Fig.~\ref{fig:sisxsec_nenergy}.
 
\begin{table}   
    \centering
    \begin{tabular}{@{\extracolsep\fill} c c c c c c} \hline\hline\noalign{\smallskip}
              && \multicolumn{4}{c} {Initial state} \\ 
    Parameter && $\nu_\mu p$ & $\nu_\mu n$ & $\bar{\nu}_\mu p$ & $\bar{\nu}_\mu n$ \\     \noalign{\smallskip}\hline\noalign{\smallskip}
    $\alpha$  && 0.40 & -0.20 & 0.02 & 0.80 \\
    $\beta$   && 1.42 &  1.42 & 1.28 & 0.95 \\
    $c$       && 7.93 &  5.22 & 5.22 & 7.93 \\   \noalign{\smallskip}\hline\hline
    \end{tabular}
     \caption{Relevant default GENIE v3 AGKY parameters for $\nu_\mu$ and $\bar{\nu}_\mu$ CC interactions on proton and neutron. }
    \label{tab:agkyParameters}
\end{table}

%
%

\section{Comprehensive model configurations in GENIE v3}
\label{sec:ComprehensiveConfigurations}

GENIE has a large degree of configuration: for each process (RES, DIS, etc.) the system offers a number of alternative models to be used for event generation. 
In previous GENIE releases, only one model-process mapping was suggested by the out-of-the-box configuration, despite the availability of alternative models.
Yet, there was no guidance on how to correctly use different configurations according to author and developers.
In fact, processes are not universal and their definitions are generator dependent.
Hence, it was easy to come up with inconsistencies between the model configuration for different processes that were supposed to be used together to get a correct comprehensive physics simulation. 

This issue was addressed in GENIE v3 by introducing the concept of comprehensive model configuration (CMC) that is a consistent process-model association.
Considering that GENIE already has about 20 different processes only for neutrinos, CMC definitions are quite complex objects and they need to be effectively named so that the community can use them unambiguously. 
For this purpose, the collaboration developed a specific naming convention, see Sec.~\ref{subsec:CMCNamingConvention}.
Sec.~\ref{sec:CMCs_in_v3} describes the models used in CMCs relevant for neutrino interactions.

\subsection{Comprehensive model configuration naming convention}
\label{subsec:CMCNamingConvention}
A comprehensive model configuration is identified by at least 7-character string in the form:
\begin{equation*}
  \text{Gdd\_MMv}
\end{equation*}
where
\begin{itemize}
  \item G is a capital letter string of arbitrary length that identifies the authors of the tune (GENIE).
  \item dd is a number describing the year during which the model configuration was first developed. 
  \item MM is a number (00, 01, 02, ...) identifying a family of model configurations. 
  \item v is a character (a, b, c, ...) enumerating different members of the given family of model configurations.
\end{itemize}

Once a comprehensive model configuration is defined, a number of different tunes may be produced. 
These may be produced, for example, by a) incorporating different combinations of experimental data, b) considering variations in different combinations of our modelling elements (e.g. bare-nucleon cross sections, nuclear model and nuclear cross sections, neutrino-induced hadronization etc), c) considering different degrees of freedom (different parameterisations) for the variation of each model, or d) incorporating different parameter priors and/or different strategies for eliminating nuisance parameters.
A tune is identified by the model configuration name, and additional information enumerating the parameters and datasets. This is at least 14-character string in the form:

\begin{equation*}
 \text{Gdd\_MMv\_PP\_xxx} 
\end{equation*}
where
\begin{itemize}
  \item Gdd\_MMv describes the model configuration (see above). 
  \item PP is a number identifying the set of tuned parameters. This parameter set is defined uniquely only in the context of a particular model configuration. 
  \item xxx is a number that identifies the dataset used for the model configuration tuning. This may include a unique set of weights associated with each component dataset. 
\end{itemize}

\subsection{CMCs available in GENIE v3}
\label{sec:CMCs_in_v3}

Several CMCs are available in GENIE v3, but they can be grouped together as their scopes are common. 
The first group of CMCs is historically motivated: it is based on the \emph{default} configuration and simply provides updates for processes that were introduced later. 
The second family is an improvement of the first group, in terms of the resonance model. 
The third one was constructed aiming to deliver the most up to date theoretical nuclear matter simulations.
Out of these main ideas, a number of CMCs can be constructed simply changing minor aspects like FSI or form factors. 
Here, we briefly summarise the modeling components used in each comprehensive model configuration available in GENIE v3.

\subsubsection{G18\_01a, G18\_01b,  G18\_01c and G18\_01d}
 These comprehensive models share an identical cross-section
    model construction, which is an adiabatic update of the 
    historical default cross-section model of GENIE v2, now named as G00\_00a CMC.
    For interactions on nucleons and nuclei, it relies on implementations of the following models: 
        the Ahrens model~\cite{Ahrens} for NC elastic,
        the Llewellyn Smith model~\cite{LLEWELLYNSMITH1972261} for CC quasi-elastic,
    the Rein-Sehgal model~\cite{REIN198179} for NC and CC resonance production, the Rein-Sehgal model~\cite{Rein:1982pf} for NC and CC coherent pion production, the Bodek-Yang model~\cite{BODEK200270} for NC and CC deep inelastic scattering and non-resonance shallow inelastic scattering, the Kovalenko model~\cite{Kovalenko:1990zi} for quasi-elastic charm production, and the Aivazis-Olness-Tung slow rescaling model~\cite{aivazis1993nexttoleading} for deep inelastic charm production.
    Nuclear cross sections are calculated within the framework of a relativistic Fermi gas model, following the approach of Bodek-Ritchie~\cite{PhysRevD.23.1070}.
    Multi-nucleon processes in neutrino scattering off nuclear targets can be optionally enabled and simulated via an empirical GENIE model~\cite{Andreopoulos:2009rq}.
    In addition, in GENIE v3, the adiabatic upgrade of the historical comprehensive model includes the simulation of processes that, previously, were either optional or missing. 
    This includes both diffractive pion production based on an implementation of the Rein model~\cite{REIN198661},
    and quasi-elastic $|\Delta S|=1$ hyperon ($\Lambda^0$, $\Sigma^-$, $\Sigma^0$) production based on the Pais model~\cite{PAIS1971361}. 
    Single Kaon production, although optionally available for neutrinos in GENIE v3~\cite{RafiAlam:2010kf}, is not yet available for antineutrinos and inclusion in any published GENIE comprehensive configurations was postponed till an antineutrino implementation is available and the Kaon content of hadronic showers produced by GENIE has been retuned following the addition of the single-Kaon generator. 
    Both G18\_01a and G18\_01b comprehensive models employ a revised resonance decay algorithm and an implementation of the AGKY~\cite{Yang:2007zzt} hadronization model that is unchanged with respect to that used at the latest releases of GENIE v2 series. 
    Four comprehensive model variations are constructed by attaching different intranuclear hadron transport models to the same underlying cross section and hadronization models~\cite{dytman2021comparison}.
    G18\_01a uses an updated INTRANUKE hA effective intranuclear rescattering model which is unique to GENIE, G18\_01b uses the new INTRANUKE hN model implementing a full intranuclear cascade including medium corrections, G18\_01c uses an interface to the GEANT4~\cite{GEANT4} Bertini intranuclear cascade~\cite{Bertini} (version 4.10.2) and G18\_01d uses an interface to the INCL++ (version 5.2.9.5) implementation of the Li\`{e}ge intranuclear model~\cite{Mancusi:2014eia}.
    
\subsubsection{G18\_02a, G18\_02b, G18\_02c and G18\_02d}
This is family of empirical models which is an evolved version of the G18\_01[a-d] ones. The general construction of the cross-section model is similar to the one discussed above, with the exception that the implementations of the Rein-Sehgal models for CC and NC resonance neutrino-production, as well as for CC and NC coherent production of mesons, were replaced with updated models by Berger-Sehgal~\cite{0709.4378}.
    Similarly to G18\_01[a-d], four comprehensive model variations are constructed by using alternative intranuclear hadron transport models on top of the same underlying cross section and hadronization models
    (a: INTRANUKE/hA, b: INTRANUKE/hN, c: GEANT4/Bertini, and d: INCL++).
    
\subsubsection{G18\_10a, G18\_10b, G18\_10c and G18\_10d}
This is a family of models derived 
    from the improved empirical ones (G18\_02[a-d]) described above, 
    by substituting both the Llewellyn Smith CC quasi-elastic model~\cite{LLEWELLYNSMITH1972261} and GENIE's empirical multinucleon model with implementations of the corresponding Valencia models by Nieves et al.~\cite{Nieves}.
    This family of comprehensive models provides a firmer theoretical basis for the simulation of neutrino-nucleus scattering around the quasi-elastic peak.
    Within this family of models, the nuclear environment is modelled using a Local Fermi Gas, matching the inputs used for the published Valencia calculations. 
    Again, four comprehensive model variations (a-d) are constructed by using alternative intranuclear hadron transport models, following the same naming convention introduced above.
    The implementation of the Valencia model in GENIE does not predict the kinematics of the outgoing hadrons and its description needs to be accompanied by one of the FSI models available in GENIE (a-d)~\cite{Schwehr2016pvn}.
    
\subsubsection{G18\_10i, G18\_10j, G18\_10k and G18\_10l}
    These comprehensive models are derived from the G18\_10[a-d] ones, by replacing the dipole axial form factor, used in the calculations of quasi-elastic cross sections, with the better-motivated z-expansion model~\cite{Meyer:2016oeg}, providing a richer set of degrees of freedom for parameterizing quasi-elastic model uncertainties. As in all previous families of models, four comprehensive model variations (i-l) are constructed by using alternative intranuclear hadron transport models, though the labelling is now different
    (i: INTRANUKE/hA, j: INTRANUKE/hN, k: GEANT4/Bertini, and l: INCL++).

\subsection{Free nucleons and CMCs}

Although a large number (16) of CMCs was summarised above, 
with respect to the cross sections for (anti)neutrino scattering off bare nucleons, there are only two different model constructions:
The one used in a) G18\_01[a-d], and the one used in b) G18\_02[a-d], G18\_10[a-d] and G18\_10[i-l]. The main difference between these two model constructions resides mainly in the treatment of the lepton mass. 
Although some differences can be expected between G18\_10[a-d] and G18\_10[i-l], because of different choices in the modelling of the axial form factor for quasi-elastic scattering, they do not manifest themselves in the context of this particular analysis.

Several variations of the tuning procedure were run and evaluated for testing purposes, before converging to the procedure presented in this paper. 
Preliminary versions of this work were released in the GENIE v3 series (v3.0.0-v3.0.6) in a series of tunes carrying the 02\_11a label.
The final results presented in the paper will be made available in GENIE v3.2 in a series of 16 tunes,
one for each of the 16 comprehensive model configurations summarised above, labelled as 02\_11b.
For example, the tune G18\_10a\_02\_11b corresponds to the G18\_10a comprehensive model with the parameters determined through the 
tuning procedure discussed in this paper (02\_11b). The GENIE tune naming convention was discussed in an earlier section.
A full list of GENIE tunes is maintained in \url{http://tunes.genie-mc.org}.  
The preliminary versions (02\_11a) of the tunes will be kept in GENIE v3.2, but they will be phased out in subsequent minor releases.

%
%

\section{Data and model uncertainties review}
\label{sec:Critial_review}
The data used in this analysis are old and a careful review of the past analysis procedure is required in order to combine all the data together in a global analysis. 
This section summarises the data details and how the models used in the fit behave in the same energy region. 

\subsection{Datasets included in the fit and their systematics}
\label{subsec:datasetifno}

In the current work, we consider Hydrogen and Deuterium data from the 
ANL 12FT, BNL 7FT, FNAL 15FT and BEBC bubble chamber experiments.
The data represent integrated cross sections for different 
incoming neutrino energy bins for
\begin{itemize}
  \item $\nu_\mu$ and $\bar{\nu}_\mu$ CC inclusive scattering \cite{Barish:1977qk, Bosetti:1977nd, Bosetti:1981ip, Baltay:1980pr, Seligman:1997fe, Jonker:1980vf, Kitagaki:1986ct, Eichten:1973cs, Morfin:1981kg, Vovenko:2002ry, Lyubushkin:2008pe, Adamson:2009ju, Barish:1978pj, Colley:1979rt, Parker:1983yi, Baker:1982ty, MacFarlane:1983ax, Allaby:1987bb, Baker:1982jf, Ciampolillo:1979wp, Asratian:1984ir, Anikeev:1995dj, Baranov:1978sx, Nakajima:2010fp, Fanourakis:1980si, Taylor:1983qj, Erriquez:1979nb, Asratian:1978rt}
  \item $\nu_\mu$ and $\bar{\nu}_\mu$ CC quasi-elastic scattering \cite{Barish:1977qk, Lyubushkin:2008pe, Fanourakis:1980si, Mann:1973pr, Allasia:1990uy, Kitagaki:1983px, Belikov:1981ut, Brunner:1989kw, Baker:1981su, Bonetti:1977cs, Belikov:1985mw, Armenise:1979zg}
  \item $\nu_\mu$ and $\bar{\nu}_\mu$ CC single-pion production \cite{Campbell:1973wg, Radecky:1981fn, Wilkinson:2014yfa, Lerche:1978cp, Allen:1980ti, Bell:1978rb, Allen:1985ti, Allasia:1990uy, Barish:1979ny, Kitagaki:1982dx} 
  \begin{itemize}
      \item $\nu_\mu+n\rightarrow \mu^-+n+\pi^+$
      \item $\nu_\mu+p\rightarrow \mu^-+p+\pi^+$
      \item $\nu_\mu+n\rightarrow \mu^-+p+\pi^0$
      \item $\bar{\nu}_\mu+p\rightarrow \mu^++p+\pi^-$
      \item $\bar{\nu}_\mu+n\rightarrow \mu^++n+\pi^-$ 
  \end{itemize}
  \item $\nu_\mu$ CC two-pion production \cite{Day:1984nf}
    \begin{itemize}
      \item $\nu_\mu+p\rightarrow \mu^-+n+2\pi^+$
      \item $\nu_\mu+p\rightarrow \mu^-+p+\pi^++\pi^0$
      \item $\nu_\mu+p\rightarrow \mu^-+n+\pi^++\pi^-$
    \end{itemize}
\end{itemize}

\begin{figure*}
    \centering
    \begin{subfigure}{7.2cm}
        \centering\includegraphics[width=\columnwidth]{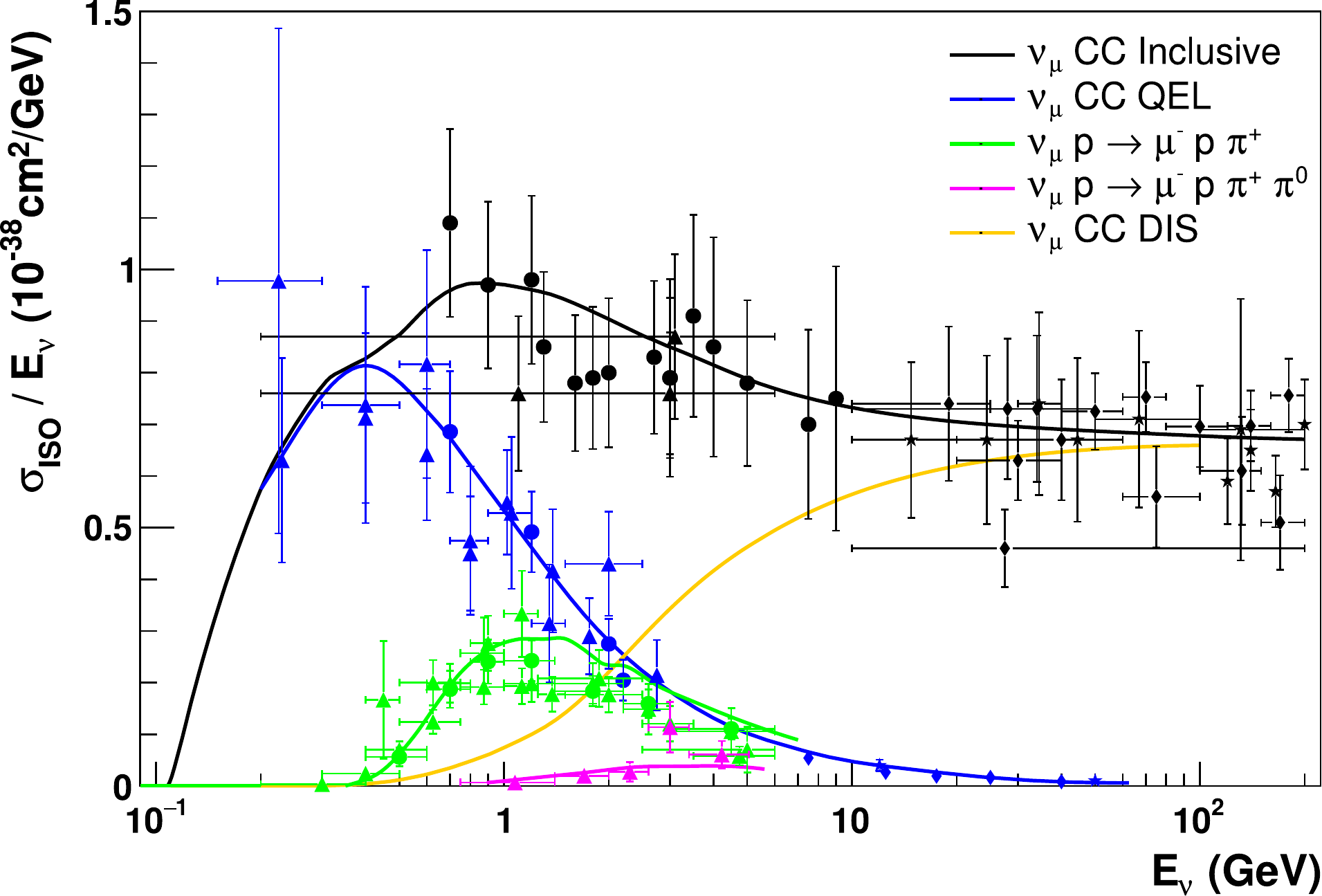}
        \caption{ $\nu_\mu$ CC cross section.   }   
    \end{subfigure} \,\,\,\,\,\,
    \begin{subfigure}{8cm}
        \centering\includegraphics[width=\columnwidth]{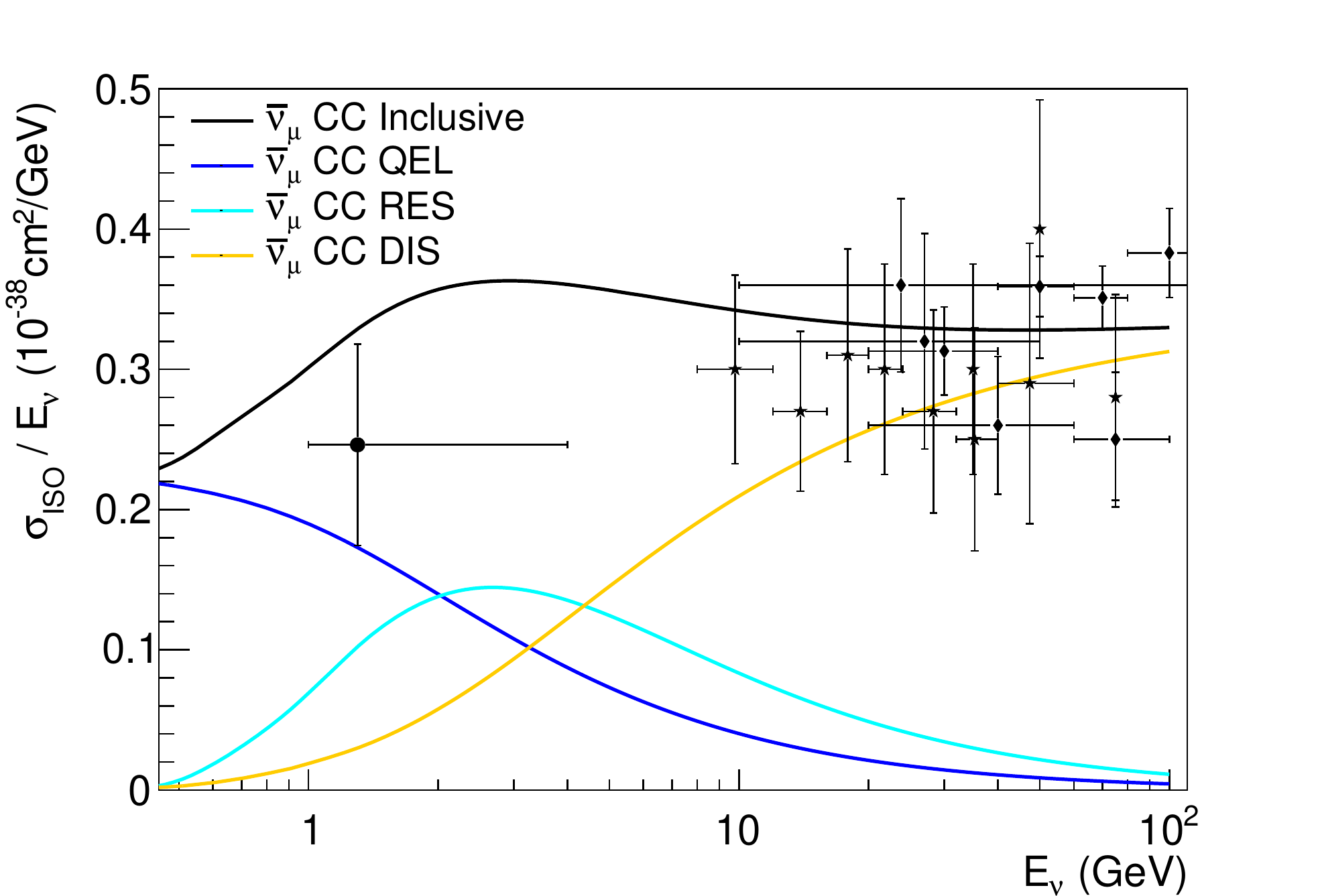}
        \caption{$\bar{\nu}_\mu$ CC cross section. }   
    \end{subfigure}
\caption{Charged current cross section on isoscalar targets as a function of the incoming neutrino energy. Breakdown of quasi-elastic, one and two pion production and deep inelastic processes is shown. The predictions are computed using the G18\_02a\_00\_000 configuration. The data on hydrogen and deuterium targets from Tab.~\ref{tab:summary_data} is shown if available from ANL 12FT ($\bigtriangleup$), BNL 7FT ($\bullet$), BEBC ($\diamond$) and FNAL ($\star$). }
\label{fig:summary_data}
\end{figure*}

Not all of the available historical data has been used for the fit, as some datasets were superseded or reanalysed, as in the case of ANL 12FT and BNL 7FT, datasets. 
The latest analysis are used.
A detailed summary of the datasets used in the fit is shown in Tab.~\ref{tab:summary_data} and in Fig.~\ref{fig:summary_data}. 
Some of the datasets included in the tune consider Hydrogen-Neon mixtures. 
The nuclear effects of the neon in the target mixture are shown to be negligible~\cite{PhysRevD.18.3905}.

Low energy bins have a higher contribution to the $\chi^2$ due to energy smearing and lack of unfolding in measurements. 
Hence, data points with $E_\nu < 0.5$~GeV are removed from the fit. 
In total, the tune is performed with 169 data points from bubble chamber experiments. 
Different analysis methods were implemented in each experiment, such as cuts applied on the $W$ invariant mass, the outgoing muon momentum or the total longitudinal momentum of the final state.
The associated GENIE prediction has been corrected by applying the same cuts to the generated events. 
Moreover, datasets from the same experiments are not independent as they share the same neutrino flux, detector, analysis methodology, etc.
Although it is clear that some correlated uncertainties exist, the data releases do not contain any information about the correlation between them. 
In the GENIE database, we added a systematic error to the datasets of 15\%.
The methodology used to include them in the fit is detailed in Sec.~\ref{subsec:Priors}.  
Other free nucleon data on heavier targets are available but used only for comparison with the GENIE prediction. 
No correction for nuclear effects is considered for deuterium targets. 

\begin{table*}
    \centering
    \begin{tabular}{@{\extracolsep\fill} c c c c c c}  \hline\hline\noalign{\smallskip}
    \textbf{Experiment}& $\mathbf{N_{p}}$ & 
    \textbf{Energy [GeV]} & \textbf{Target} & \textbf{Cuts} & 
    \textbf{Ref.} \\  \noalign{\smallskip}\hline\noalign{\smallskip}
    \multicolumn{6}{c}{$\nu_\mu + N \rightarrow \mu^- X$} \\ 
    \noalign{\smallskip}\hline\noalign{\smallskip}
     BNL 7FT  & 13 & 0.6-10 & $^2$H  &  & \cite{Baker:1982ty} \\
     BEBC     &  3 & 10-50 & H-$^{10}$Ne &  & \cite{Colley:1979rt}  \\
     FNAL     & 6 & 10-110  & $^2$H  &  & \cite{Kitagaki:1982dx}  \\ 
              & 5 & 100-110 & H-$^{10}$Ne &  & \cite{Baker:1982jf}  \\ 
     \noalign{\smallskip}\hline\noalign{\smallskip}
     \multicolumn{6}{c}{$\bar{\nu}_\mu + N \rightarrow \mu^+ X$} \\ 
     \noalign{\smallskip}\hline\noalign{\smallskip}
     BEBC     & 3 & 11-110 & $^1$H-$^{10}$Ne &    & \cite{Bosetti:1977nd}  \\ 
              & 1 & 10-50 & $^1$H-$^{10}$Ne  &    & \cite{Colley:1979rt}   \\ 
              & 6 &  30-110 & $^1$H-$^{10}$Ne &  & \cite{Bosetti:1981ip}  \\ 
              & 1 & 10-110 & $^1$H-$^{10}$Ne &    & \cite{Parker:1983yi}  \\ 
     BNL 7FT  &  1 & 1-4 &  $^1$H &  &  \cite{Fanourakis:1980si} \\
     FNAL     & 5 & 10-110 & $^2$H$-^{10}$Ne &    &  \cite{Asratian:1984ir} \\
              & 7 & 10-80 & $^2$H$-^{10}$Ne &    &  \cite{Taylor:1983qj} \\
     \noalign{\smallskip}\hline\noalign{\smallskip}
     \multicolumn{6}{c}{$\nu_\mu n \rightarrow \mu^- n \pi^+$} \\ 
     \noalign{\smallskip}\hline\noalign{\smallskip}
     ANL 12FT      &  5 & 0.3-2 & $^1$H and $^2$H  &    &  \cite{Radecky:1981fn} \\
     ANL 12FT,ReAna  &  7 & 0.3-3 & $^2$H  &  &  \cite{Wilkinson:2014yfa} \\ 
     BNL 7FT,ReAna  & 11 &  0.1-4 & $^2$H  &  &   \cite{Wilkinson:2014yfa} \\ 
     \noalign{\smallskip}\hline\noalign{\smallskip}
     \multicolumn{6}{c}{$\nu_\mu p \rightarrow \mu^- p \pi^+$} \\ 
     \noalign{\smallskip}\hline\noalign{\smallskip}
     ANL 12FT,ReAna   &  8 & 0-1.6 &$^2$H &  &  \cite{Wilkinson:2014yfa} \\
     BNL 7FT,ReAna   &  7 & 0-7 &$^2$H &  &  \cite{Wilkinson:2014yfa} \\ 
     BEBC   & 7 & 1-30 & $^1$H & $W<1.4$ GeV   &  \cite{Allen:1980ti}  \\ 
            & 6 & 5-100 & $^2$H  & $W<2$ GeV   &  \cite{Allasia:1990uy}  \\ 
            & 5 & 10-80 & $^1$H & $W<2$ GeV   &  \cite{Allen:1985ti}  \\ 
     FNAL   &  3 & 10-30 & $^1$H  & $W<1.4$ GeV   & \cite{Bell:1978qu}    \\
     \noalign{\smallskip}\hline\noalign{\smallskip}
     \multicolumn{6}{c}{$\nu_\mu n \rightarrow \mu^- p \pi^0 $}  \\ 
     \noalign{\smallskip}\hline\noalign{\smallskip}
     ANL 12FT      &   5 & 0.2-2 & $^1$H and $^2$H  &    &  \cite{Radecky:1981fn} \\
     ANL 12FT,ReAna &   7 &  0.2-2 & $^2$H &  & \cite{Wilkinson:2014yfa}\\
     BNL 7FT,ReAna  &  10 & 0.4-3 & $^2$H &  &  \cite{Wilkinson:2014yfa}  \\ 
     \noalign{\smallskip}\hline\noalign{\smallskip}
     \multicolumn{6}{c}{$\nu_\mu p \rightarrow \mu^- n \pi^+ \pi^+$} \\ 
     \noalign{\smallskip}\hline\noalign{\smallskip}
     ANL 12FT & 5 &  1-6 & $^2$H &    &  \cite{Day:1984nf}  \\
     \noalign{\smallskip}\hline\noalign{\smallskip}
     \multicolumn{6}{c}{$\nu_\mu p \rightarrow \mu^- p \pi^+ \pi^0 $}  \\ 
     \noalign{\smallskip}\hline\noalign{\smallskip}
     ANL 12FT & 5 & 1-6 & $^2$H  &    & \cite{Day:1984nf} \\
     \noalign{\smallskip}\hline\noalign{\smallskip}
     \multicolumn{6}{c}{$\nu_\mu n \rightarrow \mu^- p \pi^+\pi^-$} \\ 
     \noalign{\smallskip}\hline\noalign{\smallskip}
     ANL 12FT &  5 &  8-6 & $^2$H &   & \cite{Day:1984nf} \\
     BNL 7FT  & 10 & 0-20 &  $^2$H &    & \cite{Kitagaki:1986ct} \\ 
     \noalign{\smallskip}\hline\noalign{\smallskip}
     \multicolumn{6}{c}{$\bar{\nu}_\mu p \rightarrow \mu^+ p \pi^- $} \\ 
     \noalign{\smallskip}\hline\noalign{\smallskip}
     FNAL     &  1 &  5-70 & $^1$H & $W<1.9$ GeV   & \cite{Barish:1979ny}   \\ 
     \noalign{\smallskip}\hline\noalign{\smallskip}
     \multicolumn{6}{c}{$\nu_\mu + n \rightarrow \mu^- + p$} \\ 
     \noalign{\smallskip}\hline\noalign{\smallskip}
     ANL 12FT & 7 & 0-2  & $^2$H &    & \cite{Mann:1973pr} \\ 
              & 8 & 0-2 &  $^1$H and $^2$H &    & \cite{Barish:1977qk} \\ 
     BNL 7FT   &  4 & 0.2-2 & $^2$H &    & \cite{Baker:1981su}  \\ 
     BEBC      &  5 & 20-40 & $^2$H  &   & \cite{Allasia:1990uy} \\
     FNAL      &  2 & 0-50 & $^2$H &    &  \cite{Kitagaki:1983px} \\
     \noalign{\smallskip}\hline\hline

    \end{tabular}
    \caption{A summary of cross-section data used in this work. The number of data points released by each analysis ($N_p$), the neutrino energy range covered $E_\nu$, the type of target and the cuts applied in the analysis procedure are specified in the table. }
    \label{tab:summary_data}
\end{table*} 

\subsection{Model uncertainties}
\label{sec:model_uncertaintines}
The SIS cross section is tuned within the CMCs using either the Rein-Sehgal or Berger-Sehgal resonance models, see Sec.~\ref{sec:CMCs_in_v3}. 
The tuning main goal is the best value estimation for nine of the parameters that drive the GENIE predictions in the SIS region.
These parameters are the $W_{\text{cut}}$ as defined in Eq.~\ref{eq:total_inelastic_xsec}, the four $R_m$ coefficients for CC interactions on neutron/proton with $m=2,3$ from of the SIS region (Eq.~\ref{eq:R_parameters_definition}), the axial masses used in the dipole form factors for RES and QE interactions, and 2 global scaling factors for the RES cross section and the DIS cross section. 
For clarity, we will refer to $R_{m}$ parameters with the number of pions in the final state, namely $R_{\nu p}^{\text{CC}1\pi}$,$R_{\nu p}^{\text{CC}2\pi}$, $R_{\nu n}^{\text{CC}1\pi}$ and $R_{\nu n}^{\text{CC}2\pi}$.
Tab.~\ref{tab:fitParameters} summarises the parameter pre-fit values and the allowed ranges. 
Previous fits to data are taken into account for the determination of the ranges~\cite{Kuzmin,0712.4384}.

\begin{table*}    
    \centering
    \begin{tabular}{c c c c c c}  \hline\hline\noalign{\smallskip}
    Parameter & GENIE parameter name & \emph{Default} value & Min value  & Max value & Prior \\ 
    \noalign{\smallskip}\hline\noalign{\smallskip}
    $W_{\text{cut}}$~(GeV/c$^2$)   & \tt{Wcut} & 1.7   & 1.5 & 2.3 & \\
    $M_A^{\text{QE}}$~(GeV/c$^2$)  & \tt{QEL-Ma} & 0.999 & 0.75 & 1.10 & $1.014\pm0.014$ \cite{QELMA}\\
    $M_A^{\text{RES}}$~(GeV/c$^2$) & \tt{RES-Ma} & 1.12  & 0.8 & 1.3 & $1.12 \pm 0.03$ \cite{Kuzmin}\\
    $R_{\nu p}^{\text{CC}1\pi}$  & \tt{DIS-HMultWgt-vp-CC-m2} & 0.10  & 0.0 & 0.4& \\
    $R_{\nu p}^{\text{CC}2\pi}$  & \tt{DIS-HMultWgt-vp-CC-m3} & 1.00  & 0.0 & 2.0 & \\
    $R_{\nu n}^{\text{CC}1\pi}$  & \tt{DIS-HMultWgt-vn-CC-m2} & 0.30  & 0.0 & 0.35 & \\
    $R_{\nu n}^{\text{CC}2\pi}$  & \tt{DIS-HMultWgt-vn-CC-m3} & 1.00  & 0.8 & 3.0 & \\
    $S_{\text{RES}}$             & \tt{RES-CC-XSecScale} & 1.0 & 0.6 & 1.2 & \\
    $S_{\text{DIS}}$             & \tt{DIS-CC-XSecScale} & 1.032   & 0.9 & 1.15 & $1\pm0.05$ \\
     \noalign{\smallskip}\hline\hline\noalign{\smallskip}
    \end{tabular}
    \caption{Parameters of interest of the tunes and their statistical properties as used in the fitting procedures. The \emph{Default} values correspond to the nominal values from GENIE v2~\cite{Andreopoulos:2009rq}.}
    \label{tab:fitParameters}
\end{table*}

\begin{figure*}
    \centering
    \begin{subfigure}{8cm}
        \centering\includegraphics[width=\columnwidth]{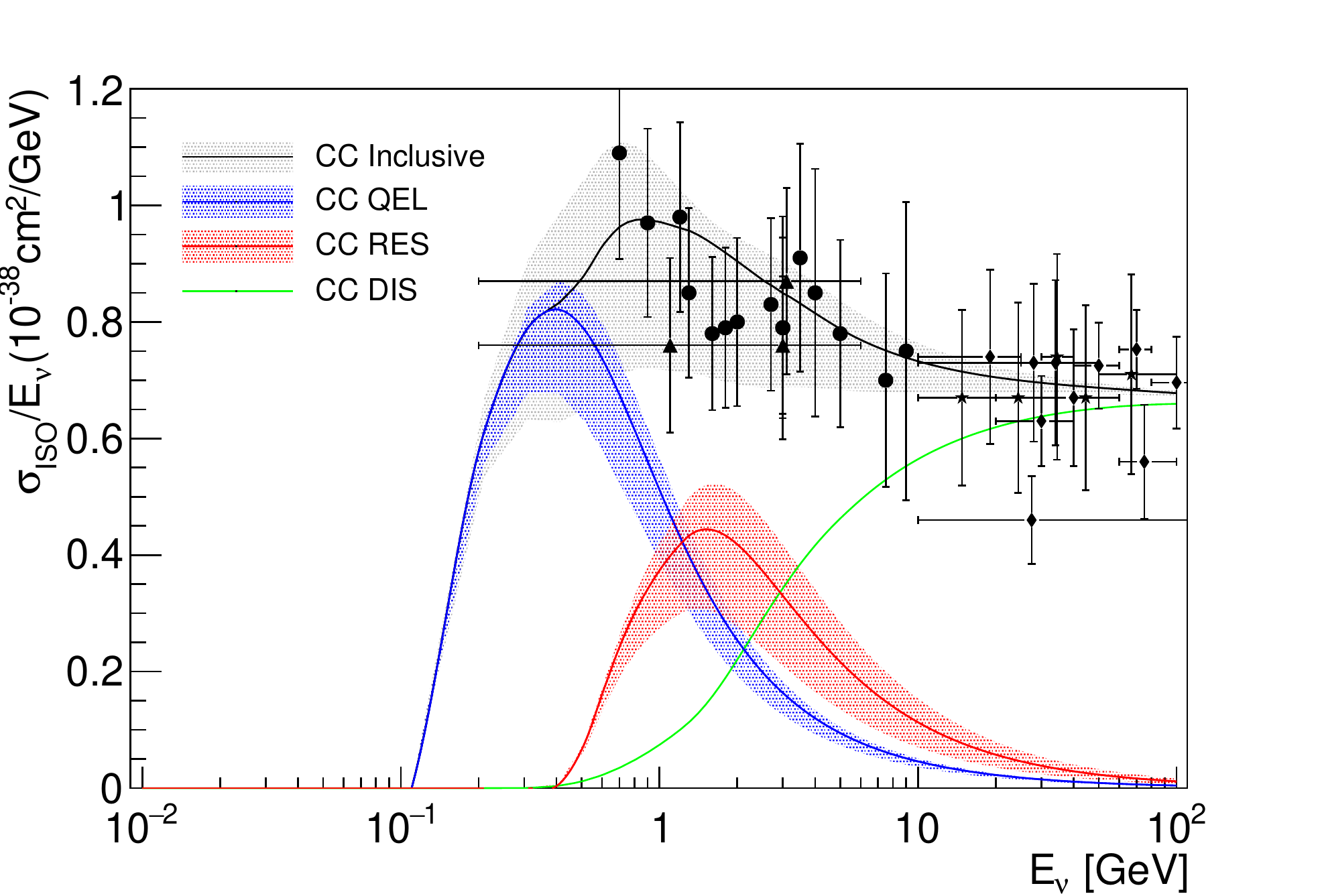}
        \caption{$M_A^{\text{RES}}$ and $M_A^{\text{QE}}$ impact.} 
        \label{fig:MAInclusive}
    \end{subfigure}
    \begin{subfigure}{8cm}
        \centering\includegraphics[width=\columnwidth]{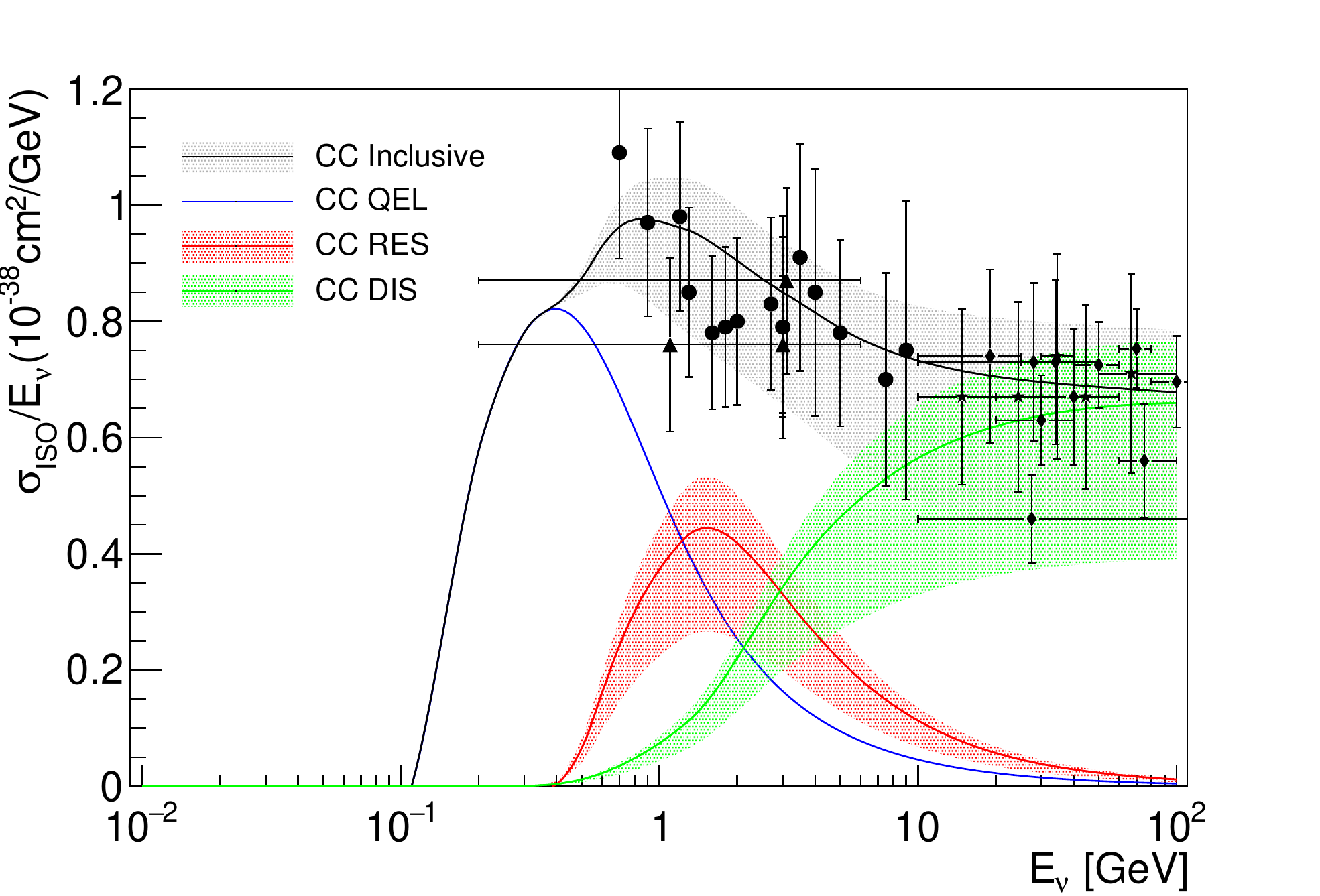}
        \caption{$S_{\text{RES}}$ and $S_{\text{DIS}}$ impact.} 
        \label{fig:DisInclusive}
    \end{subfigure}
    
    \begin{subfigure}{8cm}
        \centering\includegraphics[width=\columnwidth]{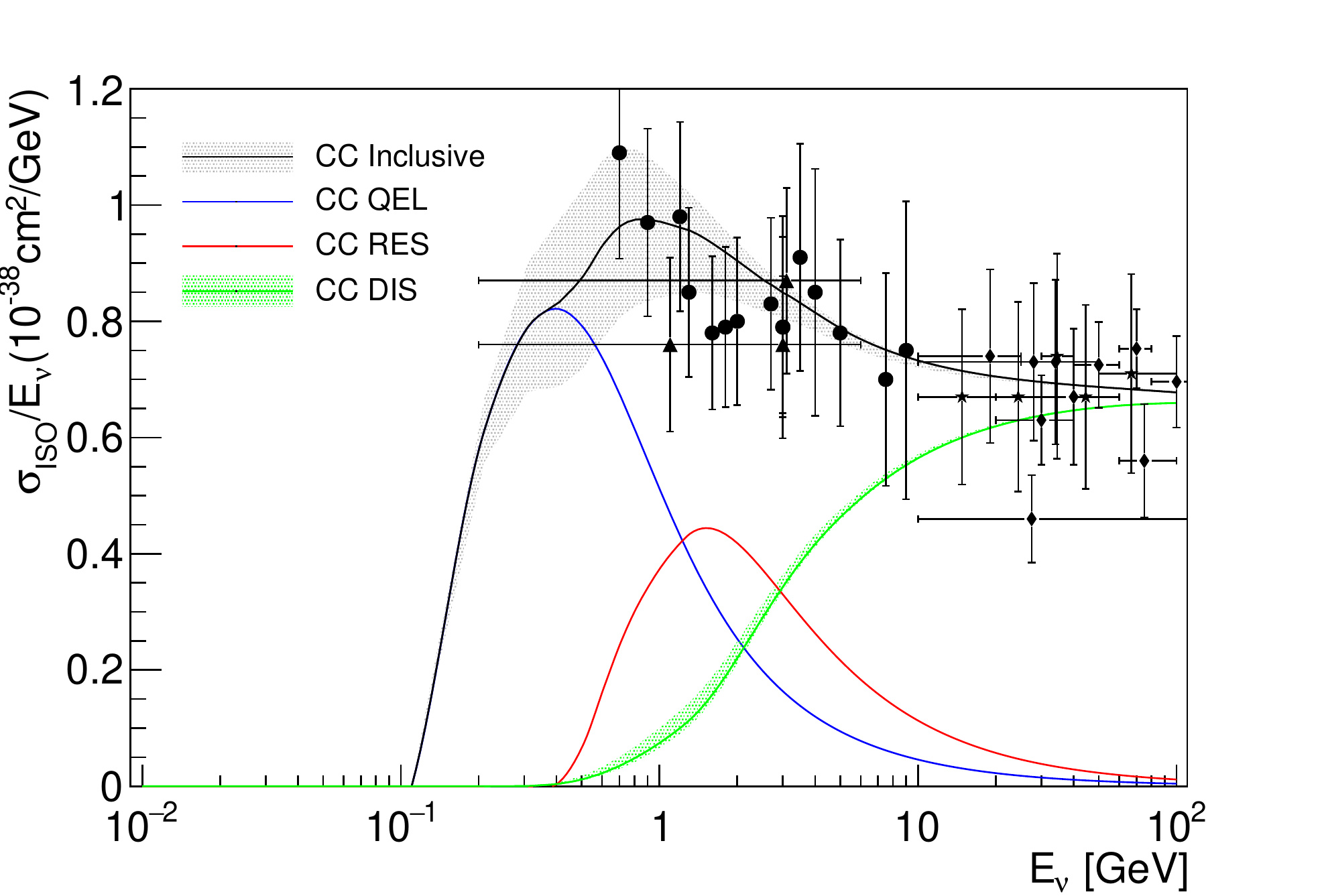}
        \caption{$R_{\nu p}^{\text{CC}1\pi}$, $R_{\nu p}^{\text{CC}2\pi}$, $R_{\nu n}^{\text{CC}1\pi}$ and $R_{\nu n}^{\text{CC}2\pi}$ impact.   } 
        \label{fig:pionprodInclusive}
    \end{subfigure}
    \begin{subfigure}{8cm}
        \centering\includegraphics[width=\columnwidth]{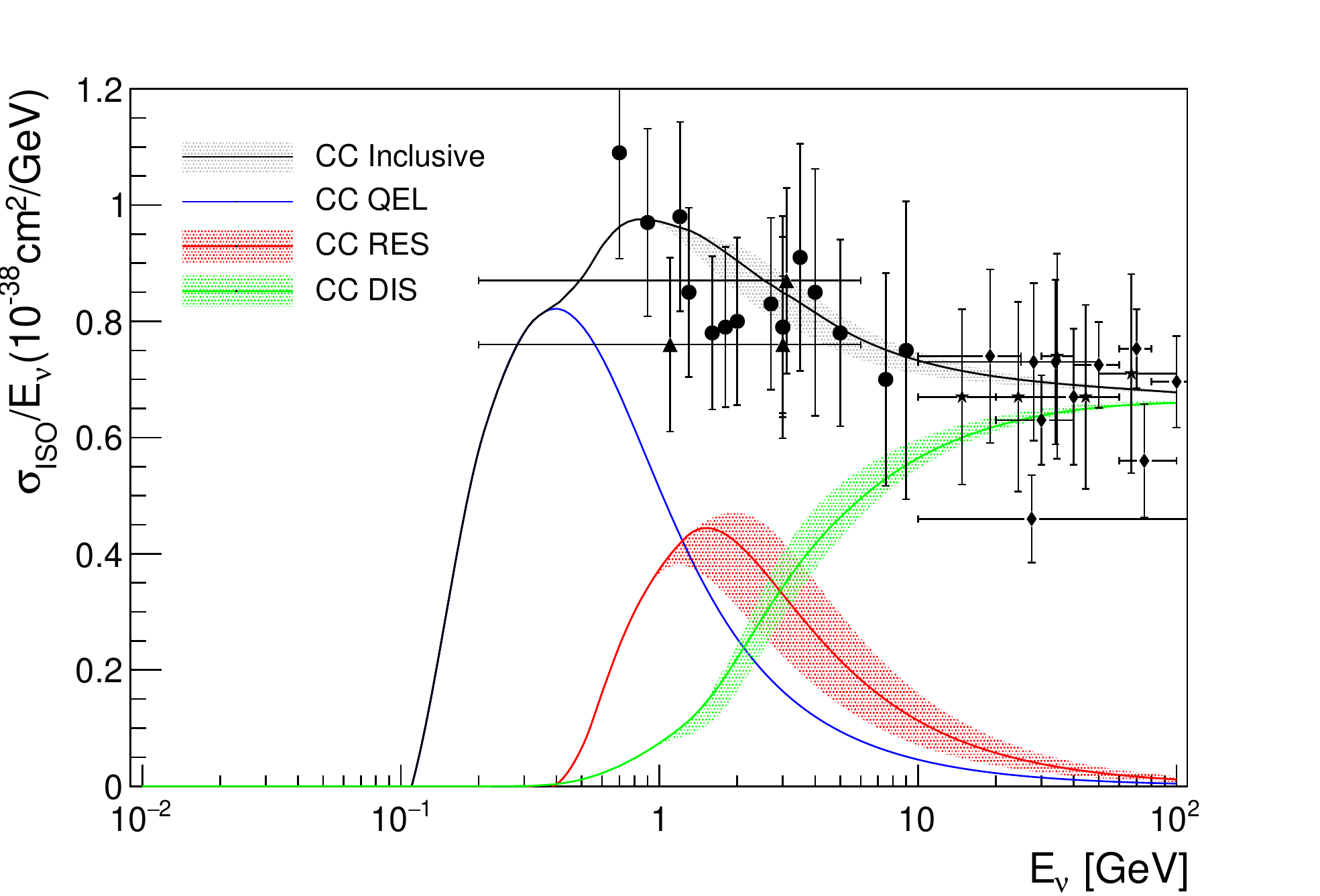}
        \caption{$W_{\text{cut}}$ impact.}  
        \label{fig:WcutInclusive}
    \end{subfigure}
    \caption{$\nu_\mu$ CC Inclusive G18\_02a prediction against hydrogen and deuterium data. Each parameter response is characterized within the tuned region, specified in Tab.~\ref{tab:fitParameters}. Data corresponding to hydrogen and deuterium targets from ANL 12FT ($\bigtriangleup$), BNL 7FT ($\bullet$), BEBC ($\diamond$) and FNAL 15FT ($\star$). }
    \label{fig:inclusiveValidation}
\end{figure*}

Each of the parameters have a different sensitivity to each dataset, as different scattering mechanisms are involved. 
The response of each parameter in the inclusive and exclusive cross sections is studied by varying each of them independently within the studied range.
In Fig.~\ref{fig:inclusiveValidation}, each parameter response is shown for inclusive and exclusive cross sections. 
When more than one parameter in the plot is impacting the same cross section, i.e.\ CC inclusive, the variations are added in quadrature. 

At the Monte Carlo level, where no correlation between the parameters is considered, the impact of each of the parameters in the cross section can be classified as influencing a variation on
\begin{enumerate}
    \item The CC Quasi-Elastic cross section 
    \item The CC RES cross section  
    \item The CC DIS cross section 
\end{enumerate}
For instance, $M_A^{\text{QE}}$ will only affect the quasi-elastic cross section prediction, as summarized in Fig.~ \ref{fig:MAInclusive}.  
Notice though that, at the tune level, this will no longer hold as the introduction of flux nuisance parameters correlates exclusive channels. 
Hence, this will introduce a correlation between $M_A^{\text{QE}}$ and the SIS parameters.  

The description of the CC RES cross section will be affected by the RES axial mass $M_{A}^{\text{RES}}$ (Fig.~\ref{fig:MAInclusive}), the resonant scaling parameter $S_{\text{RES}}$ (Fig.~\ref{fig:DisInclusive}), and $W_{\text{cut}}$ (Fig.~\ref{fig:WcutInclusive}). 
The \emph{default} G18\_01a and G18\_02a configurations overestimate one-pion production processes, and would favor a reduction in the CC RES cross section. 
Variations of $M_A^{\text{RES}}$ have a huge impact on both exclusive and inclusive CC cross sections in the few-GeV region. 
However, as it is explained in  Sec.~\ref{subsec:Priors}, this parameter should agree with the world average extracted from fits to the axial form factor~\cite{Kuzmin} and a deviation from this result is disfavoured by previous fits to data. 
Consequently, a reduction of $S_{\text{RES}}$ is expected to improve the agreement with one-pion production data. 
On the other hand, $W_{\text{cut}}$ will play an important role as it determines the number of resonances included in the CC RES calculation. 
The current default, $W_{\text{cut}}= 1.7$~GeV/$c^2$, discards the resonances contributing at~$W>W_{\text{cut}}$. 
Therefore, an increase on $W_{\text{cut}}$ will incorporate new resonances in the calculation that were not taken into account in previous tunes. 
This increase is favoured by two-pion production data, as heavier resonances producing more than one pion are incorporated. 

The SIS region is treated by combining two cross-section models, one for DIS and one for RES interactions. 
Thus, in that region, many parameters have a visible effect on the predictions as can be seen in Fig.~\ref{fig:inclusiveValidation}: regardless of the parameter considered in the plot, there is always a visible error band in the few-GeV region. 
This is a clear hint for the presence of degeneracy that must be faced by our global tunes. 
An example of this is given by the $R_n$ and the $S_{DIS}$ parameters, which act as scaling factors for the DIS contribution at $W<W{cut}$.
As mentioned above, a desired result of the tune is to reduce the one-pion prediction and increase the two pion production. 
This can be accomplished via alterations of either the $R_n$ and/or the $S_{DIS}$ parameters.

%
%

\section{Bare-nucleon cross-section tuning procedure}
\label{sec:TuningProcedure}
This section describes the core ideas behind the paper. 
Most of these are not specific for this work: they are general concepts developed within the GENIE tuning system and can therefore apply to future tune releases. 

\subsection{Likelihood construction}
\label{sec:ConstructionOfLikelihood}
The GENIE integrated cross-section prediction is denoted with
$\sigma^{i}_{\text{th}}(E_k | \boldsymbol{\theta})$,
where $E_k$ is the neutrino energy, 
$\boldsymbol{\theta}$ is a vector\footnote{
$\boldsymbol{\theta} =$ 
 $( W_{\text{cut}}$, $M_A^{\text{QE}}$, $M_A^{\text{RES}}$, 
  $R_{\nu p}^{\text{CC}1\pi}$, $R_{\nu p}^{\text{CC}2\pi}$,
  $R_{\nu n}^{\text{CC}1\pi}$, $R_{\nu n}^{\text{CC}2\pi}$, 
  $S_{\text{RES}}$, $S_{\text{DIS}})$
}
of the adjustable physics parameters introduced in Sec.~\ref{sec:model_uncertaintines},
and $i$ is any of the 10 reaction processes considered in the work presented in Tab.~\ref{tab:summary_data}. 
Using $\sigma^{i}_{\text{th}}(E | \boldsymbol{\theta})$, we produce the corresponding prediction
for the $k$-th energy bin of the $j$-th dataset for the $i$-th reaction type,
\begin{eqnarray}
  \sigma^{ij}_{\text{th}}(E_{k}|\boldsymbol{\theta}) = 
    \varepsilon^{ij}(E_{k}, \boldsymbol{\theta}) \, \sigma^{i}_{\text{th}}(E_{k}|\boldsymbol{\theta})
\end{eqnarray}

where $\varepsilon^{ij}(E_{k}, \boldsymbol{\theta})$ are dataset-dependent efficiencies expressing the fraction of events from the $i$-th process that survive the kinematical cuts imposed by the experiment, see Tab.~\ref{tab:summary_data}.
The statistical error due to the MC sample size is also evaluated and this is denoted $\delta \sigma^{ij} \left(E_k|\boldsymbol{\theta}\right)$.

\begin{figure}
    \centering
    \includegraphics[width=\textwidth]{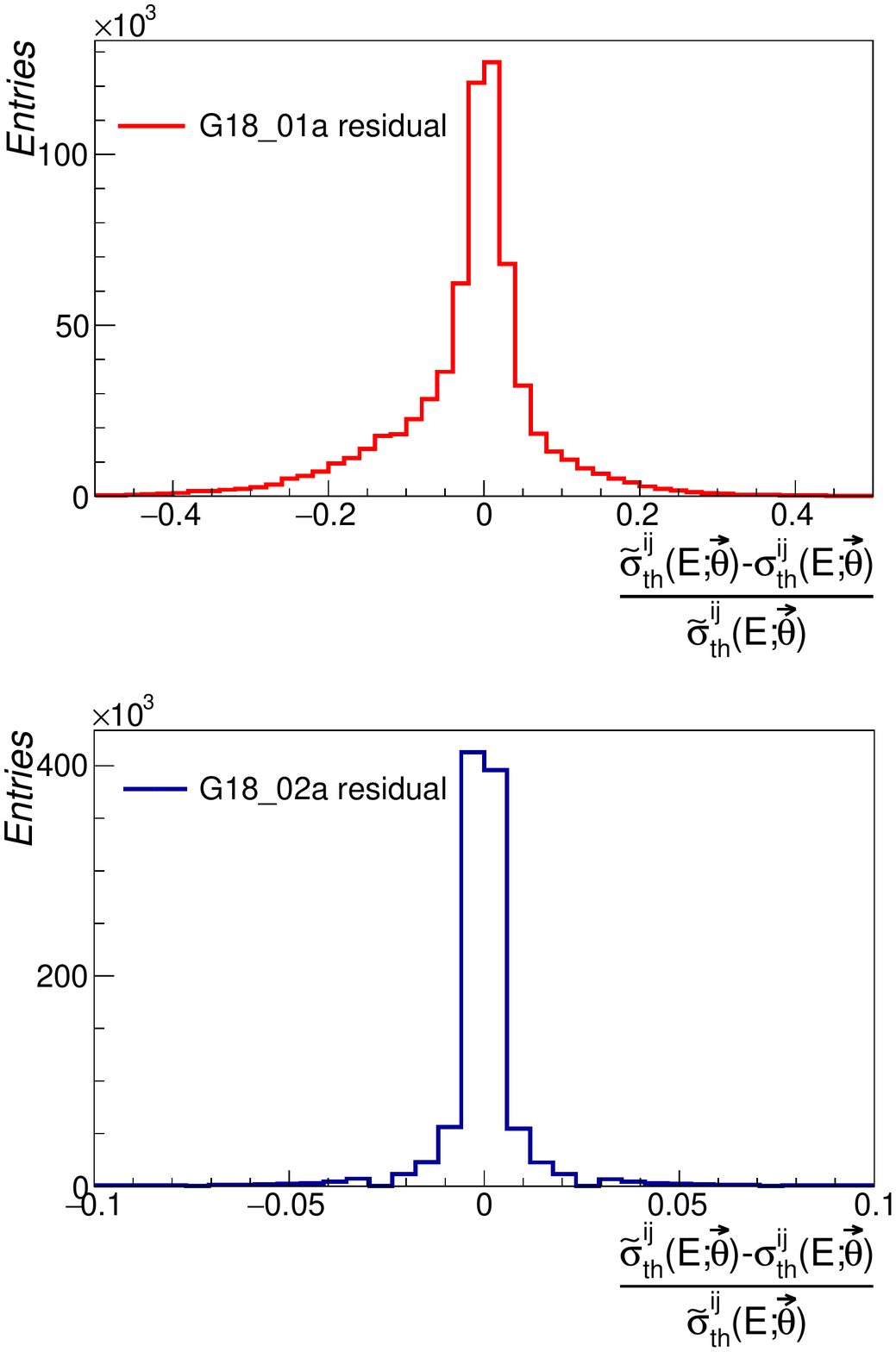}
    \caption{Fractional difference between true Monte Carlo predictions calculated with a given $\vec{\theta}$ set vs the Professor parameterisation for both tunes. The G18\_02a(b) parameterisation is improved as the number of scan points used and polynomial order $M$ are higher.}
    \label{fig:residuals}
\end{figure}

Performing a multi-parameter brute-force scan and tune using $\sigma^{ij}_{\text{th}}(E_{k}|\boldsymbol{\theta})$ is computationally inefficient.
As was highlighted in the introduction, the GENIE global analysis framework relies on Professor~\cite{Professor} to reduce the computational complexity of brute-force scans while allowing for massive parallelisation.
Using the values of $\sigma^{ij}_{\text{th}}(E_{k}|\boldsymbol{\theta})$ computed for
a number ($N_{R}$) of randomised P-dimensional vectors $\boldsymbol{\theta}$, 
produced within the P-dimensional hyper-cube defined by the parameter ranges 
given in Tab.~\ref{tab:fitParameters}, we use Professor to generate a parameterisation of $\sigma^{ij}_{\text{th}}(E_{k}|\boldsymbol{\theta})$  and $\delta \sigma^{ij}(E_{k}|\boldsymbol{\theta})$ that
will be denoted with $\widetilde{\sigma}^{ij}_{\text{th}}(E_{k}|\boldsymbol{\theta})$ and $\delta \widetilde{ \sigma}^{ij}(E_{k}|\boldsymbol{\theta})$ respectively.
As discussed in~\cite{Professor}, the parameterisation is a generic polynomial of order $M$ in the P-dimensional space, whose analytical form is 
\begin{eqnarray}
  \widetilde{\sigma}^{ij}_{\text{th}}(E_k|\boldsymbol{\theta}) & = & \alpha_{0}^{ijk} + \sum_{n=1}^P \beta_{n}^{ijk} \theta_{n} + \sum_{n \le m} \gamma_{nm}^{ijk} \theta_{n} \theta_m \nonumber \\
    & + & \ldots + \sum_{n_1 \le \ldots \le n_M } \! \! \! \! \! \! \xi ^{ijk} _{n_1 \ldots n_M } \prod _{\ell=1} ^M \theta_{n_\ell} \label{eqn:parameterisation}
\end{eqnarray}
where $\theta_{n}$ is the coordinate of the $n$-th parameter.
The polynomial order $M$ is set by the user.
The coefficients $\alpha_{0}^{ijk}$, $\beta_{n}^{ijk}, \gamma_{(nm)}^{ijk}, \ldots,\xi ^{ijk} _{(n_1 \ldots n_M)} $ are determined by Professor fitting the parameterisation against the computed  $\sigma^{ij}_{\text{th}}(E_{k}|\boldsymbol{\theta})$.
In the analysis presented here, a 4$^{\text{th}}$ order polynomial was used for the G18\_01a comprehensive model configuration while a 5$^{\text{th}}$ order polynomial was used for G18\_02a. 
Particularly, $N_R= 1500$ for G18\_01a and $N_R = 2183$ for G18\_02a. 
The accuracy of the parameterisation is demonstrated in the residual distributions shown in Fig.~\ref{fig:residuals}.
The parameterisation $\widetilde{\sigma}^{ij}_{\text{th}}(E_{k}|\boldsymbol{\theta})$ is used instead of the exact predictions in order to to estimate the best-fit parameters by minimising the $\chi^2$.

\subsection{Treatment of systematic uncertainties}
\label{subsec:Nuisance}
A number of nuisance parameters, each with a corresponding prior, can also be used to tackle the problem of correlated datasets. As seen in Sec.~\ref{subsec:datasetifno}, there are different datasets coming from the same experiments (ANL 12FT, BNL 7FT, BEBC, and FNAL 15FT). Each of these experiments share the same flux (from either a neutrino or an antineutrino beam), analysis procedure, etc. 
Therefore, there is a correlation between the datasets, even though it has not been quantified in the data releases. 
A possible approach is to add nuisance parameters that can connect datasets from experiments that used the same neutrino beam\footnote{For instance, BEBC data on $\nu_{\mu}$ will have an associated nuisance parameter, which will be different from the one applied to BEBC $\bar{\nu}_{\mu}$ data.}. 
As the main systematic uncertainty comes from the fluxes, the nuisance parameters will act as scaling factors for our predictions ($\widetilde{\sigma}^{ij}_{\text{th}}(E_k|\boldsymbol{\theta})$) and are same for datasets sharing the same flux. 

Some of the ANL 12FT and BNL 7FT data were already corrected for the flux normalization~\cite{Wilkinson:2014yfa}. 
Due to this correction, the associated systematic error is smaller and, accordingly, a more restricted nuisance parameter is applied to the re-analysed datasets. 
These restricted parameters take into account other common systematics like reconstruction procedures, so they multiply all the predictions related to the same experiment. 
At the end of this procedure, each prediction can be scaled by up to 2 nuisance parameters, one for the flux and one for the remaining systematics. 
Thus, a total of 9 independent nuisance parameters are used to account for the correlation between datasets.
They are all the available combinations of experiment and neutrino flux exposure ($\nu_\mu$ and $\bar{\nu}_\mu$) plus the restricted parameters for re-analysed data: 
\begin{eqnarray*}
  \boldsymbol{f} = & \Big(  f^{\text{AN}L}(\nu_\mu),f^{\text{ANL}}_{\text{Re}}(\nu_\mu), f^{\text{BNL}}(\nu_\mu),\\
    & f^{\text{BNL}}_{\text{Re}}(\nu_\mu),f^{\text{BNL}}(\bar{\nu}_\mu), f^{\text{BEBC}}(\nu_\mu), \\
    & f^{\text{BEBC}}(\bar{\nu}_\mu), f^{\text{FNAL}}(\nu_\mu), f^{\text{FNAL}}(\bar{\nu}_\mu)\Big)
\end{eqnarray*}

Quasi-elastic data for hydrogen and deuterium targets is included in the tune in order to constrain the nuisance parameters. 
Even though quasi-elastic data is not directly constraining the SIS parameters, it plays an important role to further constrain the fluxes of each experiment, as it is known at the 15~\% level. 

The main advantage of this method is the unbiased choice of the nuisance parameters, as their values will be determined by the minimization of the likelihood function.
For the calculation of best-fit points and the calculation of intervals, these nuisance parameters are profiled (on every instance of our fit, they are eliminated by substituting them with the value that minimises $\chi^2$).

\subsection{Discussion of priors}
\label{subsec:Priors}
The likelihood is corrected using priors on parameters of interest ($\boldsymbol{\theta}$) and nuisance parameters ($\boldsymbol{f}$).
Priors allow us to incorporate in this analysis the appropriate pre-fit uncertainties and correlations for the parameters of interest.
Only Gaussian priors are considered at present.

The priors applied to each nuisance parameter $f_j$ have a peak at 1 and different standard deviations $\delta f_j$.
In general the total scaling factor applied to non-re-analysed datasets are constrained by a conservative 15~\% $\delta f$ Gaussian prior, except for those nuisance parameters that act on the same experiment.
Thus, the BEBC and FNAL 15FT experiments have only one associated scaling factor $\delta f^{\text{BEBC}} = \delta f^{\text{FNAL}} = 0.15$ for both neutrino and anti-neutrino fluxes; the same is true for $f^{\text{BNL}}(\bar{\nu}_\mu)$.
Up to two nuisance parameters can be applied to ANL 12FT and BNL 7FT data (i.e. $f^{ANL}(\nu_\mu)$ and $f^{ANL}_{Re}(\nu_\mu)$).
The ANL 12FT and BNL 7FT restricted nuisance parameters, $f^{\text{ANL}}_{\text{Re}}(\nu)$ and $f^{\text{BNL}}_{\text{Re}}(\nu)$, have $\delta f=5\%$.
$\delta f^{\text{ANL}}$ and $\delta f^{\text{BNL}}$ are such that ANL 12FT and BNL 7FT non-re-analysed datasets data are constrained by an overall $15\%$ Gaussian. 
The full summary of the nuisance parameters is in Tab.~\ref{tab:fitNuisanceParameters}.

\begin{table}    
    \centering
    \begin{tabular}{c c c }  \hline\hline\noalign{\smallskip}
    Parameter & Prior \\ 
    \noalign{\smallskip}\hline\noalign{\smallskip}
    $f^{\text{ANL}}(\nu_\mu)$             & $1\pm0.14$ \\
    $f^{\text{ANL}}_{\text{Re}}(\nu_\mu)$ & $1\pm0.05$ \\
    $f^{\text{BNL}}(\nu_\mu)$             & $1\pm0.14$ \\
    $f^{\text{BNL}}_{\text{Re}}(\nu_\mu)$ & $1\pm0.05$ \\
    $f^{\text{BNL}}(\bar{\nu}_\mu)$       & $1\pm0.15$ \\
    $f^{\text{BEBC}}(\nu_\mu)$            & $1\pm0.15$ \\
    $f^{\text{BEBC}}(\bar{\nu}_\mu)$      & $1\pm0.15$ \\
    $f^{\text{FNAL}}(\nu_\mu)$            & $1\pm0.15$ \\
    $f^{\text{FNAL}}(\bar{\nu}_\mu)$      & $1\pm0.15$ \\
    \noalign{\smallskip}\hline\hline
    \end{tabular}
    
     \caption{Nuisance parameters, $f^{j}$, per experiment (ANL 12FT, BNL 7FT, BEBC or FNAL 15FT) and neutrino beam ($\nu_\mu$ or $\bar{\nu}_\mu$). Priors consider the systematic uncertainty applied to each dataset as $\delta f_{j}$, where $j$ is one of the datasets under study. The allowed range is $[0,2]$ for nuisance parameters considered in the tune. }
    \label{tab:fitNuisanceParameters}
    
\end{table}

Priors are applied to the parameters of interest to penalize disagreement with well-established parameter values.
For instance, the description of neutrino CC quasi-elastic cross sections and single-pion production through baryon resonances is strongly determined by the shape of the weak axial and vector transition form factors.
For the G18\_01a(/b/c/d) and G18\_02a(/b/c/d) CMCs, the axial form factors are described using the dipole parameterisation which is a function of the invariant transferred momentum ($Q^2)$:
\begin{equation}
    F_{A}(Q^2)=F_A(0)\left(1-\frac{Q^2}{M_{A}^2}\right)^{-2}
\end{equation}
with $F_{A}(0)=g_{A}=-1.2695 \pm 0.002$~\cite{ParticleDataGroup}.
The axial mass, $M_A$, is extracted from data. 
There are different masses for every interaction type: $M_{A}^{\text{QEL}}$ and $M_{A}^{\text{RES}}$. Both of these are evaluated from neutrino data on deuterium targets. 
The latest world average values for the axial masses are:
\begin{align*}
    M_{A}^{\text{QE}}  = &\ 1.014 \pm 0.014 \enskip \text{GeV}/c^2  &\ \text{\cite{QELMA}} \\
    M_{A}^{\text{RES}} = &\ 1.12  \pm 0.03  \enskip \text{GeV}/c^2  &\ \text{\cite{Kuzmin}}
\end{align*}
The extraction of these parameters requires neutrino differential cross sections as a function of $Q^2$ that are not used in this analysis. 
Our goal is not the extraction of the axial masses but the better estimation of the cross section at the SIS region. For this reason, these values are used as priors in our global fits. 

Another parameter of interest which is strongly constrained by data is the $S_{\text{DIS}}$ parameter. This parameter dominates the cross-section behaviour at high neutrino energies. Most of the data in that energy range comes from neutrino interactions with heavy nuclear targets and are therefore not included in the fit. 
A Gaussian prior is considered to ensure that agreement with these data are preserved\footnote{The best agreement with all high energy data requires $S_{\text{DIS}}\sim1$.} by our tuning procedure. 
This would not be the case otherwise as the SIS region data would prefer much higher cross-section values for the DIS contribution.
The prior on $S_{\text{DIS}}$ provides a good solution for this problem because the degeneracy between DIS and non-resonant background parameters gives us multiple ways to accommodate good agreement between data and GENIE predictions in the SIS region. 
In other words, the introduction of the $S_{\text{DIS}}$ prior breaks the degeneracy without adding more datasets to the fit. 

\subsection{Final form of the $\chi^2$ }

Including all of the contributions from the previous sections and defining $\sigma^{ijk}_{d}$ ($\delta\sigma^{ijk}_{\text{stat}}$) as the data central value (statistical error) corresponding to the $\widetilde{\sigma}^{ij}_{\text{th}}(E_{k}|\boldsymbol{\theta})$ prediction, the complete form of our $\chi^2$ distribution becomes:
\begin{eqnarray}
\chi^2 (\boldsymbol{\theta},\boldsymbol{f}) = &
  \sum_{i,j,k} w^{ijk} \frac{ (\phi_j(\boldsymbol{f})  \widetilde{\sigma}^{ij}_{\text{th}}(E_{k}|\boldsymbol{\theta})-\sigma^{ijk}_{d})^2}{(\delta\sigma^{ijk}_{\text{stat}})^2} \nonumber \\
  & + (\boldsymbol{\theta} -\boldsymbol{\theta}_{0})^T \Sigma^{-1}_{\theta} (\boldsymbol{\theta} -\boldsymbol{\theta}_{0})  \label{eqn:chi_square} \\ 
  & + \sum_{j}\frac{(f_j-1)^2}{(\delta f_{j})^2} \nonumber
\end{eqnarray}
where $\phi_j(\boldsymbol{f})$ is the product of the nuisance scaling factors that are relevant for $j$-th dataset as described in Sec.~\ref{subsec:Nuisance}.
$\boldsymbol{\theta}_0$ and $\Sigma_\theta$ are the central values and the covariance matrix of the priors for the parameters of interest, respectively.
Equation~\ref{eqn:chi_square} represents the full capability of our tuning machinery. However, the priors we applied for the present work were uncorrelated and so only the diagonal entries of $\Sigma_\theta$ were used.
The details on the priors applied in this analysis are described in Sec.~\ref{subsec:Priors}.

The contribution of each point to the likelihood can be (de-)emphasized using weights $w^{ijk}$ to set the relative importance of different datasets (or of individual data points within a dataset). 
Such weighting schemes have been used extensively in general-purpose event generator tunes for the LHC (for an example, see~\cite{Buckley:2009bj}). 
In this particular analysis, the weights are used to include or exclude datasets only ($w^{ijk} \in \{0,1\}$). 

%
%

\section{Tuning results}
\label{sec:TuningResults}
In order to properly understand the global tune, the tensions between datasets must be discussed. 
These tensions are studied by performing fits using a specific dataset to evaluate the impact of the partially-fitted predictions on the rest of the datasets included in the global tune. 

\subsection{Partial fits}
\label{sec:partial_fits}
Two main subsets were identified in the global dataset in order to study tensions: inclusive and exclusive datasets.
The fits consider the G18\_02a CMC as the base configuration and include nuisance parameters to take into account the correlation between datasets from the same experiment, see Sec.~\ref{sec:TuningProcedure} for more details. 
No priors on $M_A^{\text{RES}}$ and $M_A^{\text{QE}}$ are applied as we are interested to see the impact of each subset on the prediction.  
The fit to inclusive data only is not sensitive to the scaling multiplicity parameters for the non-resonant background, therefore those parameters are fixed to their default values during the fit.

\begin{table}   
    \centering
    \begin{tabular}{c c c } \hline\hline\noalign{\smallskip}
    Parameter & Inclusive & Exclusive \\ \noalign{\smallskip}\hline\noalign{\smallskip}
    $W_{\text{cut}}   (\text{GeV})$            & 1.52               & 2.00                \\
    $M_A^{\text{QE}}  (\text{GeV}/\text{c}^2)$ & $0.98\pm0.01$      & $1.003 \pm 0.008$   \\
    $M_A^{\text{RES}} (\text{GeV}/\text{c}^2)$ & $1.15 \pm 0.02$    & $0.88  \pm 0.02$    \\
    \noalign{\smallskip} 
    $R_{\nu p}^{\text{CC}1\pi}$                & (0.10)             & $0.30  \pm 0.02$    \\
    $R_{\nu p}^{\text{CC}2\pi}$                & (1.00)             & $1.28  \pm 0.06$    \\
    $R_{\nu n}^{\text{CC}1\pi}$                & (0.30)             & $0.294 \pm 0.002$   \\
    $R_{\nu n}^{\text{CC}2\pi}$                & (1.00)             & $3.19  \pm 0.09 $   \\
    $S_{\text{RES}}$                           & $0.87 \pm 0.03$    & $0.88  \pm 0.02 $   \\ 
    $S_{\text{DIS}}$                           & $ 1.027\pm 0.005$  & $1.026 \pm 0.007$   \\
    \noalign{\smallskip}\hline\hline
    \end{tabular}
    \caption{Parameter best-fit results for partial fits to inclusive and exclusive data using the G18\_02a CMC as base configuration. 
    Values within parentheses are kept fixed during the fit: they are the historical \emph{default} values.
    }
    \label{tab:resultsFitInclusive}
\end{table}

\begin{figure*}
    \centering
    \centering\includegraphics[width=0.8\textwidth]{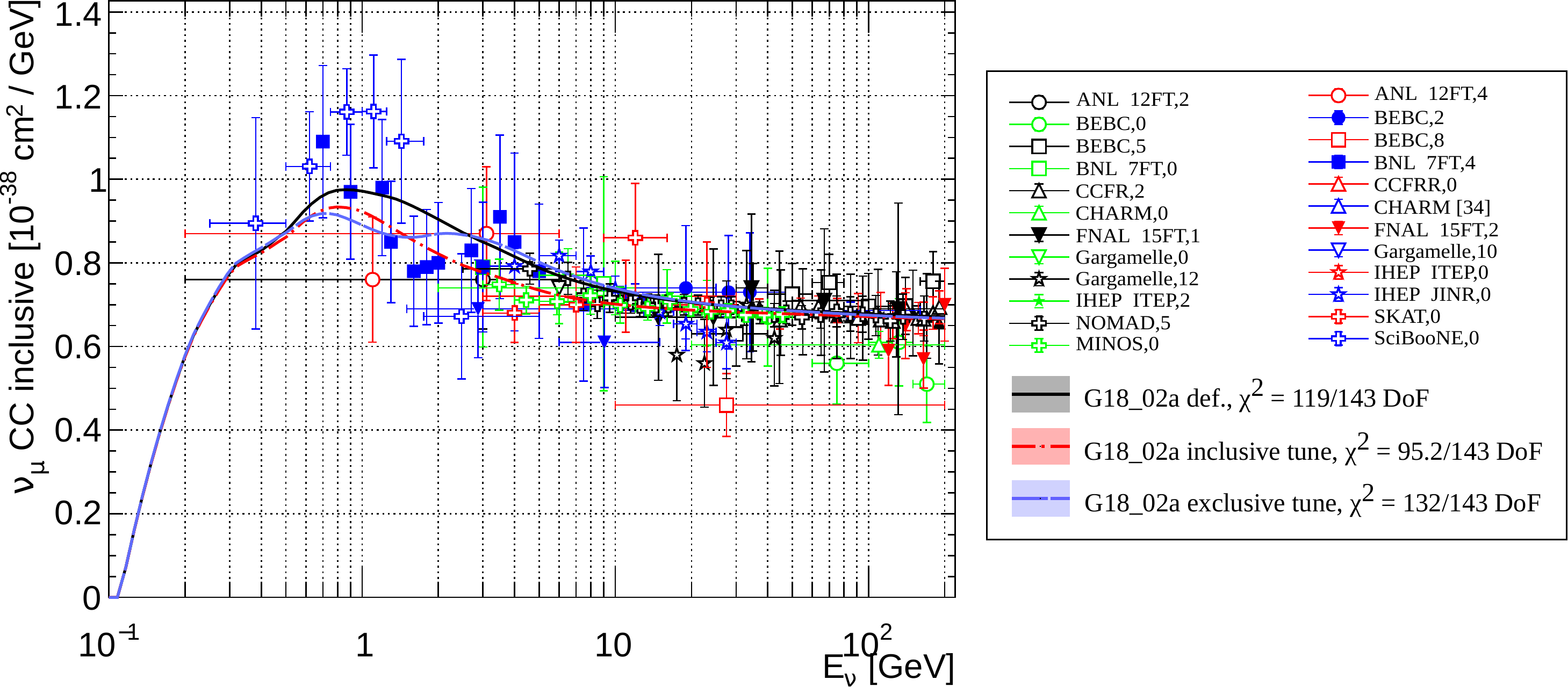}
     \caption{Comparison of $\nu_\mu$ CC Inclusive cross section against bubble chamber data. The {\em default} GENIE configuration corresponds to the G18\_02a CMC. The inclusive tune is performed using the filled datapoints only. The predictions are computed with GENIE version 3.2 using the parameters specified in Tab.~\ref{tab:resultsFitInclusive}. The $\chi^2$ values are calculated against all inclusive data available from bubble chamber experiments.  }  
        \label{fig:incl_xsec_partialfit}
\end{figure*}

Partial fit results for inclusive and exclusive data are presented in Tab.~\ref{tab:resultsFitInclusive}.
The tune against inclusive data only achieves much better agreement with inclusive data than the previous GENIE G18\_02a {\em default}, see Fig.~ \ref{fig:incl_xsec_partialfit}.
This difference between the old and new inclusive tune is due to 1) the inclusion of only hydrogen and deuterium datasets and 2) the effect of the nuisance parameters.\footnote{Note that in the \emph{default} GENIE tune, there was no method to include correlation between datasets coming from the same experiment.}
Particularly, without exclusive data, a small reduction of the resonant cross section is already observed in the CC RES region.
The result for $M_A^{\text{RES}}$ is consistent with previous results without the addition of priors~\cite{Kuzmin}.
$W_{\text{cut}}$ is pulled to the lower edge of the parameter range: the parameter uncertainty could not be estimated as the $\chi^2$ minimum was found on the contour.

\begin{figure*}
    \centering
    \begin{subfigure}{7cm}
        \centering\includegraphics[width=\columnwidth]{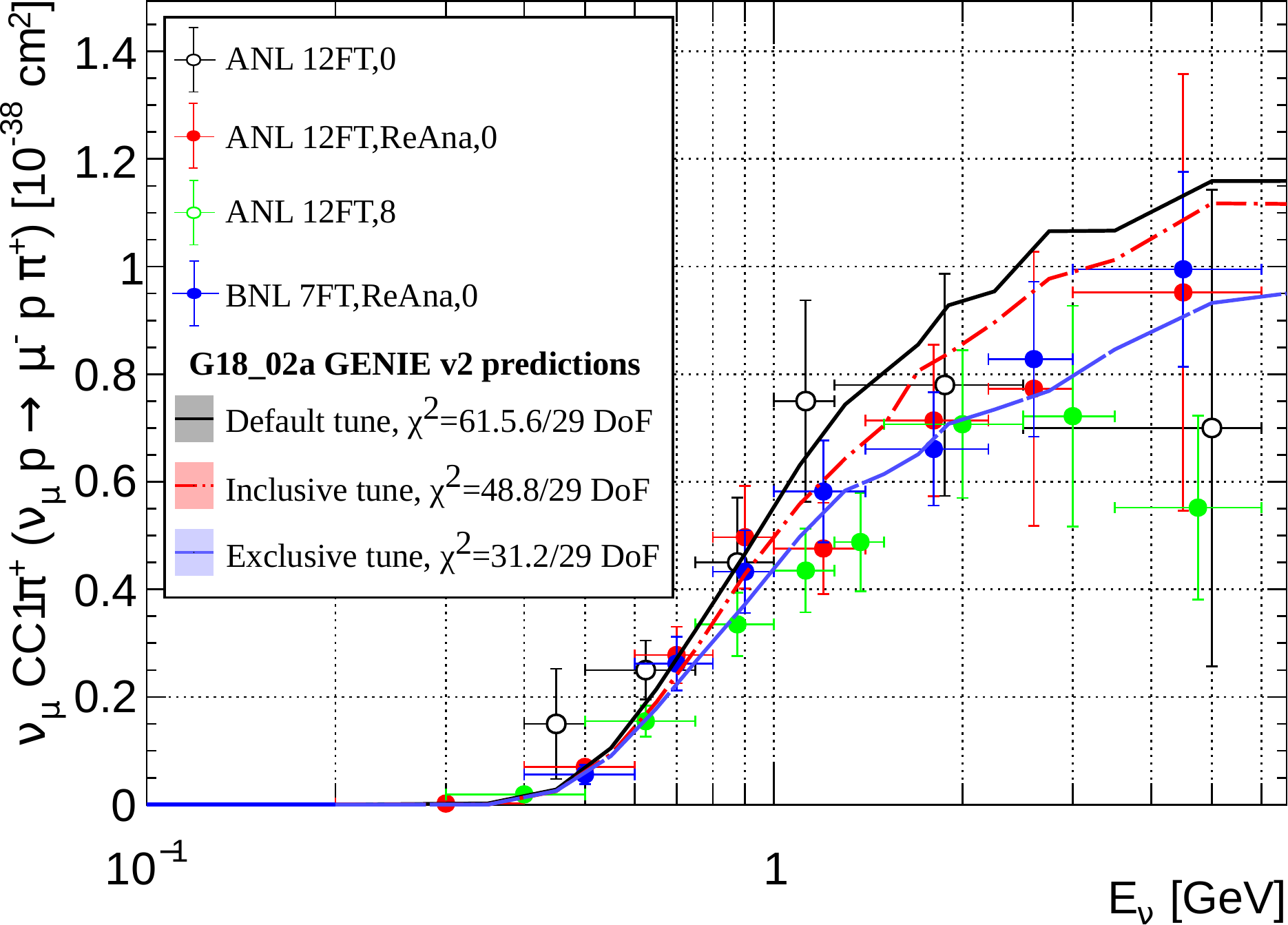}
        \caption{Comparison against $\nu_\mu CC 1\pi^+$ data.   }   
    \label{fig:excl_1pi}
    \end{subfigure}  \,\,\,\,
    \begin{subfigure}{7cm}
        \centering\includegraphics[width=\columnwidth]{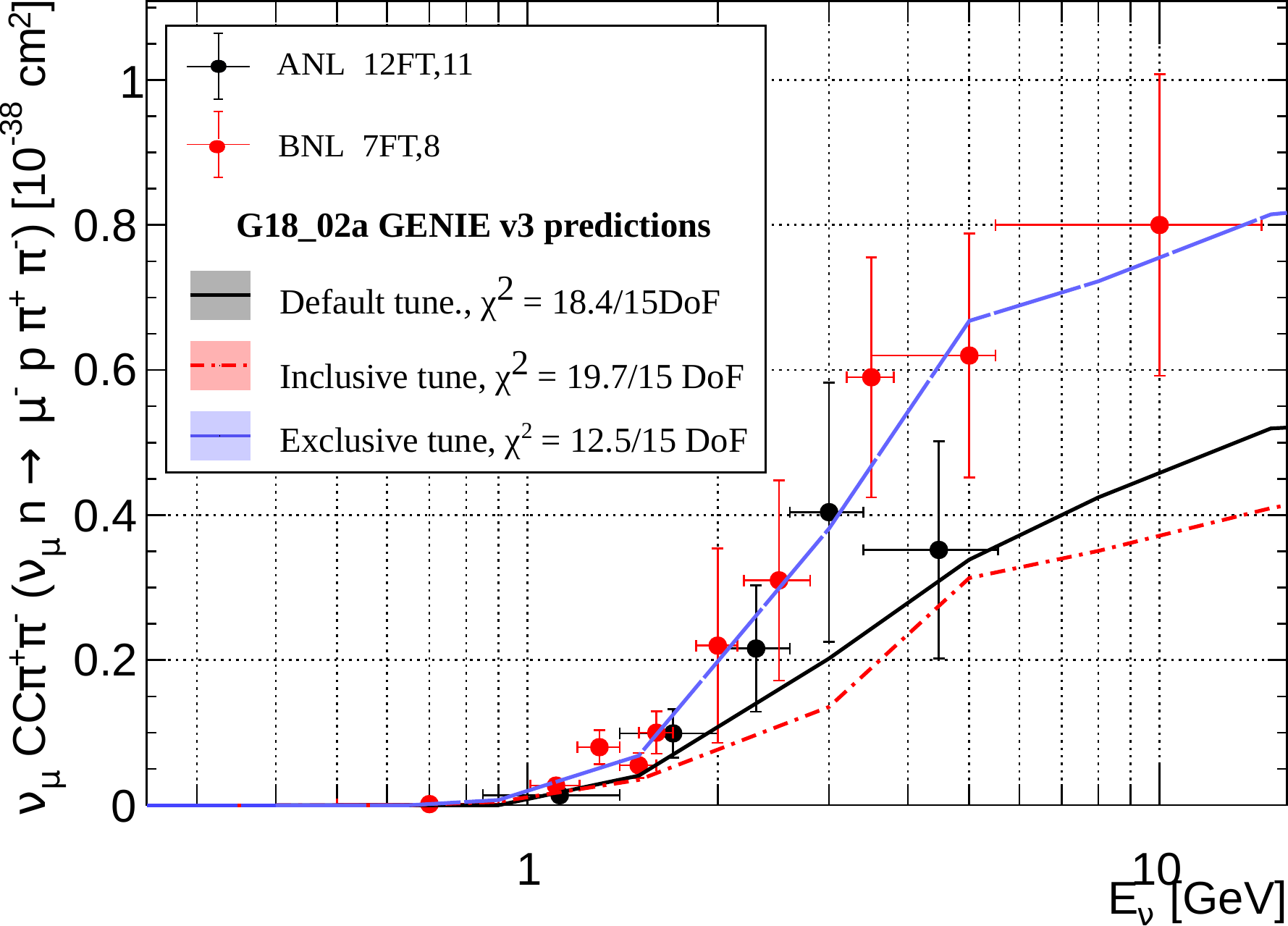}
        \caption{Comparison against $\nu_\mu CC \pi^+\pi^-$ data. }   
        \label{fig:excl_2pi}
    \end{subfigure} 
    \label{fig:excl_fit}
     \caption{Comparison of $\nu_\mu$ CC exclusive channels against bubble chamber data. The {\em default} GENIE configuration corresponds to the G18\_02a CMC. The exclusive tune is sensitive to the exclusive datasets only, see Tab.~\ref{tab:summary_data}. The predictions are computed with GENIE version 3.2 using the parameters specified in Tab.~\ref{tab:resultsFitInclusive}. The $\chi^2$ values are calculated against all exclusive data shown in each plot.  }  
\end{figure*}

As expected, the fit to exclusive data only is able to correctly describe exclusive datasets for one and two pion production.  
The low cross-section data for one pion production forces all the relevant parameters to decrease with respect to the \emph{default} values see Fig.~\ref{fig:excl_1pi}. At the same time, two pion production data forces $R_{\nu p}^{\text{CC}2\pi}$, $R_{\nu n}^{\text{CC}2\pi}$ and $W_{\text{cut}}$ to increase in order to match two pion production data, see Fig.~\ref{fig:excl_2pi}. 
The agreement with $\nu_\mu$ CC inclusive data is worse, see Fig.~\ref{fig:incl_xsec_partialfit}, but the compatibility is still acceptable given the large uncertainties on the data in that region. 
On the other hand, the partial fit does not obtain a good prediction for $M_A^{\text{RES}}$. $W_{\text{cut}}$ is fixed to its maximum value of 2~GeV to avoid nonphysical regions.

The exclusive fit clearly shows a preference for a larger total cross section in the neutrino energy region between 1 and 10~GeV due to the high value of $R_{\nu p}^{\text{CC}2\pi}$ and $R_{\nu n}^{\text{CC}2\pi}$. 
This is a tension between exclusive and inclusive datasets as the inclusive prefer a lower value in that $E_\nu$ region. Since inclusive data constitute about $40\%$ of all the data points, the inclusion of priors for the axial masses and $S_{\text{DIS}}$ becomes crucial to overcome the tension~\cite{BETANCOURT20181}.

\subsection{Global fit}
\label{sec:global_fit}
\begin{figure*}
    \centering
    \begin{subfigure}{\textwidth}
      \centering\includegraphics[width=0.85\columnwidth]{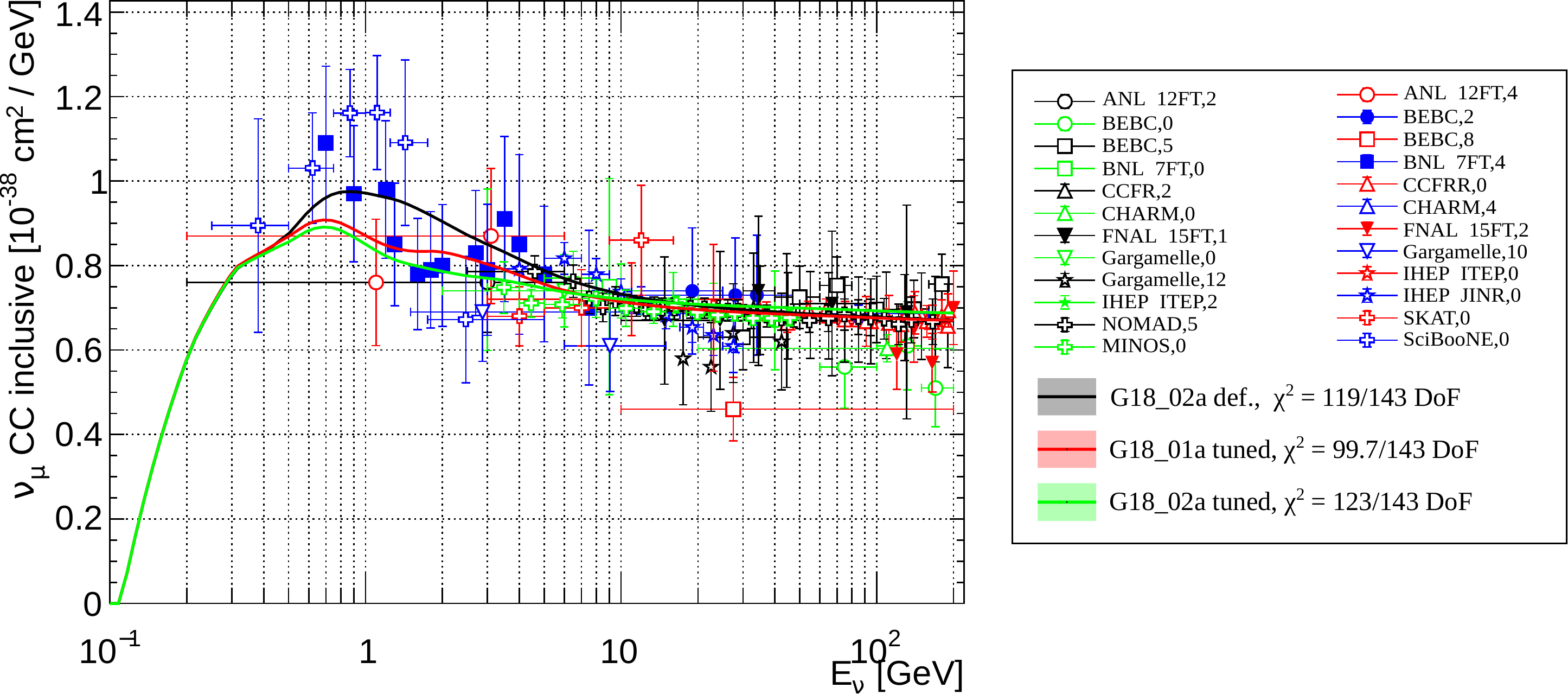}
        \caption{Comparison of $\nu_\mu$ CC Inclusive cross-section data against against the \emph{default} and tuned CMC.}
    \end{subfigure} 
    \label{fig:CCinclusive}
      \\
    \vspace{0.5cm}
    \begin{subfigure}{\textwidth}
\centering
        \centering\includegraphics[width=0.85\columnwidth]{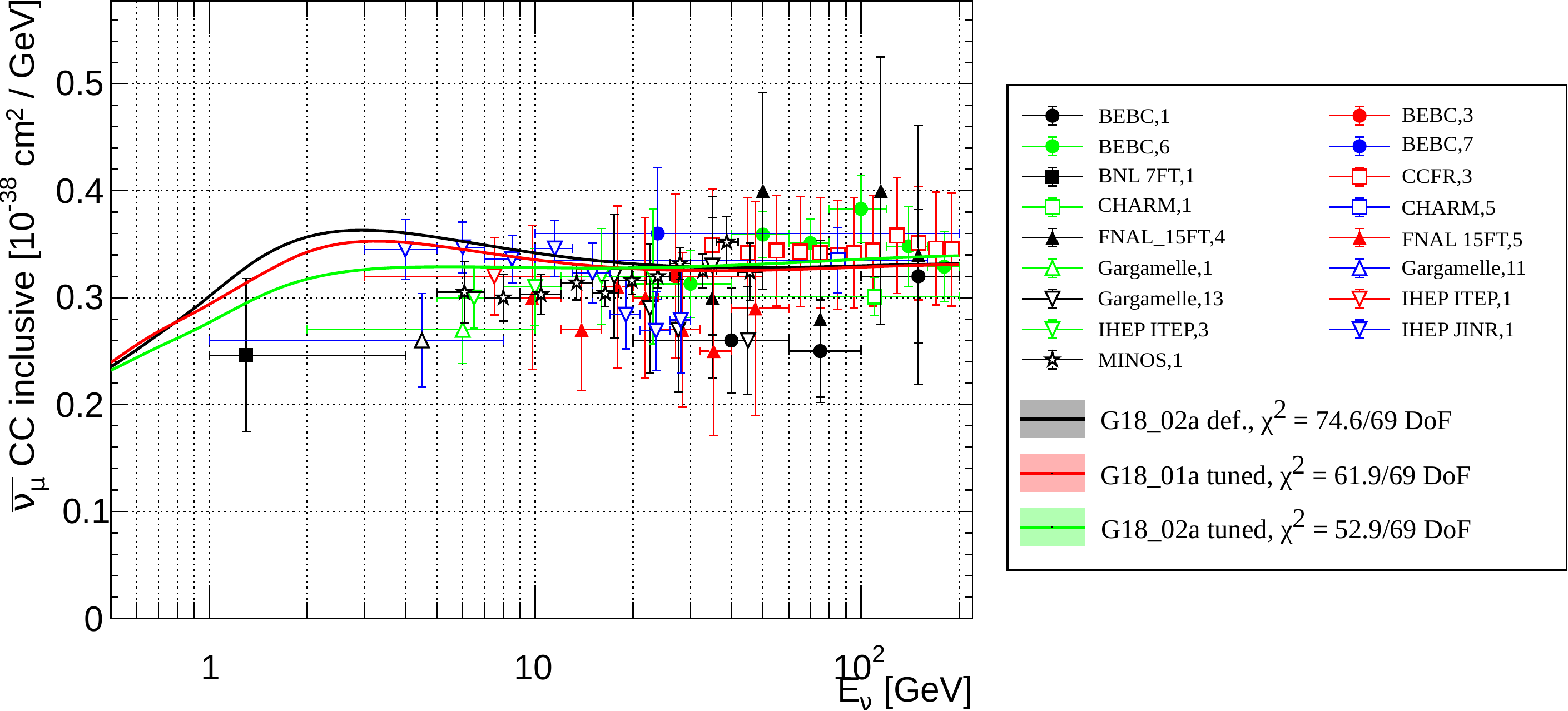}
        \caption{Comparison of $\bar{\nu}_\mu$ CC Inclusive cross-section data against against the \emph{default} and tuned CMCs. }
        \end{subfigure}
    \label{fig:barCCinclusive}
         \\
    \vspace{0.5cm}
    \caption{Best fit prediction impact on muon (anti)neutrino CC inclusive cross sections as a function of the neutrino energy ($E_\nu$). The associated predictions for the \emph{default} G18\_02a and tuned G18\_02a and G18\_02a are computed with GENIE v3.0.6. Predictions are compared against all the available data (anti)neutrino interactions on $H$, $^2$H and heavier targets. Both CMC have been tuned against some $H$, $^2$H data (filled markers). Each $\chi^2$ is computed using all data available. In Tab.~\ref{tab:resultsFitchi2}, the $\chi^2$ values per dataset are specified.}
    \label{fig:inclusivePredictions}
\end{figure*}

\begin{figure*}
    \centering
    \begin{subfigure}{\textwidth}
      \centering\includegraphics[width=0.85\columnwidth]{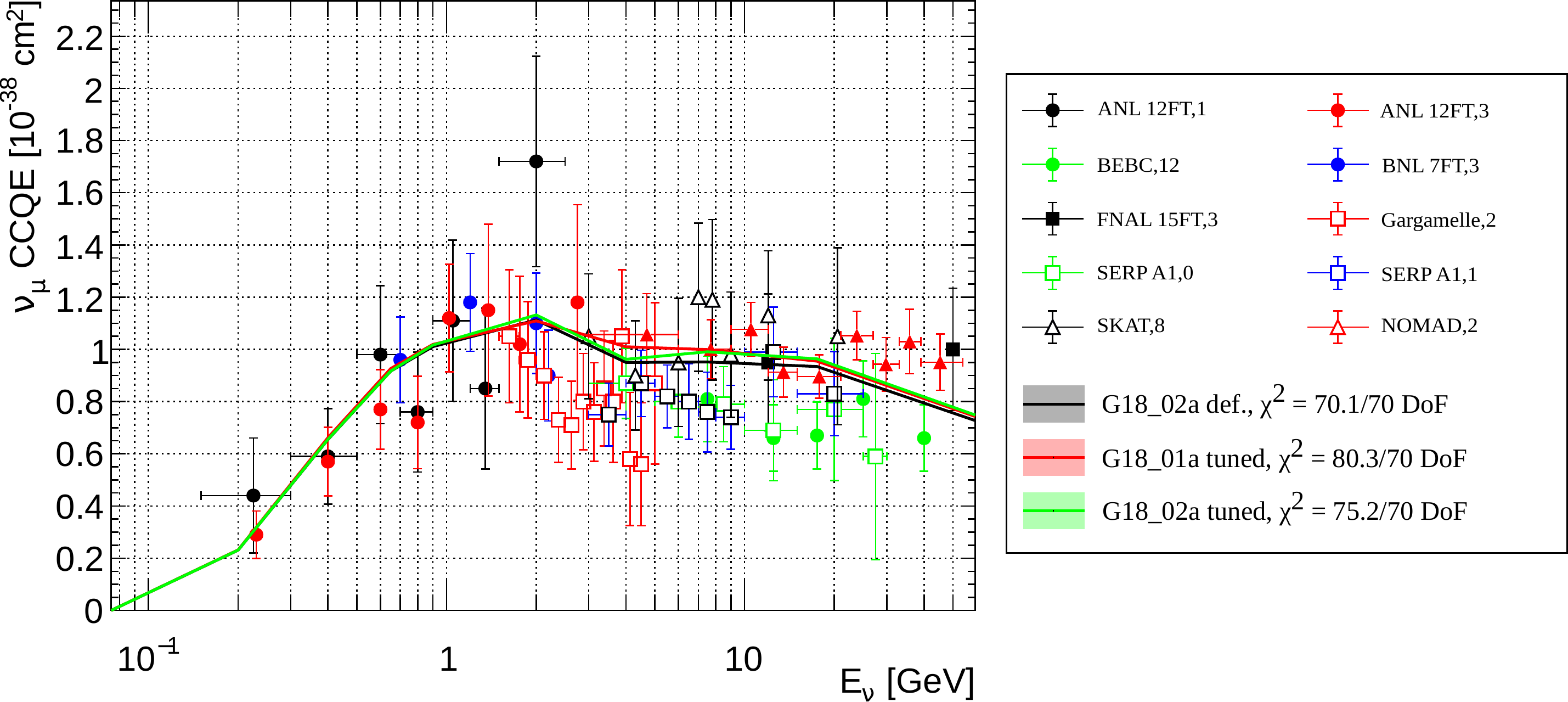}
        \caption{ Comparison of $\nu_\mu$ CC quasi-elastic cross-section data against the \emph{default} and tuned CMCs. }
    \end{subfigure} \label{fig:CCqel}
      \\
    \vspace{0.5cm}
    \begin{subfigure}{\textwidth}
\centering
        \centering\includegraphics[width=0.85\columnwidth]{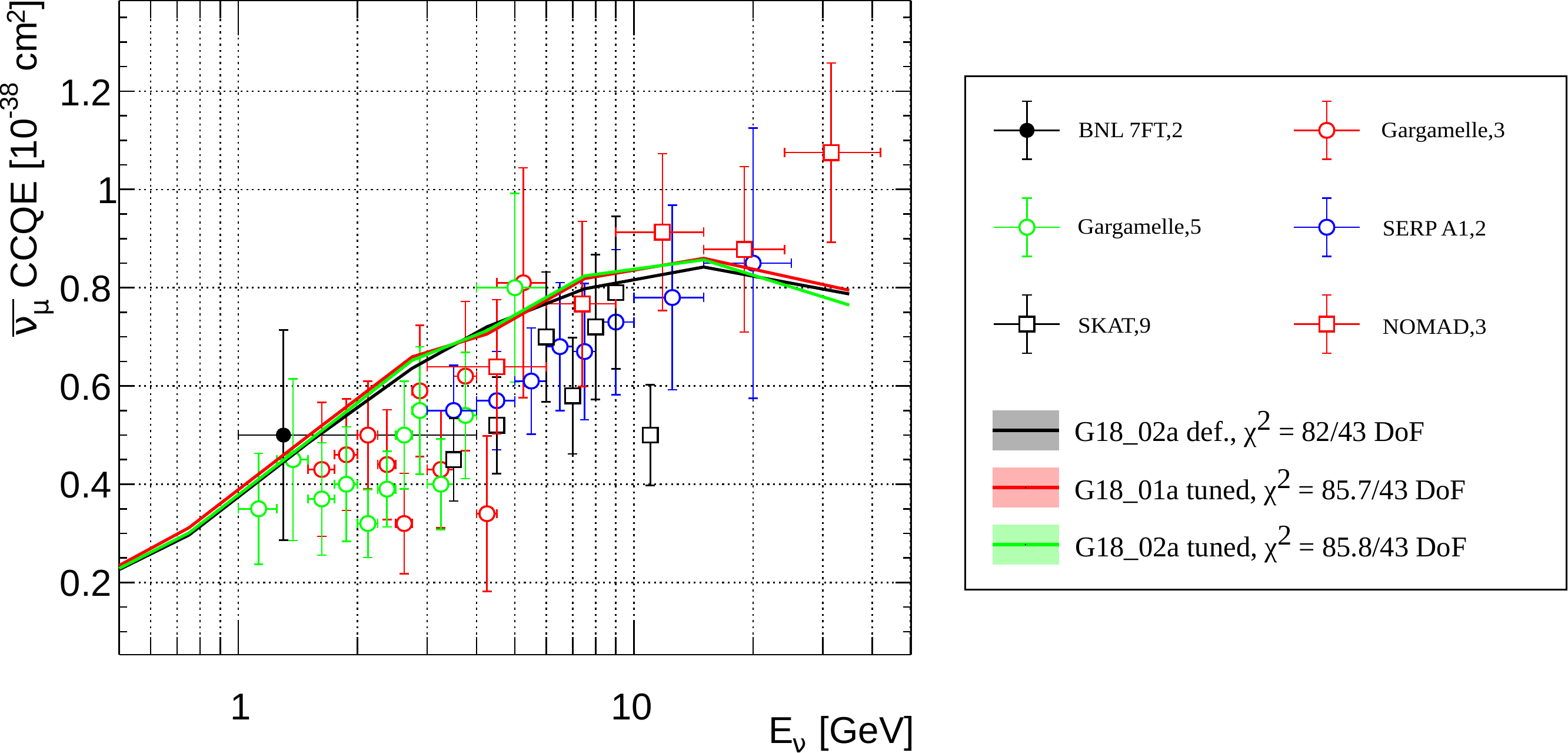}
        \caption{Comparison of $\bar{\nu}_\mu$ CC quasi-elastic cross-section data against \emph{default} and tuned CMCs.}  
    \end{subfigure}\label{barCCqel}
      \\
    \vspace{0.5cm}
    \caption{Best fit prediction impact on muon (anti)neutrino CC quasi-elastic cross sections as a function of the neutrino energy ($E_\nu$). The associated predictions for the \emph{default} G18\_02a and tuned G18\_02a and G18\_02a are computed with GENIE v3.0.6. Predictions are compared against all the available data (anti)neutrino interactions on $H$, $^2$H and heavier targets. Bot CMC have been tuned against some $H$, $^2$H data (filled markers). Each $\chi^2$ is computed using all data available. In Tab.~\ref{tab:resultsFitchi2}, the $\chi^2$ values per dataset are specified.}
    \label{fig:quasielasticPredictions}
\end{figure*}

\begin{figure*}
    \centering
    \begin{subfigure}{7.1cm}
        \centering\includegraphics[width=\columnwidth]{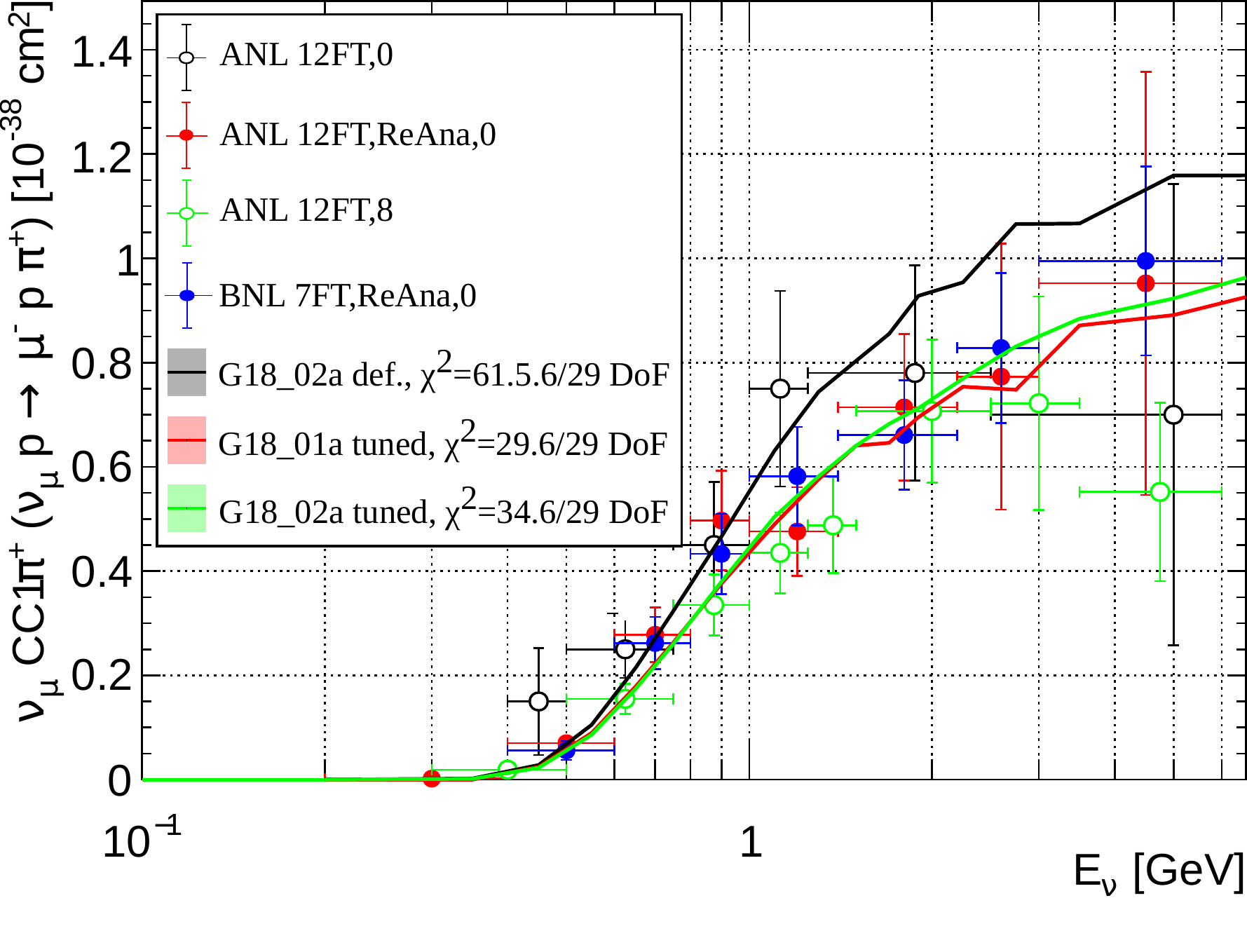}
        \caption{ Comparison of $\nu_\mu$ CC 1$\pi^+$ data on proton against the \emph{default} and tuned CMCs. }   
    \end{subfigure} \,\,\,\,\,\,
    \begin{subfigure}{7cm}
        \centering\includegraphics[width=\columnwidth]{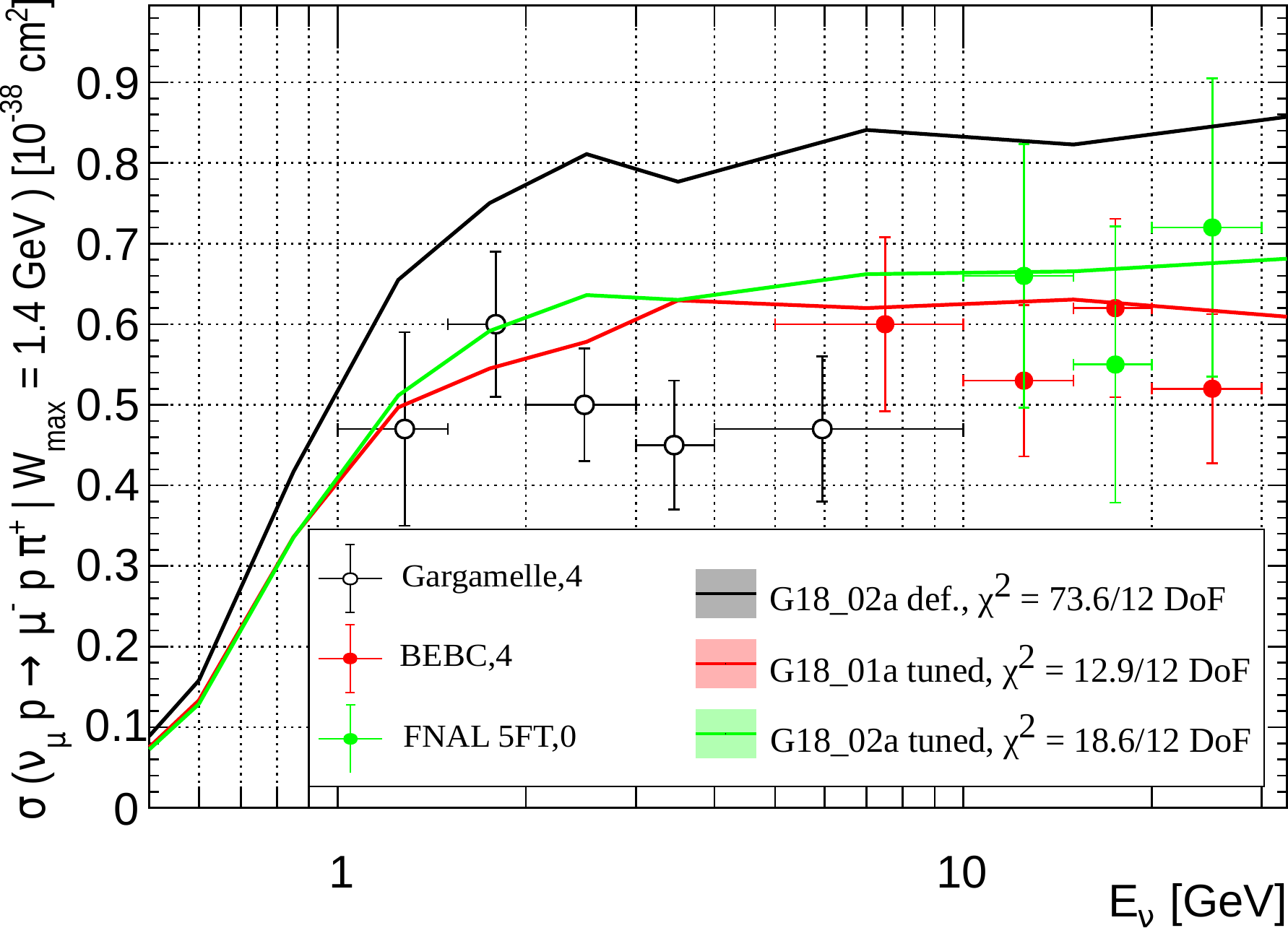}
        \caption{Comparison of $\nu_\mu$ CC 1$\pi^+$ data on proton against the \emph{default} and tuned CMCs. Experimental analysis impose a cut on $W$ at 1.4 GeV. }   
    \end{subfigure} 
      \\
    \vspace{0.5cm}
    \begin{subfigure}{7cm}
        \centering\includegraphics[width=\columnwidth]{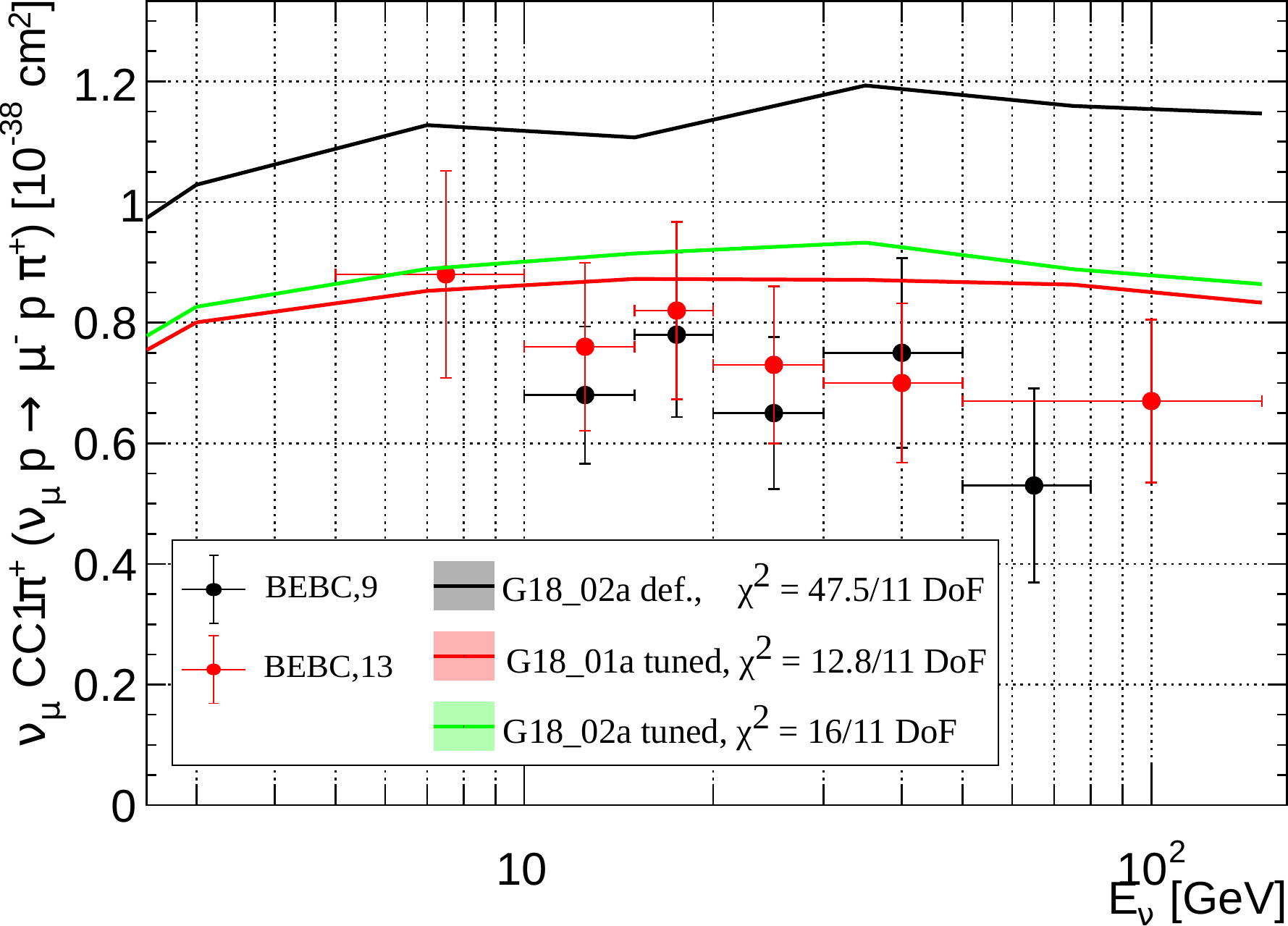}
        \caption{Comparison of $\nu_\mu$ CC 1$\pi^+$ data on proton against the \emph{default} and tuned CMCs. Experimental analysis impose a cut on $W$ at 2 GeV.}   
    \end{subfigure} \,\,\,\,\,\,
    \begin{subfigure}{7cm}
        \centering\includegraphics[width=0.8\columnwidth]{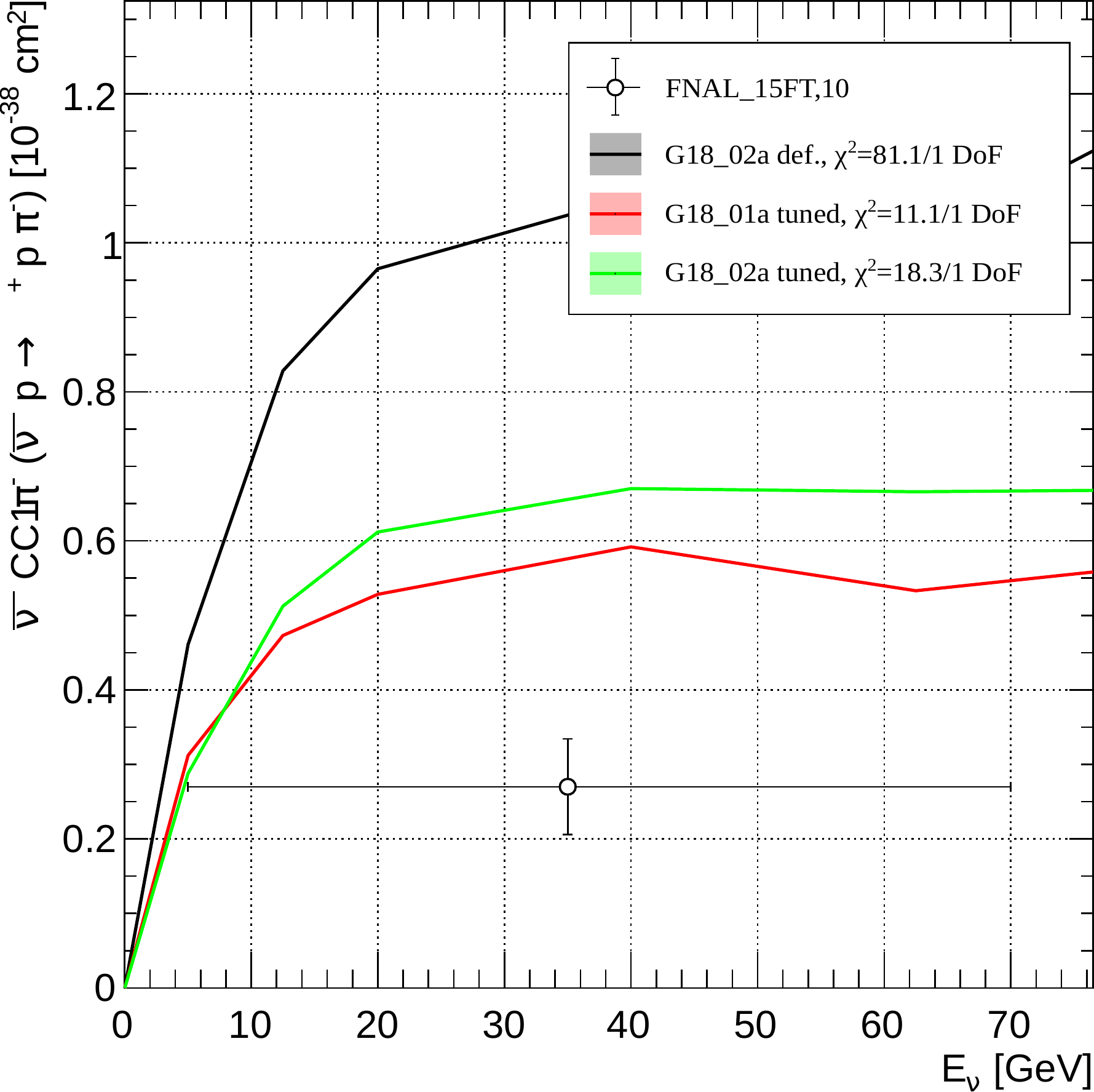}
        \caption{Comparison of $\bar{\nu}_\mu$ CC 1$\pi^+$ data on proton against the \emph{default} and tuned CMCs. The FNAL 15FT experiment applied a cut on $W$ at 1.9 GeV }   
    \end{subfigure}
      \\
    \vspace{0.5cm}
    \caption{Best fit prediction impact on muon neutrino on proton CC one pion production cross sections as a function of the neutrino energy ($E_\nu$). The associated predictions for the \emph{default} G18\_02a and tuned G18\_02a and G18\_02a are computed with GENIE v3.0.6. Experimental cuts are also applied to the predictions when needed. Predictions are compared against all the available data (anti)neutrino interactions on $H$, $^2$H and heavier targets. Both CMC have been tuned against some $H$, $^2$H data (filled markers). Each $\chi^2$ is computed using all data available. In Tab.~\ref{tab:resultsFitchi2}, the $\chi^2$ values per dataset are specified.}
    \label{fig:pPredictions}
\end{figure*}

\begin{figure*}
    \centering
    \begin{subfigure}{7cm}
        \centering\includegraphics[width=\columnwidth]{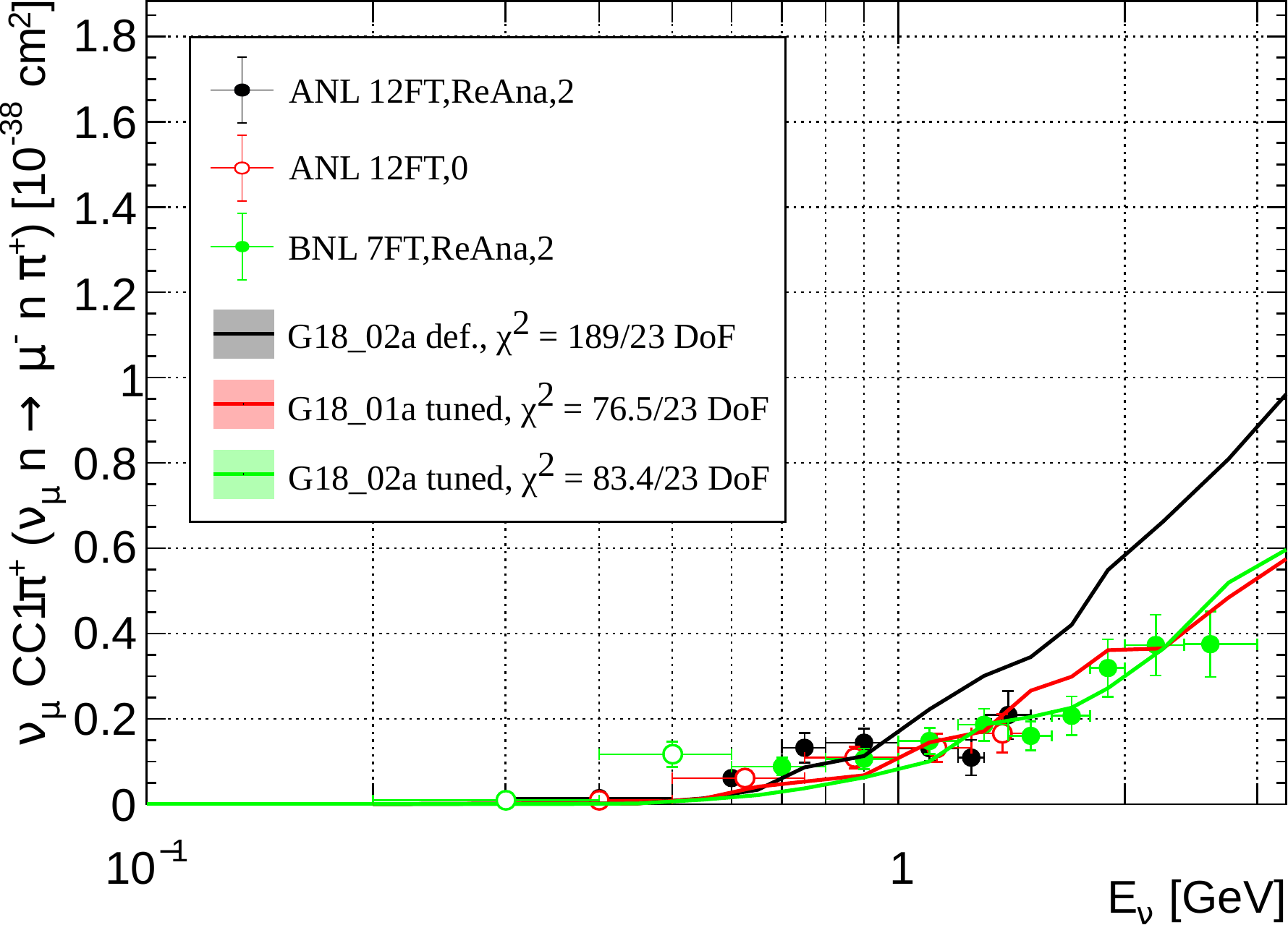}
        \caption{Comparison of $\nu_\mu$ CC 1$\pi^+$ data on neutron against the \emph{default} and tuned CMCs. }
    \end{subfigure}  \,\,\,\,\,\,
    \begin{subfigure}{7cm}
        \centering\includegraphics[width=\columnwidth]{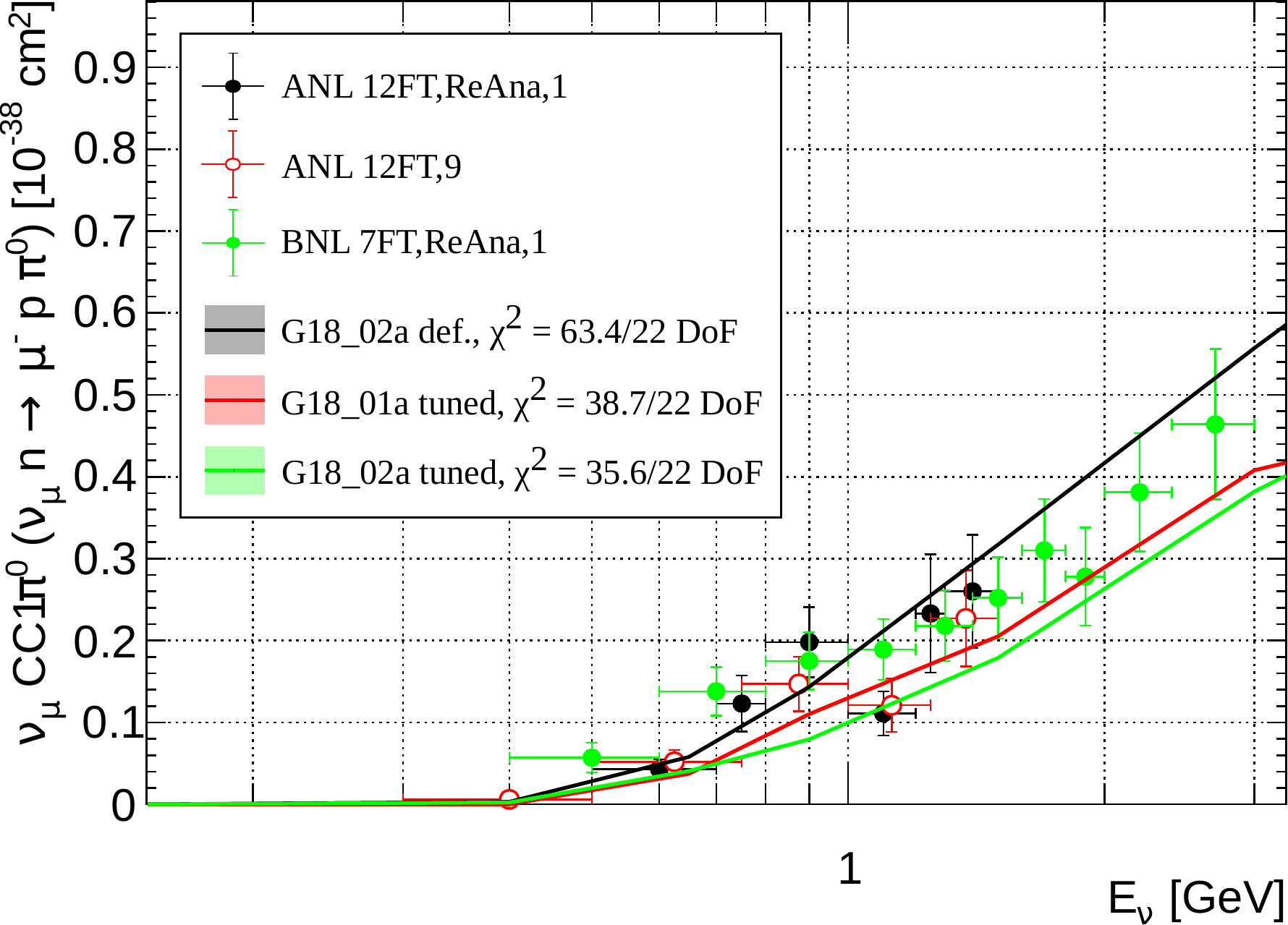}
        \caption{Comparison of $\nu_\mu$ CC 1$\pi^0$ data on neutron against the \emph{default} and tuned CMCs. }   
    \end{subfigure} 
    \caption{Best fit prediction impact on muon neutrino on neutron CC one pion production cross sections as a function of the neutrino energy ($E_\nu$). The associated predictions for the \emph{default} G18\_02a and tuned G18\_02a and G18\_02a are computed with GENIE v3.0.6. Predictions are compared against the original and reanalized ANL 12FT and BNL 7FT data \cite{Wilkinson:2014yfa,Radecky:1981fn}. Only reanalized data with $E_\nu>0.5$ GeV is used in the tune (filled markers). Each $\chi^2$ is computed using all data available. }
    \label{fig:nPredictions}
\end{figure*}

\begin{figure*}
    \centering
    \begin{subfigure}{7cm}
      \centering\includegraphics[height=\columnwidth]{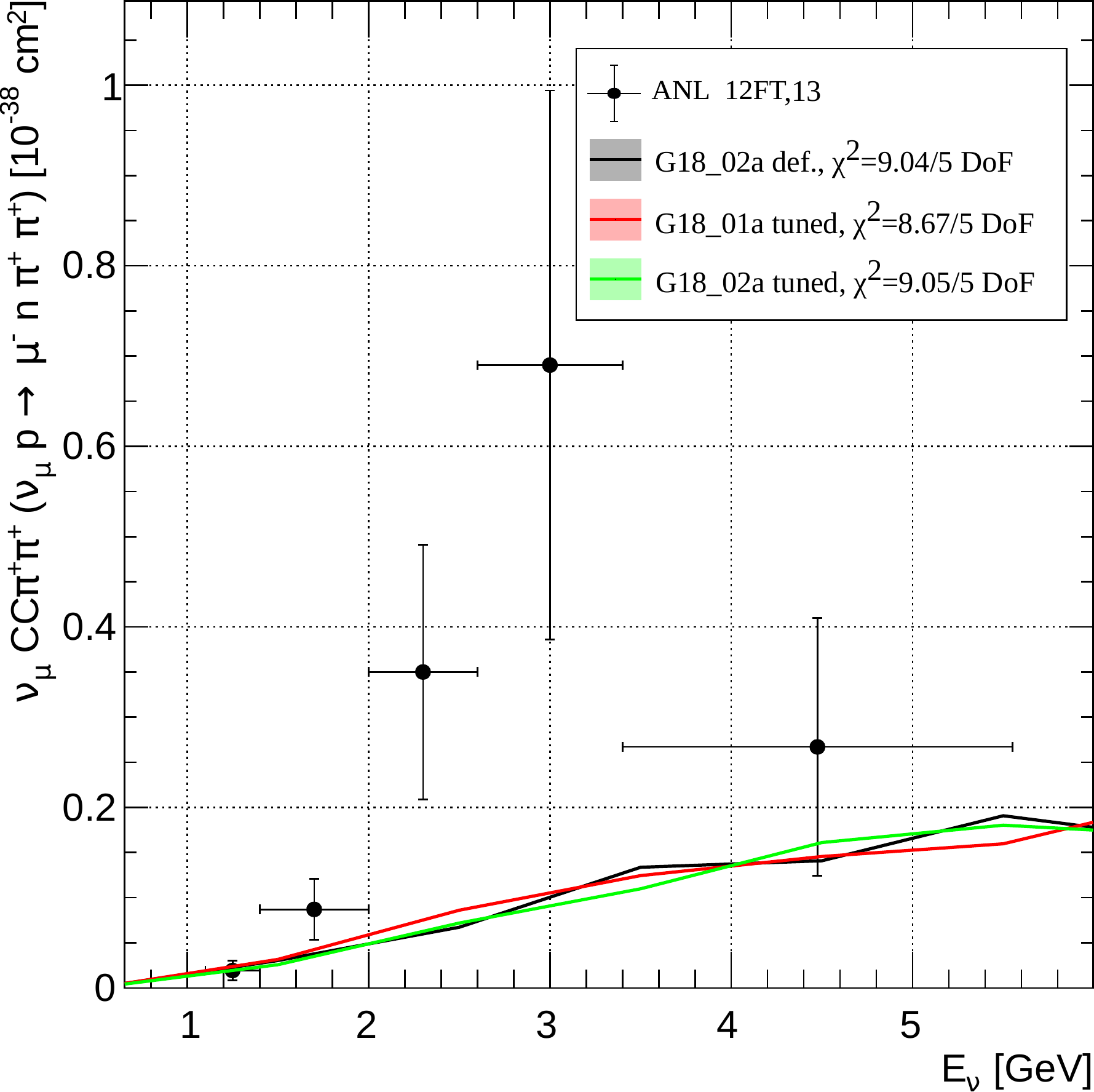}
        \caption{ Comparison of $\nu_\mu$ CC 2$\pi^+$ data on proton.}
    \end{subfigure}  \,\,\,\,\,\,
    \begin{subfigure}{7cm}
\centering
        \centering\includegraphics[height=\columnwidth]{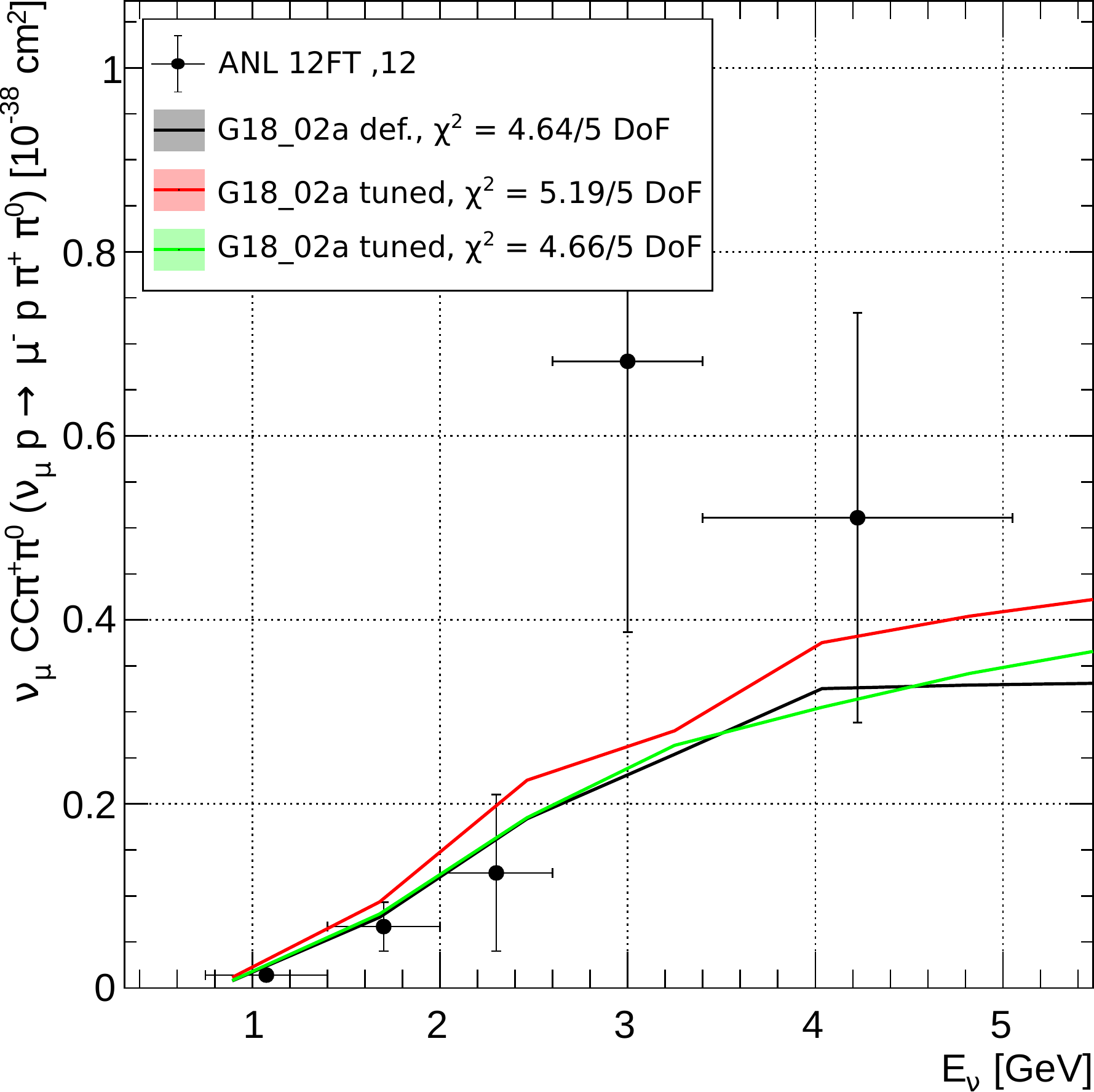}
        \caption{ Comparison of $\nu_\mu$ CC $\pi^+\pi^0$ data on proton.}

    \end{subfigure}
      \\
    \vspace{0.5cm}
    \begin{subfigure}{8cm}
        \centering\includegraphics[width=\columnwidth]{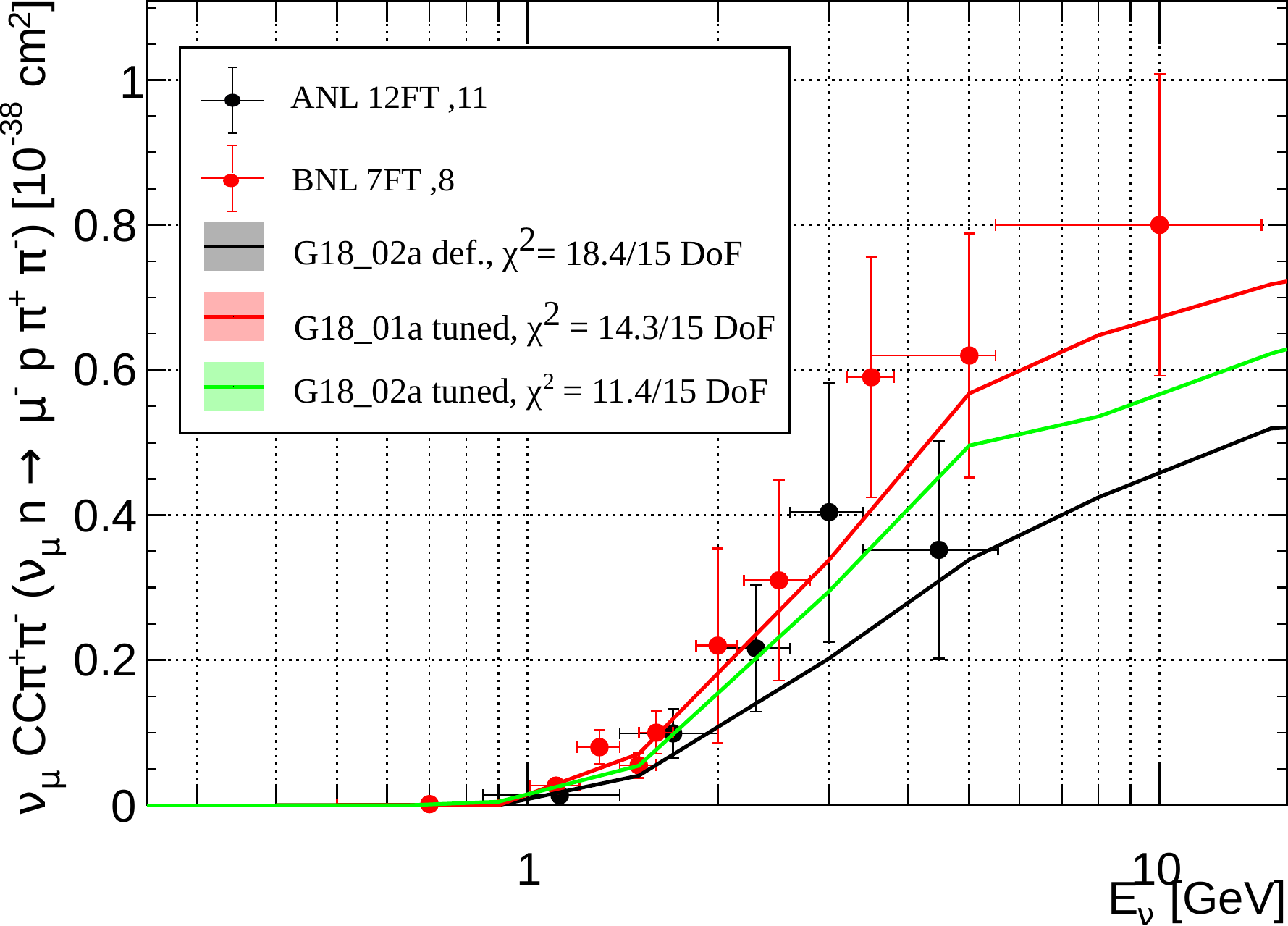}
        \caption{ Comparison of $\nu_\mu$ CC $\pi^+\pi^-$ data on neutron.}
    \end{subfigure} \caption{Best fit prediction impact on muon neutrino CC two-pion production cross sections as a function of the neutrino energy ($E_\nu$). The comparisons to two-pion production data are shown against the \emph{default} and tuned CMCs. The associated predictions for the \emph{default} G18\_02a and tuned G18\_02a and G18\_02a are computed with GENIE v3.0.6. Predictions are compared against ANL 12FT and BNL 7FT data.}
    \label{fig:twoPredictions}
\end{figure*}

The analysis procedure outlined in previous sections was applied to the comprehensive model configurations listed in Sec.~\ref{sec:ComprehensiveConfigurations}. 
The best-fit parameter values obtained from the GENIE analysis for each alternative CMC are shown in Tab.~\ref{tab:BestFitValuesAndErrors} and Tab.~\ref{tab:BestFitNuisanceParameters}. 
The GENIE v3 cross-section curves that correspond to the two sets of tuned parameters are shown in Figs.~\ref{fig:inclusivePredictions}, \ref{fig:quasielasticPredictions}, \ref{fig:pPredictions}, \ref{fig:twoPredictions} and \ref{fig:nPredictions}. 
For reference, we also show the cross-section predictions made by the \emph{default} G18\_02a CMC that is available in the last public release of the GENIE v3 series (3.2).

\begin{table}
    \centering
    \begin{tabular}{c c c}  \hline\hline\noalign{\smallskip}
    Parameter                    & G18\_01a(/b)  & G18\_02a(/b)  \\ 
    \noalign{\smallskip}\hline\noalign{\smallskip}
    $W_{\text{cut}}$             & $1.94$        & $1.81$ \\
    \noalign{\smallskip} 
    $M_A^{\text{QE}}$            & $1.00\pm0.01$ & $1.00\pm0.013$ \\
    \noalign{\smallskip} 
    $M_A^{\text{RES}}$           & $1.09\pm0.02$ & $1.09\pm0.014$ \\
    \noalign{\smallskip}
    $R_{\nu p}^{\text{CC}1\pi}$  & $0.06\pm0.03$ & $0.008$        \\
    $R_{\nu p}^{\text{CC}2\pi}$  & $1.1\pm0.2$   & $0.94\pm0.075$ \\
    $R_{\nu n}^{\text{CC}1\pi}$  & $0.14\pm0.03$ & $0.03\pm0.010$ \\
    $R_{\nu n}^{\text{CC}2\pi}$  & $2.8\pm0.4$   & $2.3\pm0.12$   \\
    $S_{\text{RES}}$             & $0.89\pm0.04$ & $0.84\pm0.028$ \\
    $S_{\text{DIS}}$             & $1.03\pm0.02$ & $1.06\pm0.01$  \\
    \noalign{\smallskip}\hline\noalign{\smallskip}
    $\chi^2$/157 DoF      & $1.84$ & $1.64$ \\
    \noalign{\smallskip}\hline\hline
    \end{tabular}
    \caption{Best-fit parameter values and 
     parameter ranges obtained by requiring that 
     ${\Delta\chi^2_{\text{profiled}}<\Delta\chi^2_{\text{critical}}=1}$.
         Results are shown for all alternative CMCs considered in this analysis. The best-fit values obtained for the G18\_02a(/b) CMC can be used for the G18\_10a(/b) as the same bare-nucleon underlying models are used. Moreover, for the G18\_10i(/j) CMCs, the best-fit values from the G18\_02a(/b) tune can also be used, except for $M_A^{\text{QE}}$, as the quasi-elastic axial form factor is parametrised with the z-expansion model instead of a dipole and the corresponding z-expansion parameters are kept to the \emph{default} values.
     }
    \label{tab:BestFitValuesAndErrors}
\end{table} 

\begin{table}    
    \centering
    \begin{tabular}{c c c c}  \hline\hline\noalign{\smallskip}
    Parameter                    & G18\_01a(/b)  & G18\_02a(/b)  \\ 
    \noalign{\smallskip}\hline\noalign{\smallskip}
    $f^{\text{ANL}}(\nu_\mu)$             & $0.98\pm0.05$ & $0.89\pm0.05$ \\
    $f^{\text{ANL}}_{\text{Re}}(\nu_\mu)$ & $1.12\pm0.05$ & $1.2\pm0.05$  \\
    $f^{\text{BNL}}(\nu_\mu)$             & $1.01\pm0.04$ & $1.06\pm0.04$ \\
    $f^{\text{BNL}}_{\text{Re}}(\nu_\mu)$ & $1.08\pm0.05$ & $1.03\pm0.04$ \\
    $f^{\text{BNL}}(\bar{\nu}_\mu)$       & $1.00\pm0.10$ & $0.99\pm0.10$ \\
    $f^{\text{BEBC}}(\nu_\mu)$            & $0.91\pm0.04$ & $0.86\pm0.03$ \\
    $f^{\text{BEBC}}(\bar{\nu}_\mu)$      & $1.04\pm0.04$ & $0.99\pm0.03$ \\
    $f^{\text{FNAL}}(\nu_\mu)$            & $0.97\pm0.04$ & $0.94\pm0.04$ \\
    $f^{\text{FNAL}}(\bar{\nu}_\mu)$      & $0.95\pm0.05$ & $0.92\pm0.05$ \\
    \noalign{\smallskip}\hline\hline
    \end{tabular}
    
     \caption{Best-fit nuisance parameters, $f^{j}$, per experiment (ANL 12FT, BNL 7FT, BEBC or FNAL 15FT) and neutrino beam ($\nu_\mu$ or $\bar{\nu}_\mu$). 
     The nuisance parameters included in the fit are independent of GENIE. }
    \label{tab:BestFitNuisanceParameters}
\end{table}

For all CMCs the tune has the most impact on the SIS region. 
In the inclusive cross-section prediction, this translates into a decrease of both $\nu_\mu$ and $\bar{\nu}_\mu$ CC inclusive cross section in the 0.5-10 GeV region, see Fig.~\ref{fig:inclusivePredictions}.
At the same time, the cross section at higher neutrino energies has barely changed, respecting the constraints of high-energy data. 
The agreement with quasi-elastic data, included in the tune in order to constrain the fluxes of each experiment, remained the same, see Fig.~\ref{fig:quasielasticPredictions}. 

As discussed in Sec.~\ref{sec:TuningResults}, this decrease of the inclusive cross section at the SIS region is driven mainly by one pion production data. 
The impact on one pion exclusive channels is shown for (anti)neutrino on proton, Fig.~\ref{fig:pPredictions}, and neutrino on neutron, Fig.~\ref{fig:nPredictions}. 
The reduction of the one pion production cross section for neutrino on proton and neutron shows an improvement on $\nu_\mu \text{CC}1\pi^+$, $\nu_\mu$ and $\nu_\mu \text{CC}1\pi^-$ and $\nu_\mu \text{CC}1\pi^0$ channels when comparing it with the available data.

Two pion production exclusive cross sections are summarized in Fig.~\ref{fig:twoPredictions}. 
This is the first time that two pion production data are used to tune the SIS region, allowing the $R_{\nu p}^{\text{CC}2\pi}$ and $R_{\nu n}^{\text{CC}2\pi}$ parameters to be constrained. 
In this case, the two pion exclusive cross section was underestimated by the \emph{default} tune. 
For this particular exclusive process, comparisons are made against $\nu_\mu \text{CC}\pi^+\pi^+$, $\nu_\mu \text{CC}\pi^+\pi^0$ and $\nu_\mu \text{CC}\pi^+\pi^-$ data. 
The shape of the GENIE prediction for the $\nu_\mu \text{CC}\pi^+\pi^+$ and  $\nu_\mu \text{CC}\pi^+\pi^0$ channels differs strongly from data, and the models are not able to accommodate this behaviour. 
However, the agreement with $\nu_\mu \text{CC}\pi^+\pi^-$ data has improved by increasing the cross section with respect to the \emph{default} cross-section model.

Despite the tensions between inclusive and exclusive data discussed in Sec.~\ref{sec:model_uncertaintines}, the overall agreement for both cross-section model constructions has improved, see Tab.~\ref{tab:resultsFitchi2}. 
Particularly, $\bar{\nu}_\mu$ CC inclusive predictions show better agreement after the tune, and the same is observed for $\nu_\mu$ CC inclusive predictions for the G18\_01a free nucleon tune.
Although the impact on the cross-section prediction of the tune is similar for the existing configurations, the response of each model at the parameter level is not expected to be the same. 
Therefore, each tune is strongly affected by how the model is able to accommodate the data by modifying the tuned parameters. 
This reflects on the $R_m$ parameters and $W_{\text{cut}}$ which best fit values are incompatible in some cases, such as for $R_{\nu n}^{\text{CC}1\pi}$, see Tab.~\ref{tab:BestFitValuesAndErrors}. 
Particularly, the behaviour of $R_{\nu p}^{\text{CC}1\pi}$ on the G18\_02a(/b) tune was showing preference for nonphysical regions of the tune, forcing us to fix this value to $R_{\nu p}^{\text{CC}1\pi} = 0.008$. 
On the other hand, the remaining parameters, such as $M_{A}^{\text{RES}}$ and $M_{A}^{\text{QE}}$, show agreement between the tunes and respect the applied priors. 
 
\subsection{Parameter error estimation}
\begin{figure*}
\centering
    \begin{subfigure}{7cm}
        \centering\includegraphics[width=\columnwidth]{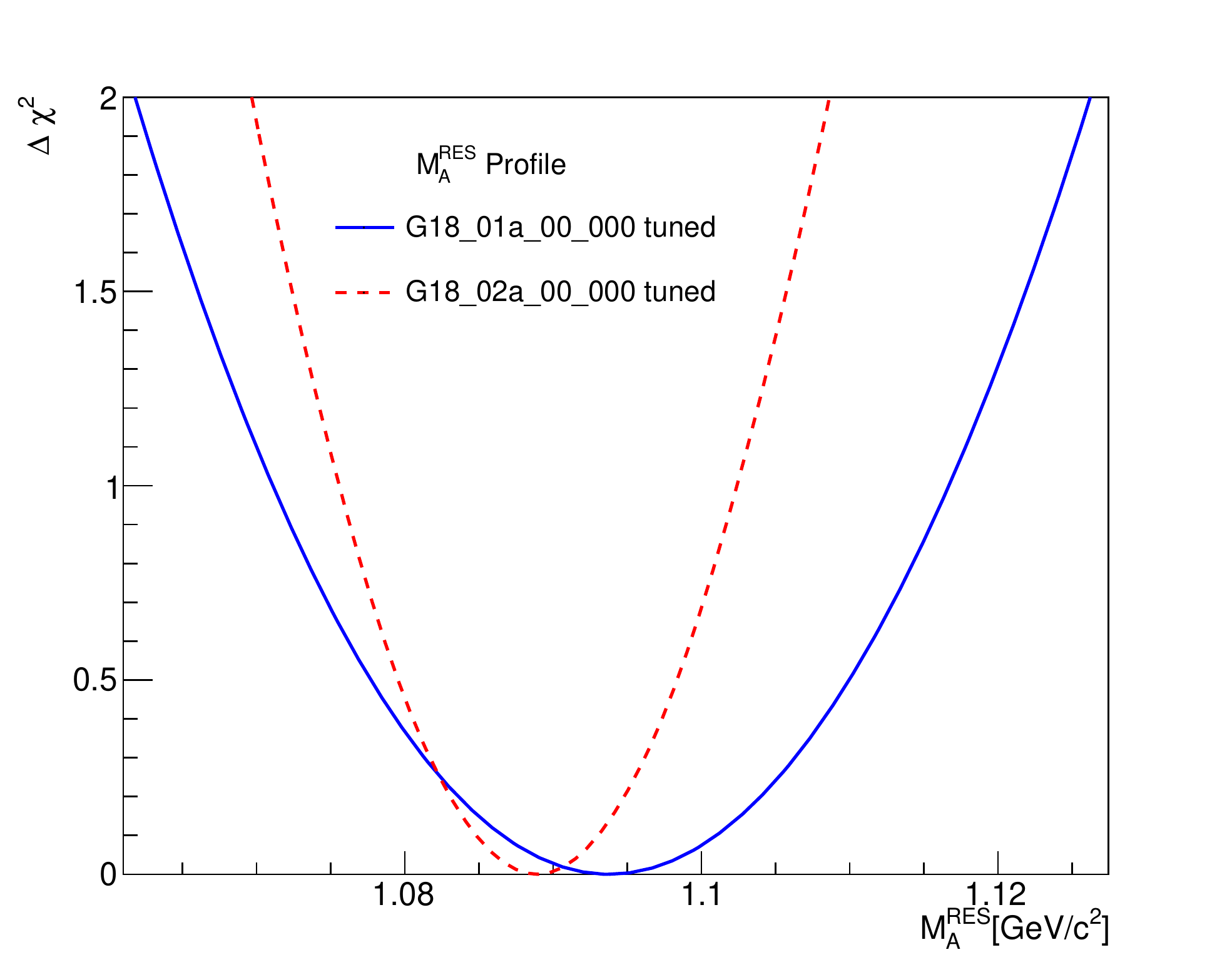}
        \caption{Profile for $M_{A}^{\text{RES}}.$}    
    \end{subfigure}
    \begin{subfigure}{7cm}
        \centering\includegraphics[width=\columnwidth]{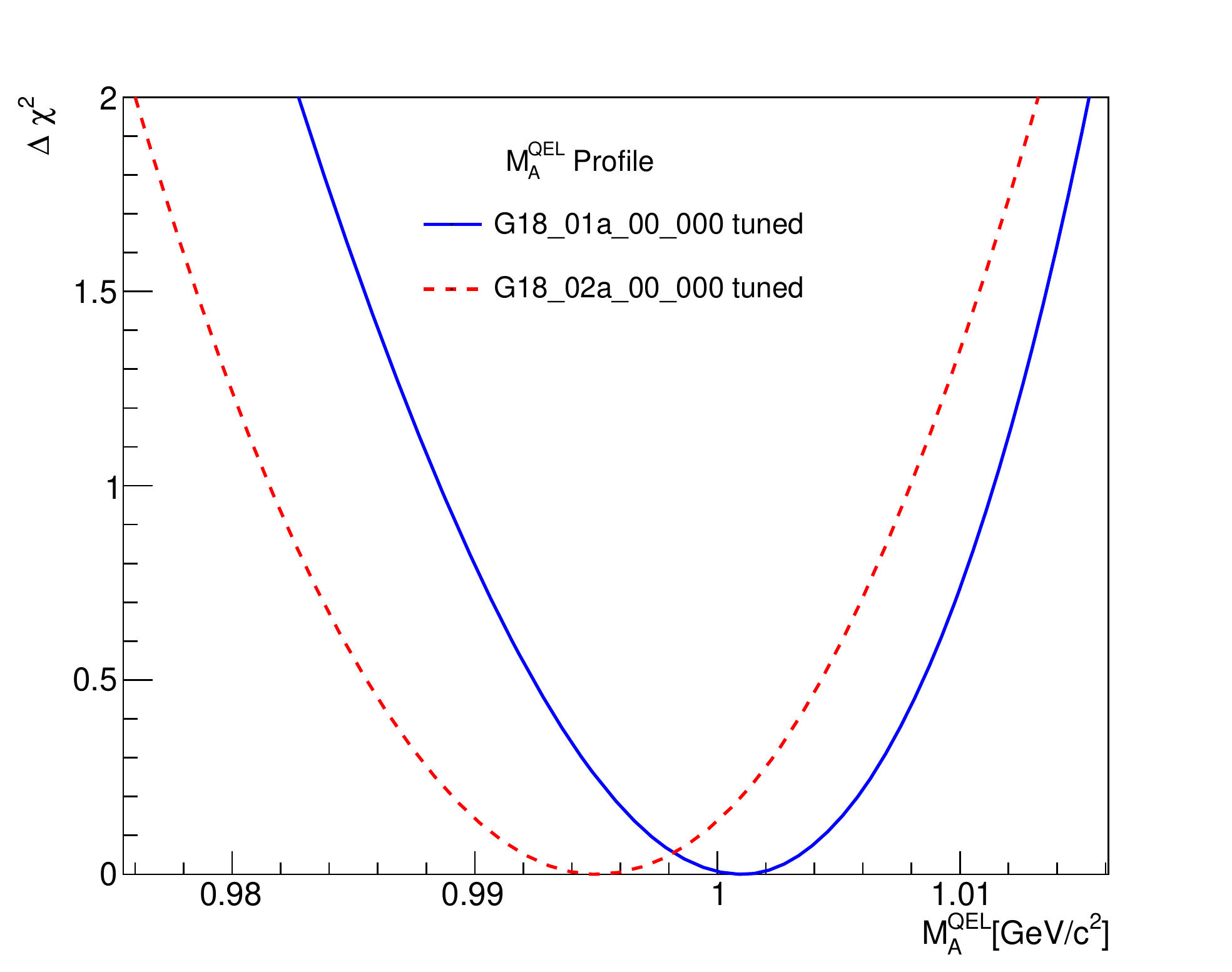}
        \caption{Profile for $M_{A}^{\text{QE}}.$}   
    \end{subfigure}   
    
    \begin{subfigure}{7cm}
        \centering\includegraphics[width=\columnwidth]{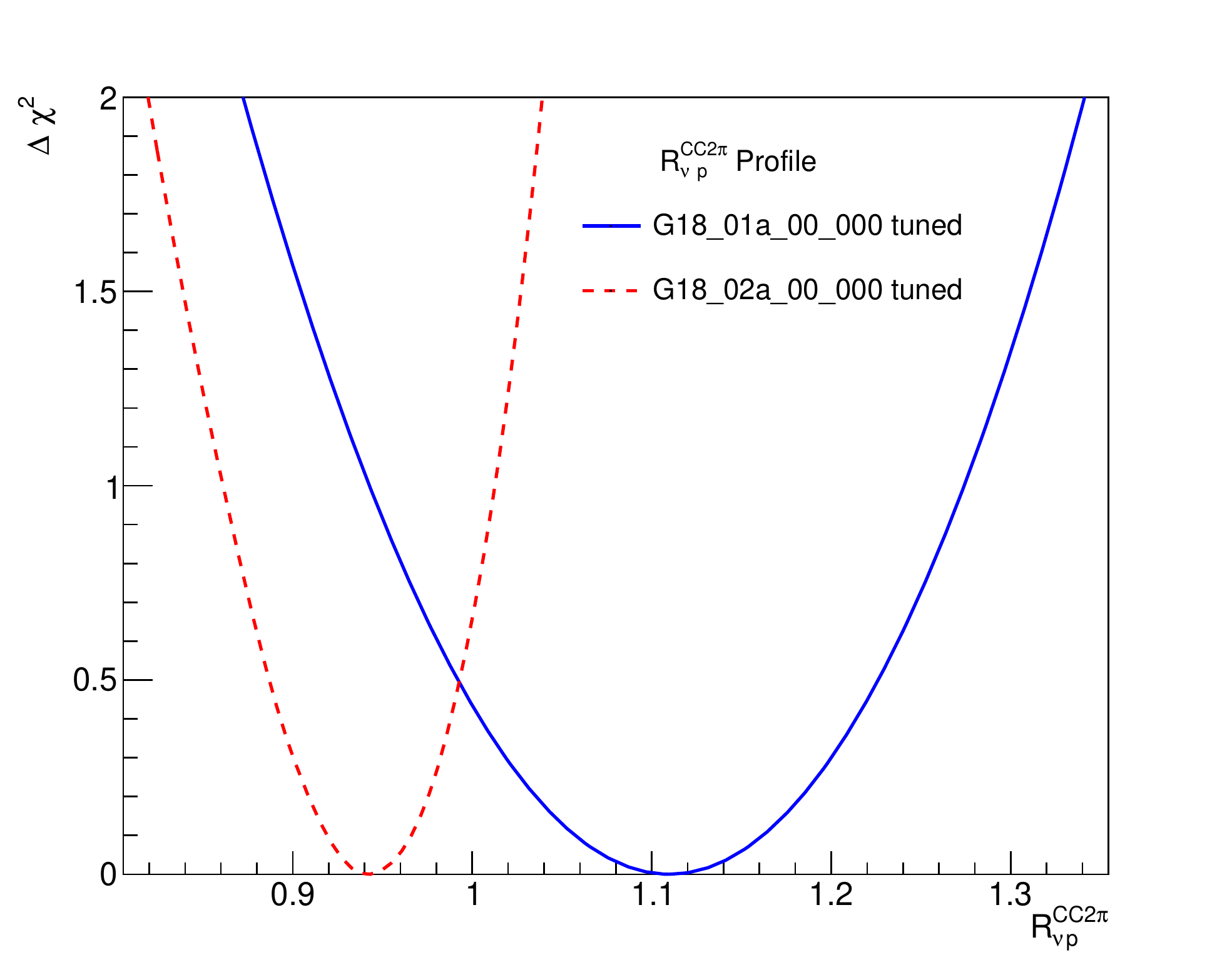}
        \caption{Profile for $R_{\nu p}^{\text{CC}2\pi}$.}    
    \end{subfigure}  
    \begin{subfigure}{7cm}
        \centering\includegraphics[width=\columnwidth]{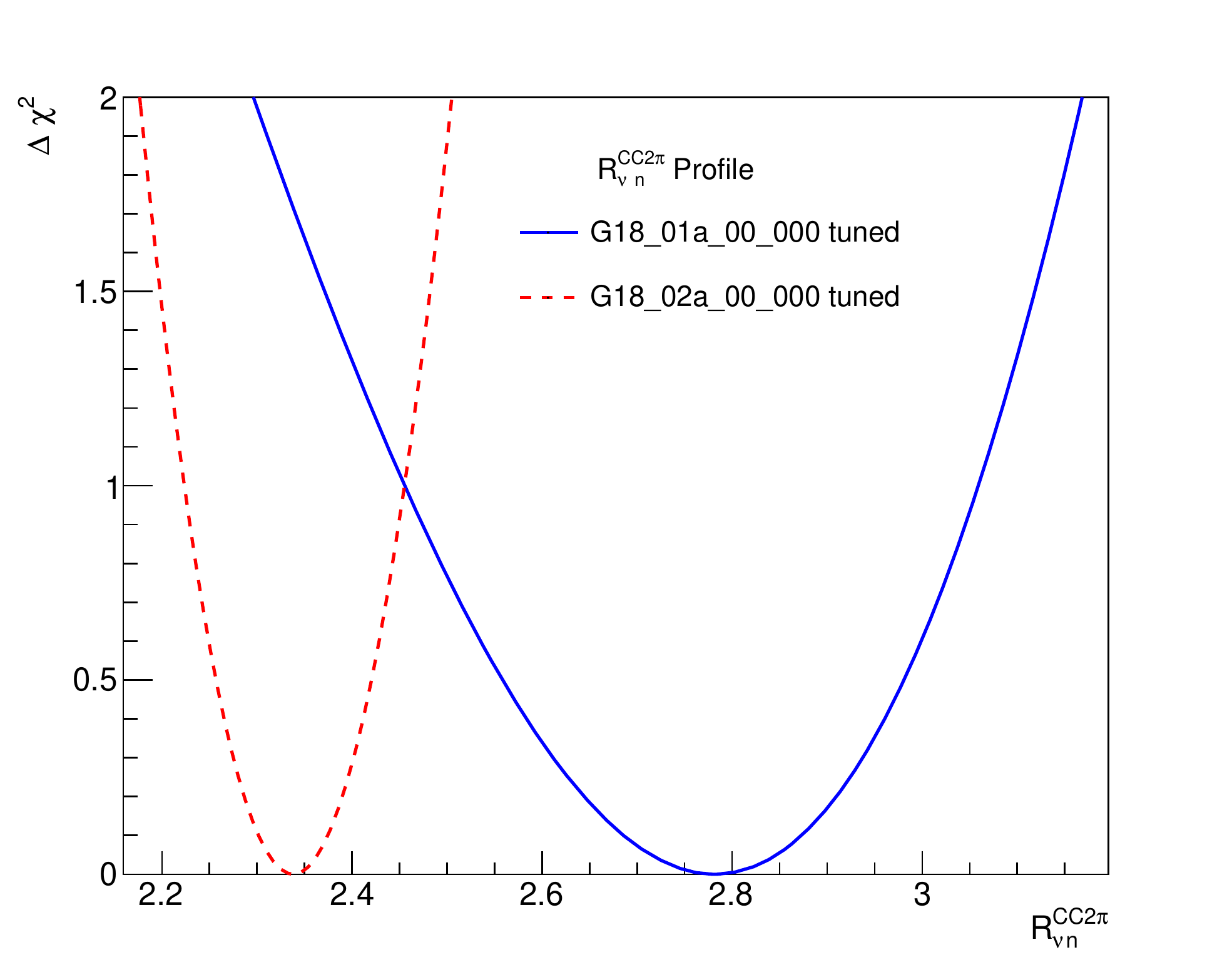}
        \caption{Profile for $R_{\nu n}^{\text{CC}2\pi}$.}    
    \end{subfigure}
    
    \begin{subfigure}{7cm}
        \centering\includegraphics[width=\columnwidth]{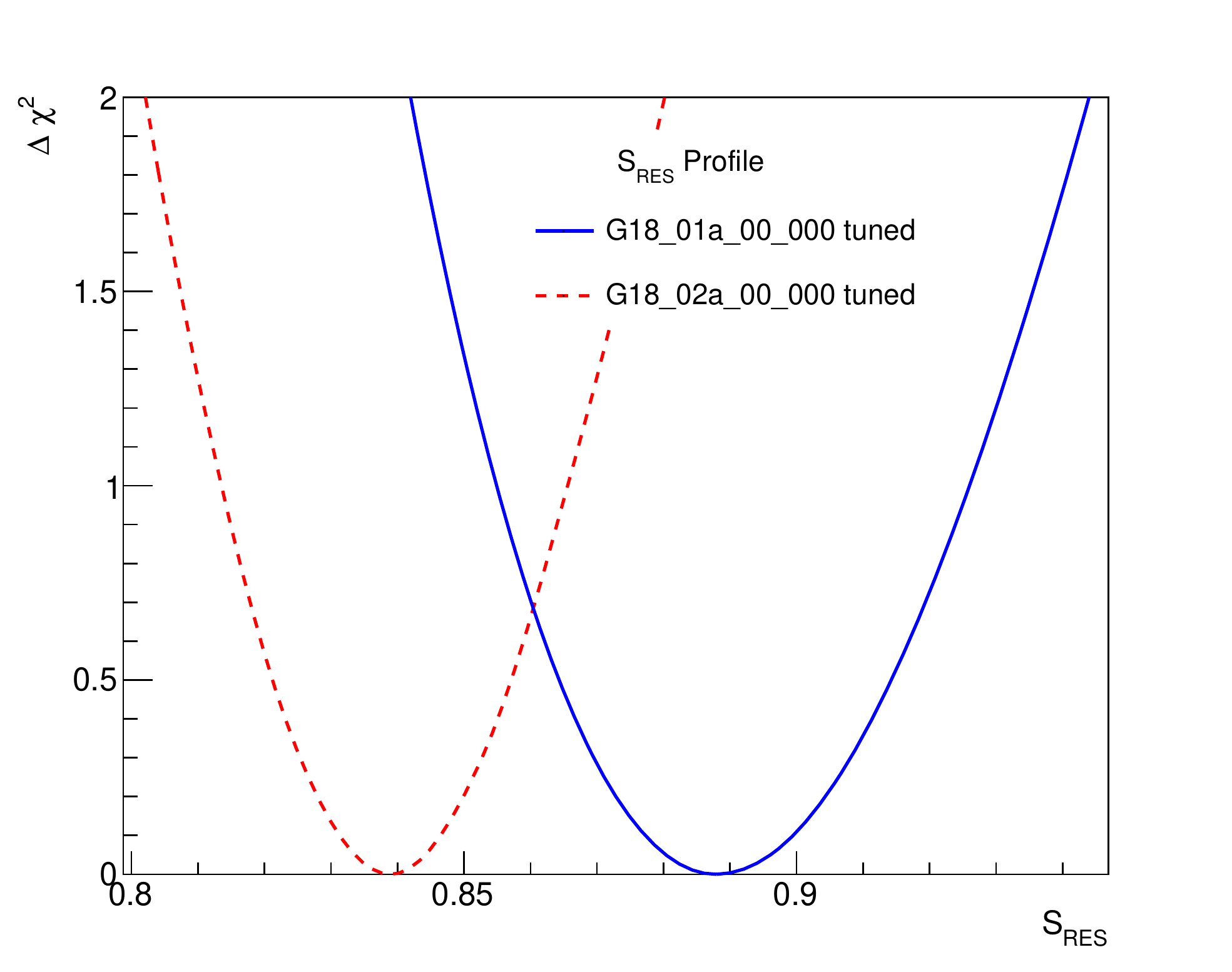}
        \caption{Profile for $S_{\text{RES}}$.}   
    \end{subfigure}
    \begin{subfigure}{7cm}
        \centering\includegraphics[width=\columnwidth]{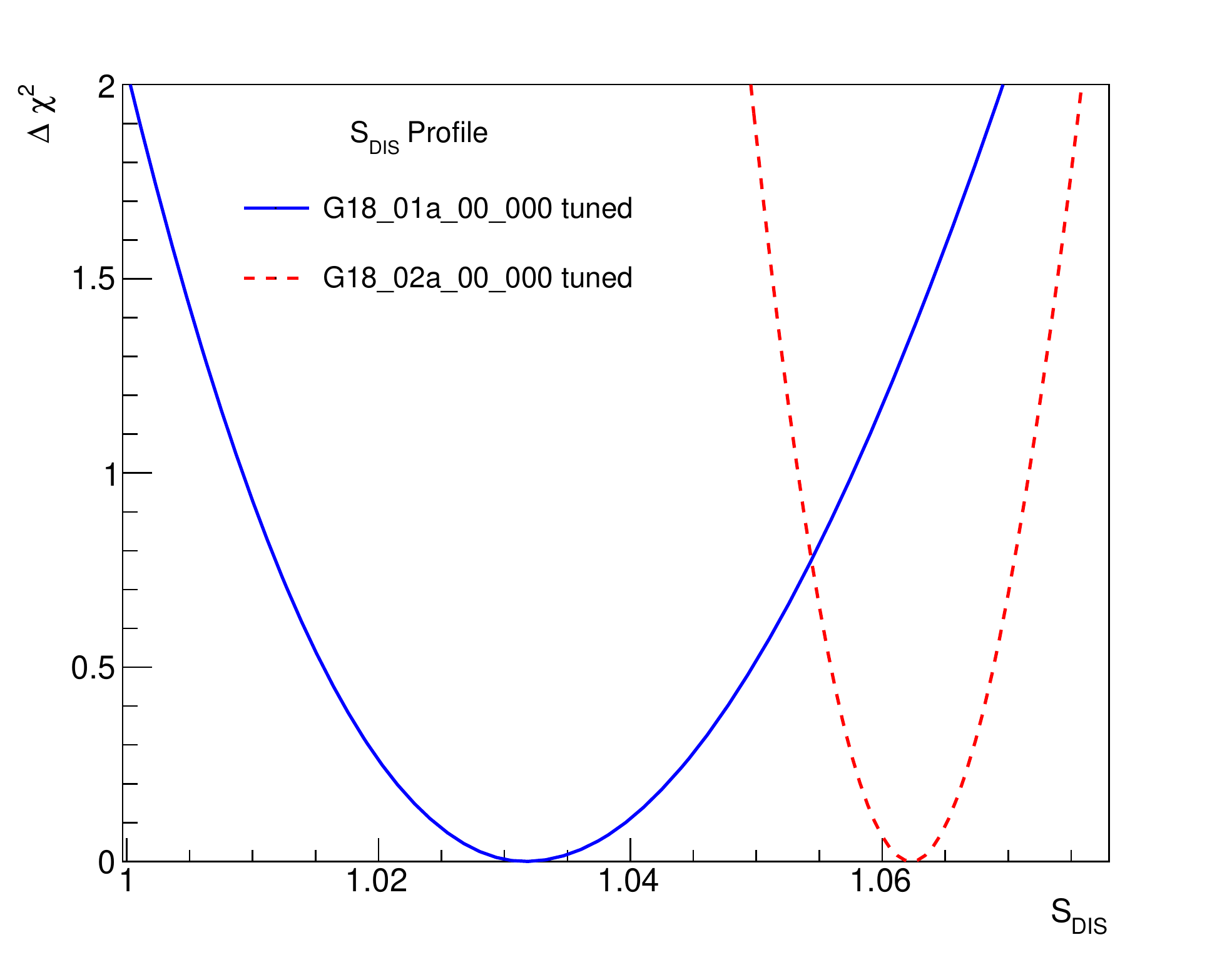}
        \caption{Profile for $S_{\text{DIS}}$.}  
    \end{subfigure} 
    \caption{$\Delta\chi^2_{\text{profile}}(\theta_i)$ functions obtained fixing the parameter under study and minimizing $\Delta\chi^2(\theta_i) = \chi^2(\vec{\theta},\vec{f}) - \chi^2_{\text{min}} $ respect the other parameters in the tune. The profiles for both tunes are shown for each case. The G18\_02a profiles show that this configuration is less able to accommodate in the model variations on each parameter best fit values.  } 
    \label{fig:profiles}
\end{figure*}

An estimate of the parameter uncertainties is shown in Tab.~\ref{tab:BestFitValuesAndErrors}. 
For each parameter of interest allowed to float in the fit, the table shows the range of values that satisfies the condition ${\Delta\chi^2_{\text{profile}}(\theta_i) < \Delta\chi^2_{\text{critical}} = 1}$. 
In the previous expression, the function $\Delta\chi^2_{\text{profile}}(\theta_i)$ is constructed by fixing $\theta_i$ to a desired value and minimising the quantity $\Delta\chi^2(\boldsymbol{\theta},\boldsymbol{f}) = 
\chi^2(\boldsymbol{\theta},\boldsymbol{f}) - \chi^2_{\text{min}}$ with respect to all other parameters that were allowed to float in the fit. 
See Sec.~\ref{sec:ConstructionOfLikelihood} for the definition of $\chi^2(\boldsymbol{\theta},\boldsymbol{f})$. 
The constant $\chi^2_{\text{min}}$ corresponds to the minimum value of $\chi^2(\boldsymbol{\theta},\boldsymbol{f})$ obtained from the global fit. 
The $\Delta\chi^2_{\text{profile}} (\theta_i)$ functions we derive from our analysis are shown in Fig.~\ref{fig:profiles}, for all parameters $\theta_i$ that were allowed to float in the fit, up to $\Delta\chi^2_{\text{profile}}$ values of 2. 
Particularly, $W_{\text{cut}}$ is fixed to the best fit value during this approach, as it is an ad-hoc parameter introduced by the generator: by fixing it, its uncertainty will be reflected on the other parameters. 
It is important to emphasize that the uncertainties quoted relate only to $\Delta\chi^2_{\text{critical}}=1$. 
However, this region is strongly determined by the underlying model used in the tune. 

\begin{figure}
    \centering
    \includegraphics[width=\textwidth]{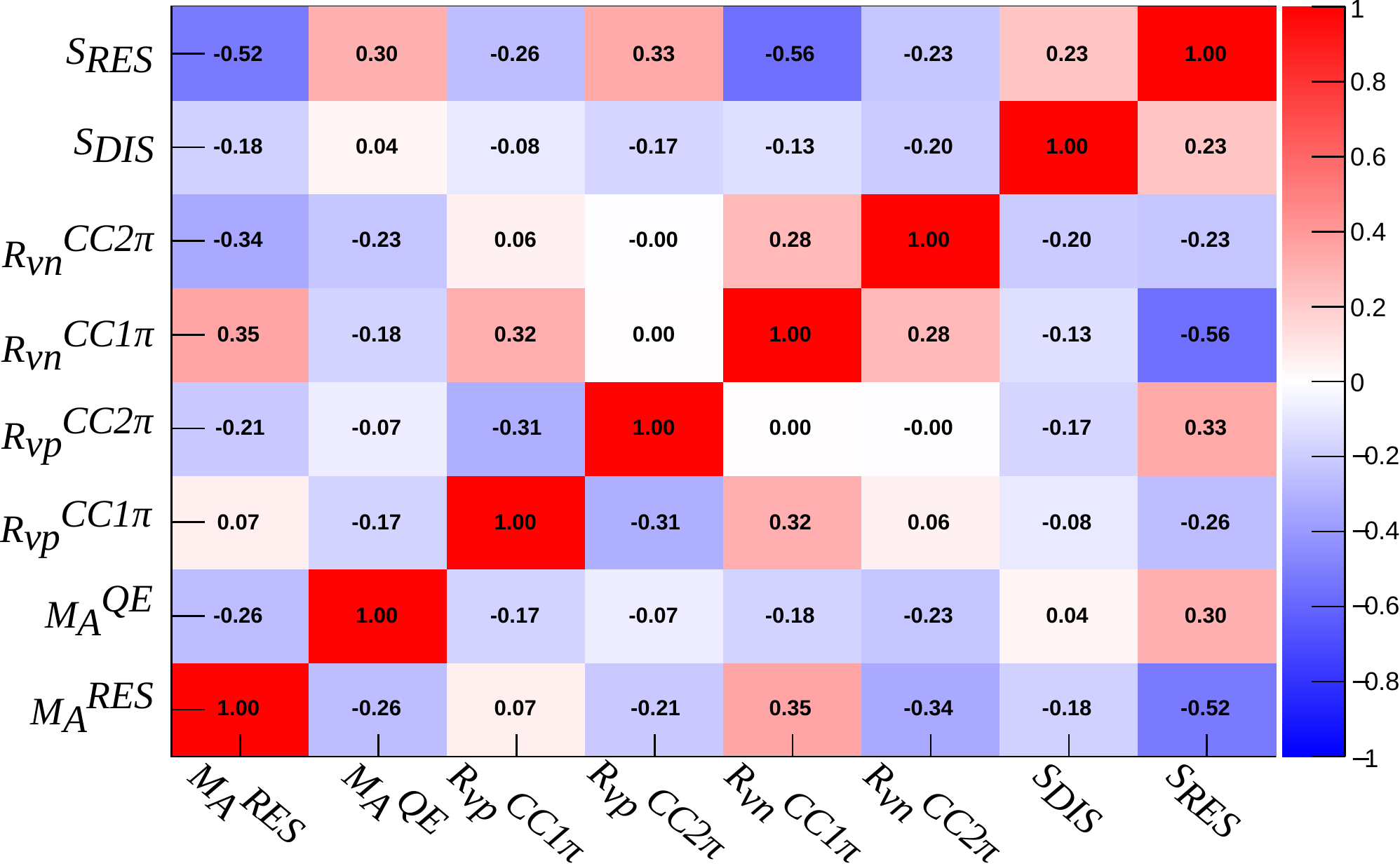}
    \caption{Parameter correlation matrix from the GENIE fit using the G18\_01a(/b) CMC correlation matrix.}
    \label{fig:correlation1}
\end{figure}

\begin{figure}
    \includegraphics[width=\textwidth]{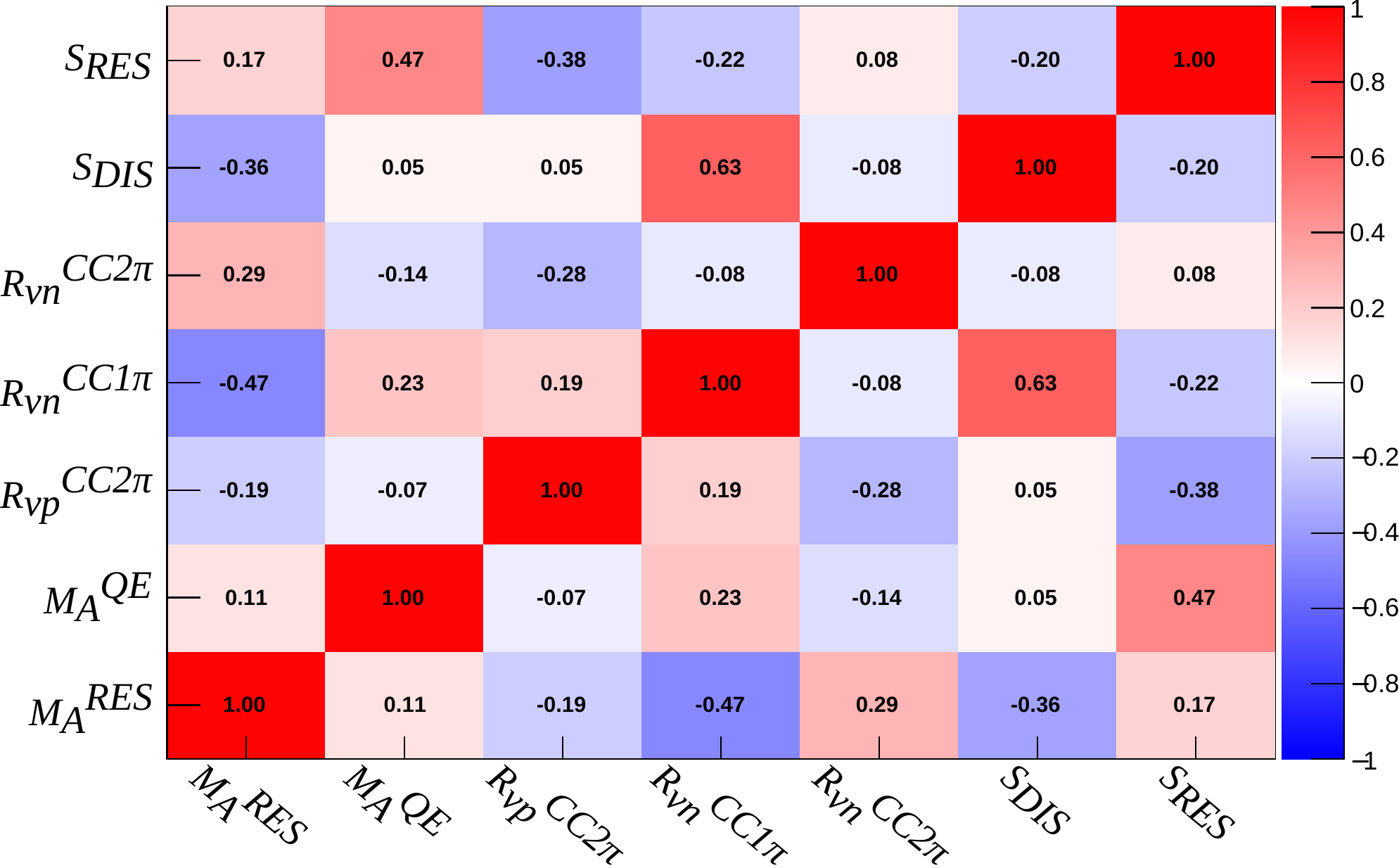}
    \caption{Parameter correlation matrix from the GENIE fit using the G18\_02a(/b) CMC correlation matrix.}
    \label{fig:correlation2}
\end{figure}

\begin{figure*}
    \centering
    \begin{subfigure}{7cm}
        \centering\includegraphics[width=\columnwidth]{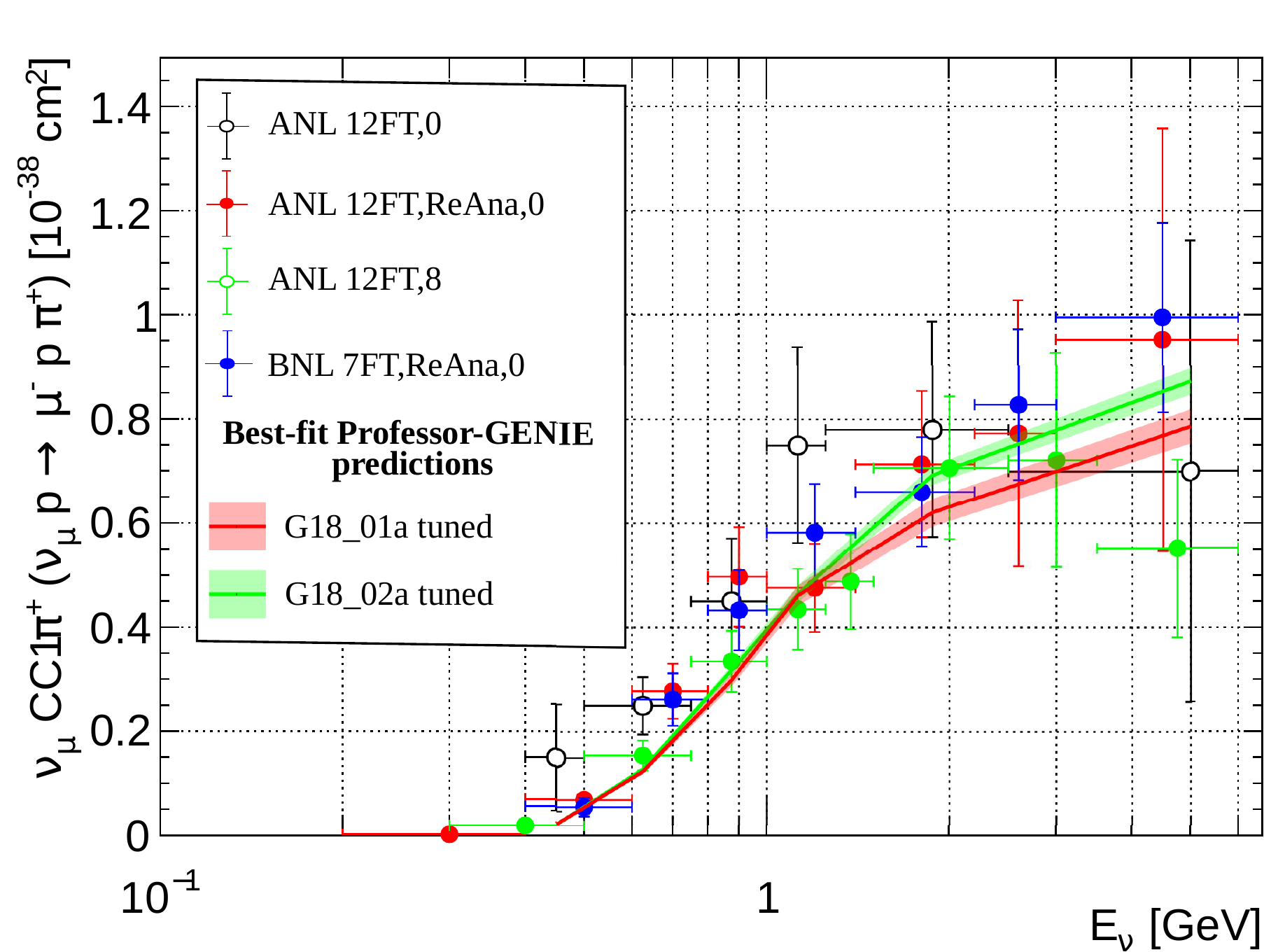}
        \caption{$\nu_\mu \text{CC}1\pi^+$ comparison.}   
    \end{subfigure}  \,\,\,\,\,\,
    \begin{subfigure}{7cm}
        \centering\includegraphics[width=0.8\columnwidth]{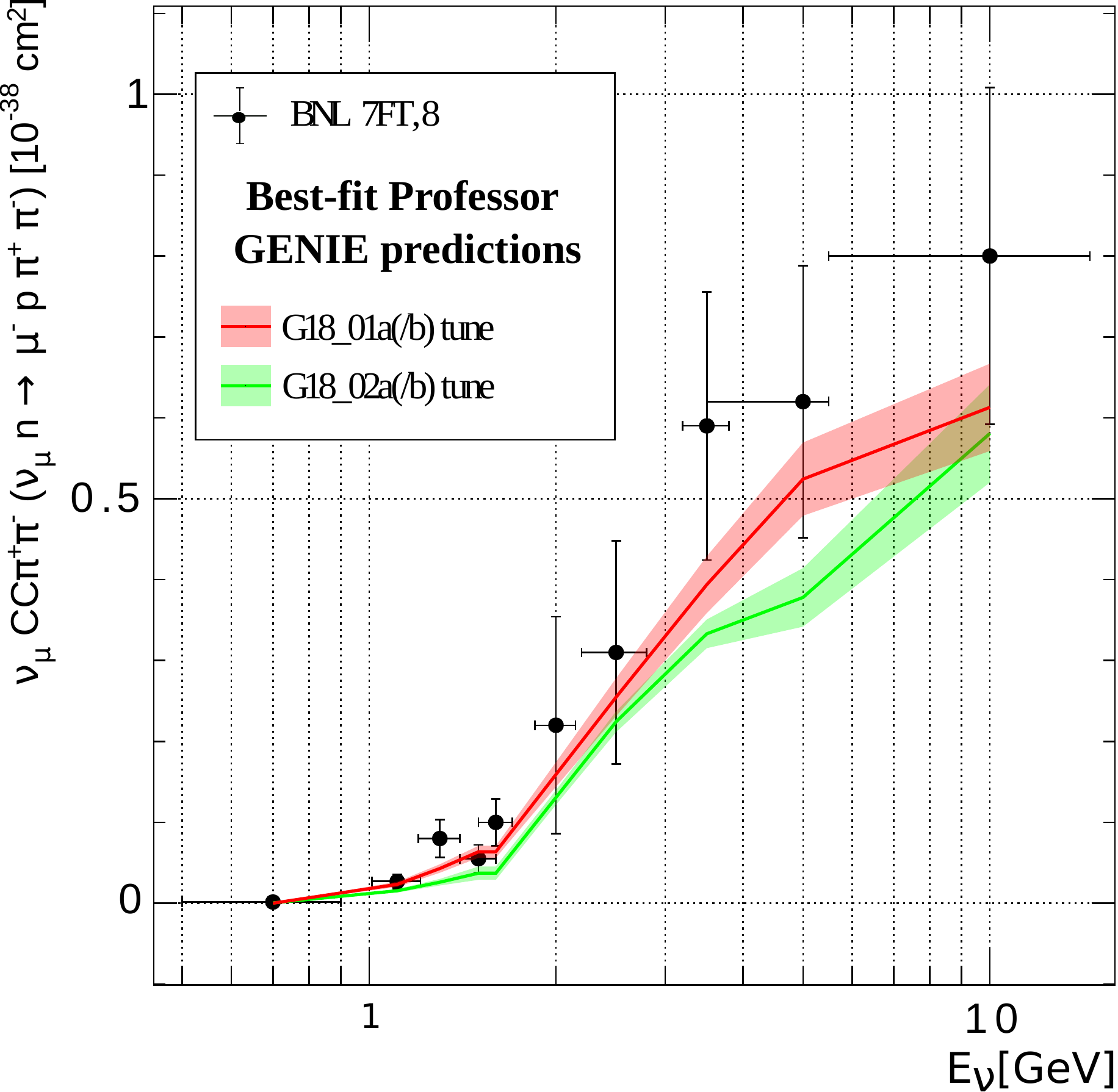}
        \caption{$\nu_\mu \text{CC}\pi^+\pi^-$ comparison.}   
    \end{subfigure} 
    \caption{Comparisons of neutrino data against predictions obtained from the Professor-GENIE parametrization at the best-fit value. 
    The uncertainties of the tune are propagated to the prediction considering the full covariance matrix. \label{fig:ProfessorPred} }
\end{figure*}

A covariance matrix is also obtained through the inversion of the the Hessian of the log-likelihood function at the best-fit parameter point. 
The corresponding correlation matrices are presented in Tab.~\ref{tab:cov1} and Tab.~\ref{tab:cov2} for the tunes of all 4 different cross-section model constructions used in this work (see the correlation matrices in Fig.~\ref{fig:correlation1} and Fig.~\ref{fig:correlation2} for a graphical interpretation). 
An example of the propagation of model uncertainties from the Professor output to the GENIE Comparisons framework is shown in Fig.~\ref{fig:ProfessorPred}. 

\begin{table*}
\centering
\begin{subtable}{\textwidth}
\centering
    \centering
    \begin{tabular}{c c c c c c c c c}  \hline\hline\noalign{\smallskip}
                                 & $M_A^{\text{RES}}$ & $M_A^{\text{QE}}$ & $R_{\nu p}^{\text{CC}1\pi}$
                                 & $R_{\nu p}^{\text{CC}2\pi}$ & $R_{\nu n}^{\text{CC}1\pi}$
                                 & $R_{\nu n}^{\text{CC}2\pi}$ & $S_{\text{RES}}$ &  $S_{\text{DIS}}$  \\ 
    \noalign{\smallskip}\hline\noalign{\smallskip}
    $M_A^{\text{RES}}$           &  5.3E-4 & -7E-5   &  5E-5   & -8E-4   &  2.2E-4 & -2.4E-3 & -4.3E-4 & -9E-5   \\
    $M_A^{\text{QE}}$            & -7E-5   &  1.2E-4 & -6E-5   & -1.2E-4 & -5E-5   & -7.6E-4 &  1.2E-4 &  1E-5   \\
    $R_{\nu p}^{\text{CC}1\pi}$  &  5E-5   & -6E-5   &  9.3E-4 & -1.6E-3 &  2.6E-4 &  5.4E-4 & -2.8E-4 & -6E-5   \\
    $R_{\nu p}^{\text{CC}2\pi}$  & -8E-4   & -1.2E-4 & -1.6E-3 &  2.7E-2 &  2.0E-5 & -2.5E-4 &  2E-3   & -6.2E-4 \\
    $R_{\nu n}^{\text{CC}1\pi}$  &  2.2E-4 & -5E-5   &  2.6E-4 &  2E-5   &  7.1E-4 &  2.3E-3 & -5.3E-4 & -8E-5   \\
    $R_{\nu n}^{\text{CC}2\pi}$  & -2.4E-3 & -7.6E-4 &  5.4E-4 & -2.5E-4 &  2.3E-3 &  9.6E-2 & -2.5E-3 & -1.4E-3 \\
    $S_{\text{RES}}$             & -4.3E-4 &  1.2E-4 & -2.8E-4 &  2E-3   & -5.3E-4 & -2.5E-3 &  1.3E-3 &  1.8E-4 \\
    $S_{\text{DIS}}$             & -9E-5   &  1E-5   & -6E-5   & -6.2E-4 & -8E-5   & -1.4E-3 &  1.8E-4 &  5.1E-4 \\
    \noalign{\smallskip}\hline\hline
    \end{tabular}
    \caption{ G18\_01a(/b)  covariance matrix.}
    \label{tab:cov1}

\end{subtable}%
\\
\begin{subtable}{\textwidth}
\centering
    \centering
    \begin{tabular}{c c c c c c c c} \hline\hline\noalign{\smallskip}
                                 & $M_A^{\text{RES}}$ & $M_A^{\text{QE}}$ & $R_{\nu p}^{\text{CC}2\pi}$
                                 & $R_{\nu n}^{\text{CC}1\pi}$ & $R_{\nu n}^{\text{CC}2\pi}$
                                 & $S_{\text{RES}}$ & $S_{\text{DIS}}$                                 \\ 
    \noalign{\smallskip}\hline\noalign{\smallskip}
    $M_A^{\text{RES}}$           &  1.7E-4 &  2.0E-5 & -1.9E-4 & -6.0E-5 &  4.4E-4 &  6.0E-5 & -4.0E-5 \\
    $M_A^{\text{QE}}$            &  2.0E-5 &  1.8E-4 & -7.0E-5 &  3.0E-5 & -2.1E-4 &  1.5E-4 &  1.0E-5 \\
    $R_{\nu p}^{\text{CC}2\pi}$  & -1.9E-4 & -7.0E-5 &  5.5E-3 &  1.5E-4 & -2.4E-3 & -6.9E-4 &  3.0E-5 \\
    $R_{\nu n}^{\text{CC}1\pi}$  & -6.0E-5 &  3.0E-5 &  1.5E-4 &  1.1E-4 & -1.0E-4 & -6.0E-5 &  6.0E-5 \\
    $R_{\nu n}^{\text{CC}2\pi}$  &  4.4E-4 & -2.1E-4 & -2.4E-3 & -1.0E-4 &  1.3E-2 &  2.3E-4 & -8.0E-5 \\
    $S_{\text{RES}}$             &  6.0E-5 &  1.5E-4 & -6.9E-4 & -6.0E-5 &  2.3E-4 &  6.0E-4 & -4.0E-5 \\
    $S_{\text{DIS}}$             & -4.0E-5 &  1.0E-5 &  3.0E-5 &  6.0E-5 & -8.0E-5 & -4.0E-5 &  8.0E-5 \\
    \noalign{\smallskip}\hline\hline
    \end{tabular}
    \caption{ G18\_02a(/b) covariance matrix.}
    \label{tab:cov2}
\end{subtable}

\caption{Parameter covariance matrices extracted the GENIE fit for the tuned CMCs.}
\end{table*}

\begin{figure*}
    \centering
    \begin{subfigure}{8cm}
        \centering\includegraphics[width=\columnwidth]{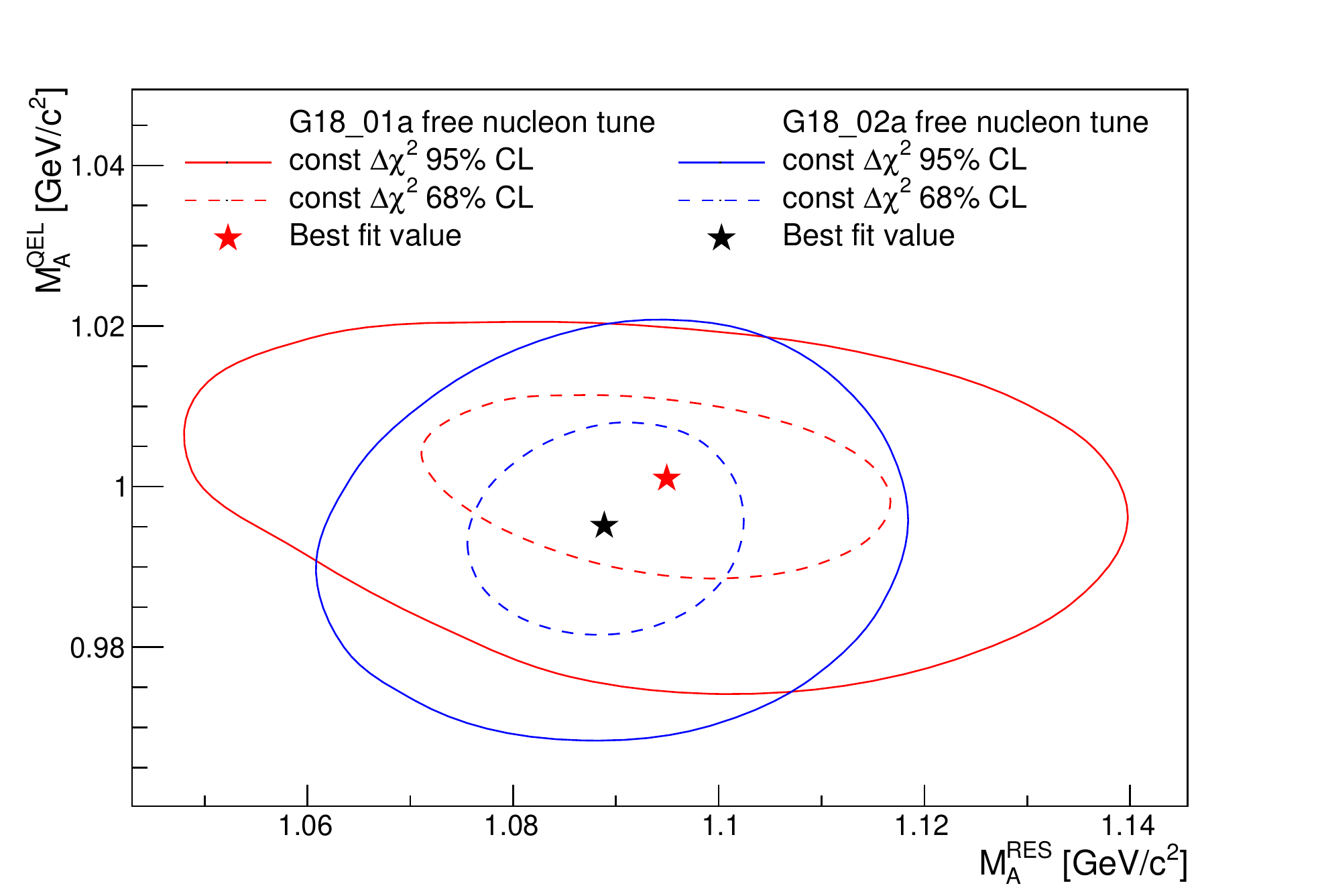}
        \caption{Contour $M_A^{\text{RES}}$ vs $M_A^{\text{QE}}$}   
    \end{subfigure}
    \begin{subfigure}{8cm}
        \centering\includegraphics[width=\columnwidth]{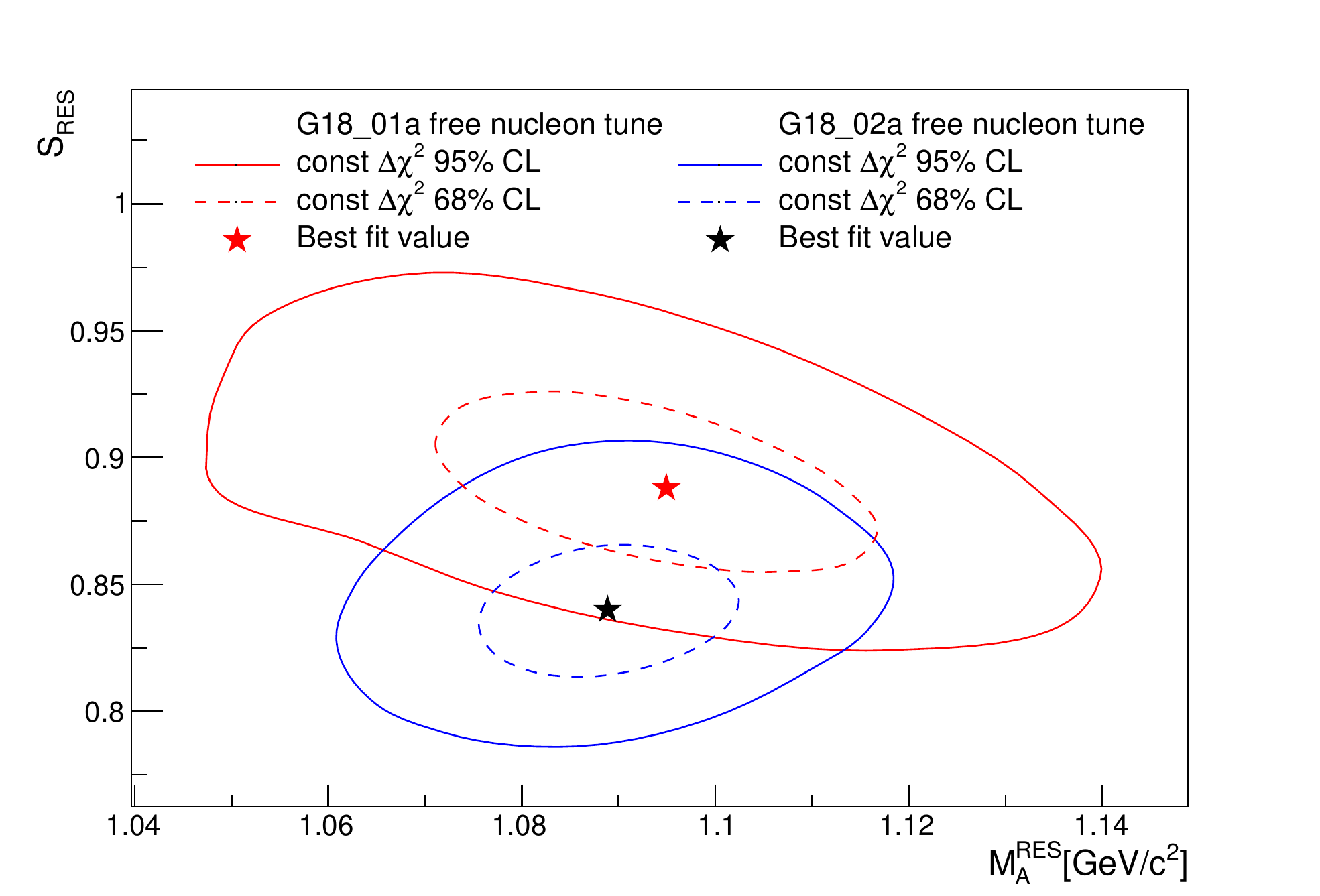}
        \caption{Contour $M_A^{\text{RES}}$ vs $S_{\text{RES}}$}   
    \end{subfigure}
    
    \begin{subfigure}{8cm}
        \centering\includegraphics[width=\columnwidth]{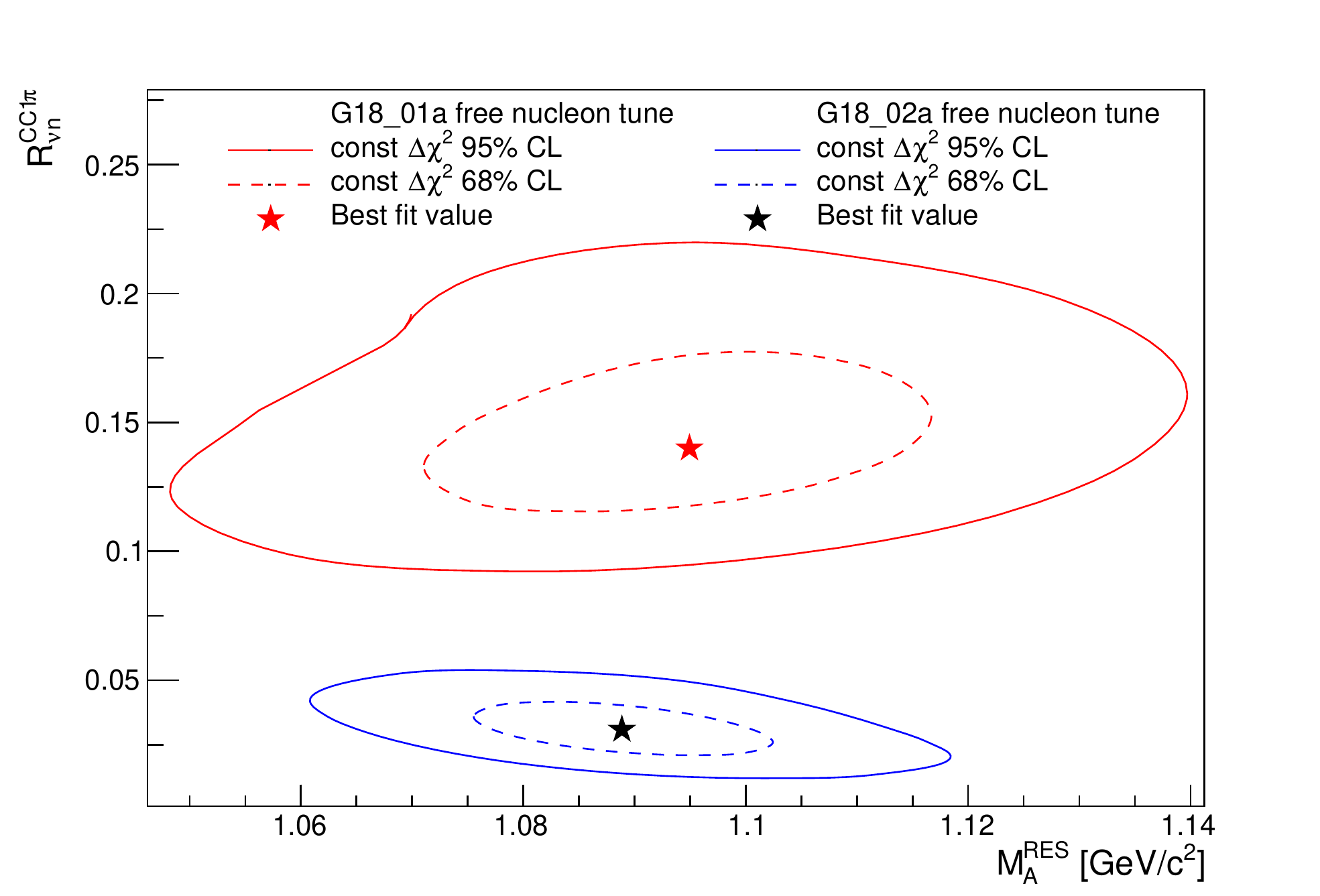}
        \caption{Contour $M_A^{\text{RES}}$ vs $R_{\nu n}^{\text{CC}1\pi}$}   
    \end{subfigure}
        \begin{subfigure}{8cm}
        \centering\includegraphics[width=\columnwidth]{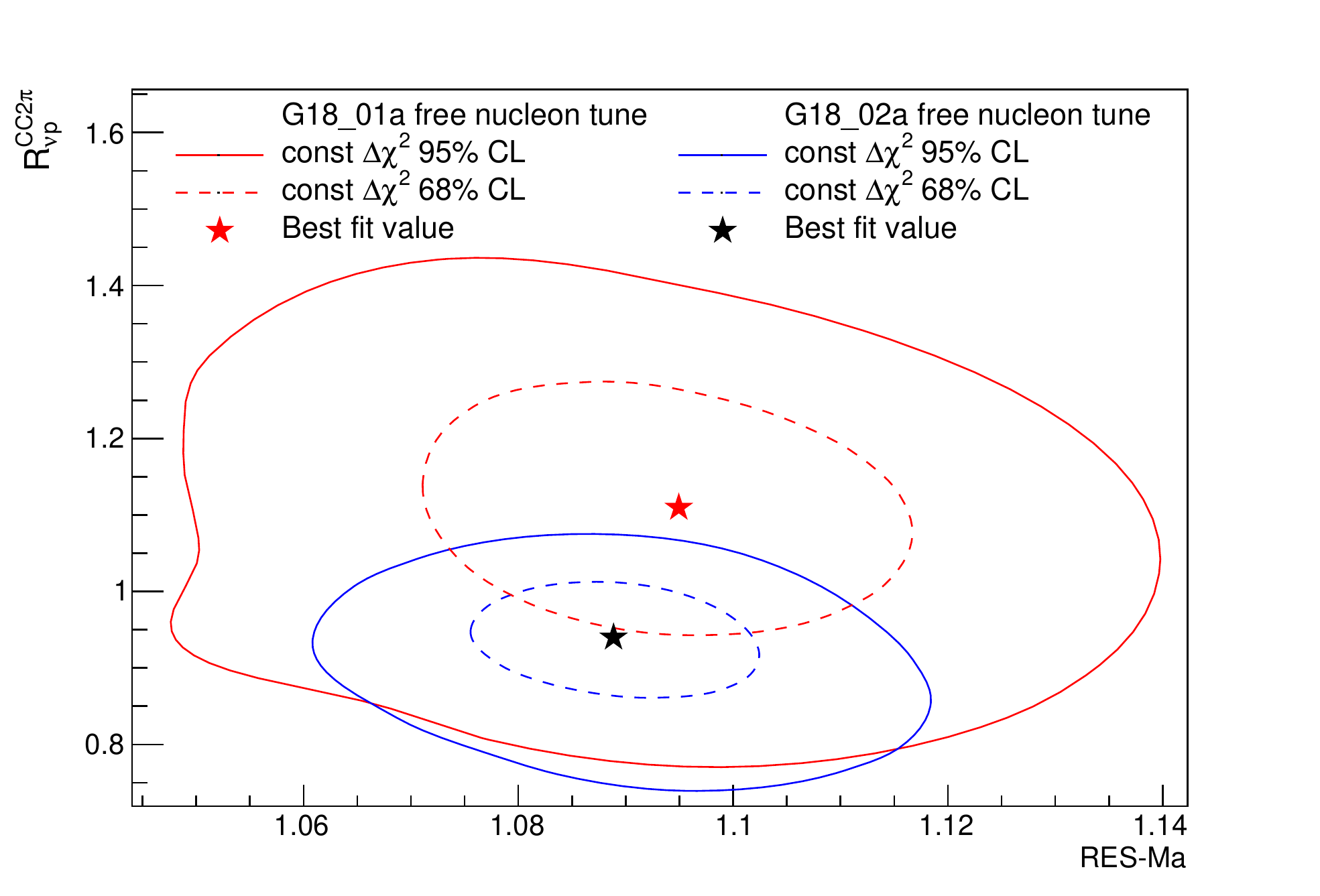}
        \caption{Contour $M_A^{\text{RES}}$ vs $R_{\nu p}^{\text{CC}2\pi}$}   
    \end{subfigure}
    
\caption{Joint $\Delta\chi^2_{\text{profile}}(\theta_i,\theta_j)$ functions obtained fixing the two parameters under study and minimizing $\Delta\chi^2(\vec{\theta},\vec{f})$  respect the other parameters in the tune. The contours for both tunes are shown for each case as well as the best fit values of each tune.} 
\label{fig:countours}
\end{figure*}

Joint $\Delta\chi^2_{\text{profile}}(\theta_i,\theta_j)$ functions, constructed by fixing two parameters at a grid of values and minimizing and $\Delta\chi^2(\boldsymbol{\theta},\boldsymbol{f})$ with respect to all other new parameters, are shown in Fig.~ \ref{fig:countours} for selected sets of parameters. 
In Figs.\ref{fig:countours}, we can see that the coverage of the parameter space for the 68\% and 95\% confidence level lines is wider for the G18\_01a(/b) tunes. 
This characteristic is again not related with how well we can constrain the parameters from the data, but with the capability of the models to accommodate this data in each model implementation. 

%
%

\section{Conclusions}

GENIE has released a number of comprehensive model configurations (CMCs) which consist of different modelling aspects combined altogether. 
In previous GENIE versions, there was a preferred {\em default} comprehensive model which failed to describe both inclusive and exclusive channels due to unresolved tensions between the data.
These tensions, which are crucial to understand for the new generation of neutrino experiments, motivated a careful investigation and retune of the bare-nucleon cross-section model for all GENIE comprehensive models available in GENIE v3. 
Best-fit values and correlations for several parameters influencing the GENIE bare-nucleon cross sections are released in this paper. 

In GENIE v3, we focus on improving understanding of the SIS region by tuning the GENIE CMC predictions on hydrogen and deuterium data from ANL 12FT, BNL 7FT, BEBC and FNAL 15FT bubble chamber experiments. 
The tuning of the non-resonant background takes a central stage in this work in order to remove double counting issues. 
The SIS region has been tuned against $\nu_\mu$ and $\bar{\nu}_\mu$ CC inclusive, quasi-elastic, one pion and two pion integrated cross sections as a function of $E_\nu$. 
Quasi-elastic data has been introduced to the fit to better constrain the flux of each experiment.

The global fit is able to describe both inclusive and exclusive cross sections simultaneously. 
Tensions between inclusive and exclusive data have been re-encountered, and, as a consequence, the inclusive cross section at the 1--10~GeV energy region decreased with respect to the historical default value. 
The systematic treatment of correlations between datasets and the inclusion of priors were crucial to address tensions. 
After the tune, GENIE predictions of one pion production cross sections on free nucleons ($\nu_\mu$CC p$\pi^+$, n$\pi^+$, p$\pi^0$ and p$\pi^+\pi^-$) show a decrease in the non-resonant background contribution, improving the agreement with the data. 
The prediction for two pion production mechanisms is also in better agreement with data for the $\nu_\mu n \rightarrow \mu^- p \pi^+ \pi^-$ channel by increasing the two pion production non-resonant background contribution. 

%
%

\section{Acknowledgements}
We would like to thank Andy Buckley (University of Glasgow, UK) and Holger Schultz (Institute of Particle Physics Phenomenology, University of Durham, UK)
for their support interfacing the Professor tool with the software products that underpin the GENIE global analysis. 
We are also grateful with Luis Alvarez-Ruso for insightful discussions and the review of this paper.
We would like to thank the CC-IN2P3 Computing Center, as well as the Particle Physics Department at Rutherford Appleton Laboratory for providing computing resources and for their support. 
This work, as well as the ongoing development of several other GENIE physics tunes was enabled through a PhD studentship funded by STFC through LIV.DAT, 
the Liverpool Big Data Science Centre for Doctoral Training (project reference: 2021488).
The initial conceptual and prototyping work for the development of the GENIE / Professor interfaces, as well as for the development of the GENIE global analysis framework that, currently, underpins several analyses, was supported in part through an Associateship Award
by the Institute of Particle Physics Phenomenology, University of Durham.

This document was prepared by the GENIE collaboration using the resources of the Fermi National Accelerator Laboratory (Fermilab), a U.S. Department of Energy, Office of Science, HEP User Facility. Fermilab is managed by Fermi Research Alliance, LLC (FRA), acting under Contract No. DE-AC02-07CH11359.

\begin{table*}
    \centering
    \begin{tabular}{c c c c | c c c c } \hline\hline\noalign{\smallskip}
     & & \multicolumn{2}{c|}{$\mathbf{\chi^2}$ \textbf{values for G18\_01a}}&  \multicolumn{2}{c}{$\mathbf{\chi^2}$ \textbf{values for G18\_02a}}  \\ 
     \noalign{\smallskip}
     \textbf{Dataset} &  $\mathbf{N_{DOF}}$ & \textbf{\emph{default}} & \textbf{Best Fit}  &\textbf{\emph{default}}  & \textbf{Best Fit} \\  \noalign{\smallskip}\hline\noalign{\smallskip}
     $\nu_\mu$ CC Inclusive \\ 
     \noalign{\smallskip}\hline\noalign{\smallskip}
     BNL 7FT \cite{Baker:1982ty} & 13  & 11.1 & 9.95 & 14.7 & 7.75\\
     BEBC ~\cite{Colley:1979rt} & 3  & 0.215 &0.101 & 0.067 & 0.045\\
     FNAL 15FT ~\cite{Kitagaki:1982dx,Baker:1982jf} & 10 & 3.85 & 3.92& 4.04 & 4.98 \\
     \noalign{\smallskip}\hline\noalign{\smallskip}
     $\bar{\nu}_\mu$ CC Inclusive \\
     \noalign{\smallskip}\hline\noalign{\smallskip}
     BEBC~\cite{Colley:1979rt,Bosetti:1981ip,Parker:1983yi} & 11 & 11.17& 11.5 & 9.79 & 9.8 \\ 
     BNL 7FT~\cite{Fanourakis:1980si} & 1  &1.83 &1.51 & 1.96 & 0.827 \\
     FNAL 15FT~\cite{Asratian:1984ir,Taylor:1983qj} & 13 &3.86 & 4.12& 4.32 & 4.13\\
     \noalign{\smallskip}\hline\noalign{\smallskip}
     $\nu_\mu n \rightarrow \mu^- n \pi^+$ \\
     \noalign{\smallskip}\hline\noalign{\smallskip}
     ANL 12FT~\cite{Radecky:1981fn} & 5 & 11.6& 9.88 & 27.3 & 14.3\\
     ANL 12FT,ReAna~\cite{Wilkinson:2014yfa} & 7 & 31.3& 21.0 & 48.8 & 25.2\\ 
     BNL 7FT,ReAna~\cite{Wilkinson:2014yfa} & 11 & 103& 45.7& 112 & 43.9\\
     \noalign{\smallskip}\hline\noalign{\smallskip}
     $\nu_\mu p \rightarrow \mu^- p \pi^+$ \\
     \noalign{\smallskip}\hline\noalign{\smallskip}
     ANL 12FT,ReAna ~\cite{Wilkinson:2014yfa} & 8 & 11& 8.71& 17.8 & 9.64\\
     BNL 7FT,ReAna ~\cite{Wilkinson:2014yfa} & 7 & 6.16& 3.11& 9.71 & 3.9\\ 
     BEBC~\cite{Allen:1980ti,Allasia:1990uy,Allen:1985ti} & 15 & 33.98& 15.9& 82.6 & 21.0\\ 
     FNAL~\cite{Bell:1978qu} & 3  &1.11 &0.74 & 2.87 & 0.66\\
     \noalign{\smallskip}\hline\noalign{\smallskip}
     $\nu_\mu n \rightarrow \mu^- p \pi^0 $  \\
     \noalign{\smallskip}\hline\noalign{\smallskip}
     ANL 12FT~\cite{Radecky:1981fn} & 5 &4.89 & 4.98& 7.57 & 4.63\\
     ANL 12FT,ReAna~\cite{Wilkinson:2014yfa} & 7 & 12.6& 12.0& 17.4& 11.5\\
     BNL 7FT,ReAna~\cite{Wilkinson:2014yfa} & 10 & 31.8&21.7& 38.4 & 19.4\\
     \noalign{\smallskip}\hline\noalign{\smallskip}
     $\nu_\mu p \rightarrow \mu^- n \pi^+ \pi^+$ \\
     \noalign{\smallskip}\hline\noalign{\smallskip}
     ANL 12FT~\cite{Day:1984nf} & 5 & 9.23&8.67 & 9.04 & 9.05\\
     \noalign{\smallskip}\hline\noalign{\smallskip}
     $\nu_\mu p \rightarrow \mu^- p \pi^+ \pi^0 $  \\
     \noalign{\smallskip}\hline\noalign{\smallskip}
     ANL 12FT~\cite{Day:1984nf} & 5& 4.28& 5.19& 4.64 & 4.66 \\
     \noalign{\smallskip}\hline\noalign{\smallskip}
     $\nu_\mu n \rightarrow \mu^- p \pi^+\pi^-$ \\
     \noalign{\smallskip}\hline\noalign{\smallskip}
     ANL 12FT~\cite{Day:1984nf} & 5  &8.24 & 8.36& 8.09 & 4.95\\
     BNL 7FT~\cite{Kitagaki:1986ct} & 10 &11.6 & 5.96& 10.3 & 6.46\\
     \noalign{\smallskip}\hline\noalign{\smallskip}
     $\nu_\mu$ CC QE \\
     \noalign{\smallskip}\hline\noalign{\smallskip}
     ANL 12FT~\cite{Mann:1973pr,Barish:1977qk}& 15 & 11.7& 12.2 & 11.75 & 11.58\\ 
     BNL 7FT~\cite{Baker:1981su} & 4 &6.88 & 6.91& 6.98 & 7.58\\ 
     BEBC~\cite{Allasia:1990uy} & 5  &8.18 &9.45 & 8.21 & 9.54\\
     FNAL~\cite{Kitagaki:1983px} & 2 &0.886 & 0.951& 0.992 & 0.893\\
     \noalign{\smallskip}\hline\noalign{\smallskip}
    $\bar{\nu}_\mu$ CC QE \\
     \noalign{\smallskip}\hline\noalign{\smallskip}
    BNL 7FT~\cite{Fanourakis:1980si} & 1  & 0.161 & 0.135 & 0.078 & 0.106\\
     \noalign{\smallskip}\hline\noalign{\smallskip}
    \textbf{Total} & \textbf{182} &\textbf{400.6} & \textbf{229.5}& \textbf{459.4} & \textbf{236.5} \\
    \noalign{\smallskip}\hline\hline
    \end{tabular}
    \caption{ Contributions to the \emph{default} and best fit $\chi^2$ for the datasets included. The data points with ${E_\nu < 0.5}$ GeV, a total of 10 points, are considered in the $\chi^2$ calculations of this table, but were not used in the fit. 
    For  the  calculation  of  the $\chi^2$,  the  covariance  matrix  between  the  datasets is used instead of Eq.~\ref{eqn:chi_square}, which incorporates nuisance parameters which are not implemented in GENIE.
    This explains the difference when comparing with the $\chi^2$ out of Professor from Tab.~\ref{tab:BestFitValuesAndErrors}.  }
    \label{tab:resultsFitchi2}  
\end{table*}

\begin{table*}
    \centering
     \parbox{.45\linewidth}{
     \begin{tabular}{c c c}  \hline\hline\noalign{\smallskip}
     \textbf{Experiment}& \textbf{Tag }& \textbf{Ref.} \\  
     \noalign{\smallskip}\hline\noalign{\smallskip}
     \multicolumn{3}{c}{$\nu_\mu$ CC Inclusive} \\ 
     \noalign{\smallskip}\hline\noalign{\smallskip}
     ANL 12 FT   & ANL 12 FT,2   & \cite{Barish:1977qk} \\
     BEBC        & BEBC,0        & \cite{Bosetti:1977nd} \\
     BEBC        & BEBC,5        & \cite{Bosetti:1981ip} \\
     BNL 7FT     & BNL 7FT,0     & \cite{Baltay:1980pr}\\
     CCFR        & CCFR,2        & \cite{Seligman:1997fe}\\
     CHARM       & CHARM,0       & \cite{Jonker:1980vf}\\
     FNAL 15FT   & FNAL 15FT,1   & \cite{Kitagaki:1986ct} \\
     Gargamelle  & Gargamelle,0  & \cite{Eichten:1973cs} \\
     Gargamelle  & Gargamelle,12 & \cite{Morfin:1981kg} \\
     IHEP\_ITEP  & IHEP\_ITEP,2  & \cite{Vovenko:2002ry} \\
     NOMAD       & NOMAD,5       & \cite{Lyubushkin:2008pe} \\
     MINOS       & MINOS,0       & \cite{Adamson:2009ju}\\
     ANL 12FT    & ANL 12FT,4    & \cite{Barish:1978pj}\\
     BEBC        & BEBC,2        & \cite{Colley:1979rt} \\
     BEBC        & BEBC,8        & \cite{Parker:1983yi} \\
     BNL 7FT     & BNL 7FT,4     & \cite{Baker:1982ty} \\
     CCFRR       & CCFRR,0       & \cite{MacFarlane:1983ax} \\
     CHARM       & CHARM,4       & \cite{Allaby:1987bb} \\
     FNAL 15FT   & FNAL 15FT,2   & \cite{Baker:1982jf}\\
     Gargamelle  & Gargamelle,10 & \cite{Ciampolillo:1979wp}\\
     IHEP\_ITEP  & IHEP\_ITEP,0  & \cite{Asratian:1984ir}\\
     IHEP\_JINR  & IHEP\_JINR,0  & \cite{Anikeev:1995dj}\\
     SKAT        & SKAT,0 & \cite{Baranov:1978sx}\\
     SciBooNE    & SciBooNE,0    & \cite{Nakajima:2010fp} \\
    \noalign{\smallskip}\hline\noalign{\smallskip}
    \multicolumn{3}{c}{$\bar{\nu}_\mu$ CC Inclusive} \\ 
    \noalign{\smallskip}\hline\noalign{\smallskip}
    BEBC         & BEBC,1        & \cite{Bosetti:1977nd}\\
    BEBC         & BEBC,6        & \cite{Bosetti:1981ip} \\
    BNL 7FT      & BNL 7FT,1     & \cite{Fanourakis:1980si} \\
    CHARM        & CHARM,1       & \cite{Jonker:1980vf}\\
    FNAL 15FT    & FNAL 15FT,4   & \cite{Taylor:1983qj} \\
    Gargamelle   & Gargamelle,1  & \cite{Eichten:1973cs} \\
    Gargamelle   & Gargamelle,13 & \cite{Morfin:1981kg}\\
    IHEP\_ITEP   & IHEP\_ITEP,3  & \cite{Vovenko:2002ry} \\
    MINOS        & MINOS,1       & \cite{Adamson:2009ju} \\
    BEBC         & BEBC,3        & \cite{Colley:1979rt} \\
    BEBC         & BEBC,7        & \cite{Parker:1983yi} \\
    CCFR         & CCFR,3        & \cite{Seligman:1997fe} \\
    CHARM        & CHARM,5       & \cite{Allaby:1987bb} \\
    FNAL 15FT    & FNAL 15FT,5   & \cite{Asratian:1984ir} \\
    Gargamelle   & Gargamelle,11 & \cite{Erriquez:1979nb}\\
    IHEP\_ITEP   & IHEP\_ITEP,1  & \cite{Asratian:1978rt} \\
    IHEP\_JINR   & IHEP\_JINR,1  & \cite{Anikeev:1995dj} \\
    \noalign{\smallskip}\hline\hline    
    \end{tabular}
    }
    \parbox{.45\linewidth}{
     \begin{tabular}{c c c} \hline\hline\noalign{\smallskip}
     \textbf{Experiment}& \textbf{Tag} & \textbf{Ref.} \\  
     \noalign{\smallskip}\hline\noalign{\smallskip}
     \multicolumn{3}{c}{$\nu_\mu$ CC Quasi-elastic} \\ 
     \noalign{\smallskip}\hline\noalign{\smallskip}
     ANL 12FT    & ANL 12FT,1       & \cite{Mann:1973pr} \\
     BEBC        & BEBC,12          & \cite{Allasia:1990uy} \\
     FNAL 15FT   & FNAL 15FT,3      & \cite{Kitagaki:1983px} \\
     SERP A1     & SERP A1,0        & \cite{Belikov:1981ut} \\
     SKAT        & SKAT,8           & \cite{Brunner:1989kw} \\
     ANL 12FT    & ANL 12FT,3       & \cite{Barish:1977qk} \\
     BNL 7FT     & BNL 7FT,3        & \cite{Baker:1981su} \\
     Gargamelle  & Gargamelle,2     & \cite{Bonetti:1977cs} \\
     SERP A1     & SERP A1,1        & \cite{Belikov:1985mw} \\
     NOMAD       & NOMAD,2          & \cite{Lyubushkin:2008pe} \\
     \noalign{\smallskip}\hline\noalign{\smallskip}
     \multicolumn{3}{c}{$\bar{\nu}_\mu$ CC Quasi-elastic} \\ 
     \noalign{\smallskip}\hline\noalign{\smallskip}
     BNL 7FT     & BNL 7FT,2        & \cite{Fanourakis:1980si}\\
     Gargamelle  & Gargamelle,5     & \cite{Armenise:1979zg}\\
     SKAT        & SKAT,9           & \cite{Brunner:1989kw} \\
     Gargamelle  & Gargamelle,3     & \cite{Bonetti:1977cs} \\
     SERP A1     & SERP A1,2        & \cite{Belikov:1985mw} \\
     NOMAD       & NOMAD,3          & \cite{Lyubushkin:2008pe}\\
     \noalign{\smallskip}\hline\noalign{\smallskip}
     \multicolumn{3}{c}{$\nu_\mu \text{CC}1\pi^+$ ($\nu_\mu p\rightarrow \mu^-p\pi^+$)} \\ 
     \noalign{\smallskip}\hline\noalign{\smallskip}
     ANL 12FT    & ANL 12FT,0       & \cite{Campbell:1973wg} \\
     ANL 12FT    & ANL 12FT,ReAna,0 & \cite{Radecky:1981fn} \\
     ANL 12FT    & ANL 12FT,8       & \cite{Wilkinson:2014yfa} \\
     BNL 7FT     & BNL 7FT,ReAna,0  & \cite{Wilkinson:2014yfa} \\
     Gargamelle  & Gargamelle,4     & \cite{Lerche:1978cp} \\
     BEBC        & BEBC,4           & \cite{Allen:1980ti} \\
     FNAL 15FT   & FNAL 15FT,0      & \cite{Bell:1978rb} \\
     BEBC        & BEBC,9           & \cite{Allen:1985ti}\\
     BEBC        & BEBC,13          & \cite{Allasia:1990uy} \\
     \noalign{\smallskip}\hline\noalign{\smallskip}
     \multicolumn{3}{c}{$\bar{\nu}_\mu \text{CC}1\pi^-$ ($\bar{\nu} p\rightarrow \mu^+p\pi^-$)} \\ 
     \noalign{\smallskip}\hline\noalign{\smallskip}
     FNAL 15FT   & FNAL 15FT,10     & \cite{Barish:1979ny} \\
     \noalign{\smallskip}\hline\noalign{\smallskip}
     \multicolumn{3}{c}{$\nu_\mu \text{CC}1\pi^+$ ($\nu n\rightarrow \mu^-n\pi^+$)} \\ 
     \noalign{\smallskip}\hline\noalign{\smallskip}
     ANL 12FT    & ANL 12FT,ReAna,2 & \cite{Wilkinson:2014yfa} \\
     BNL 7FT     & BNL 7FT,ReAna,2  & \cite{Wilkinson:2014yfa}\\
     ANL 12FT    & ANL 12FT,10      & \cite{Radecky:1981fn} \\
     \noalign{\smallskip}\hline\noalign{\smallskip}
     \multicolumn{3}{c}{$\nu_\mu \text{CC}1\pi^0$ ($\nu n\rightarrow \mu^-p\pi^0$)} \\ 
    \noalign{\smallskip}\hline\noalign{\smallskip}    
     ANL 12FT    & ANL 12FT,ReAna,1 & \cite{Wilkinson:2014yfa}\\
     BNL 7FT     & BNL 7FT,ReAna,1  & \cite{Wilkinson:2014yfa}\\
     ANL 12FT    & ANL 12FT,9       & \cite{Radecky:1981fn}\\
     \noalign{\smallskip}\hline\noalign{\smallskip}
     \multicolumn{3}{c}{$\nu_\mu \text{CC}1\pi^+\pi^+$ ($\nu p\rightarrow \mu^-n\pi^+\pi^+$)} \\ 
     \noalign{\smallskip}\hline\noalign{\smallskip}
     ANL 12FT    & ANL 12FT,13      & \cite{Day:1984nf} \\
     \noalign{\smallskip}\hline\noalign{\smallskip}
     \multicolumn{3}{c}{$\nu_\mu \text{CC}1\pi^+\pi^0$ ($\nu p\rightarrow \mu^-p\pi^+\pi^0$)} \\ 
     \noalign{\smallskip}\hline\noalign{\smallskip}
     ANL 12FT    & ANL 12FT,12      & \cite{Day:1984nf} \\
     \noalign{\smallskip}\hline\noalign{\smallskip}
     \multicolumn{3}{c}{$\nu_\mu \text{CC}1\pi^+\pi^-$ ($\nu n\rightarrow \mu^-p\pi^+\pi^-$)} \\ 
     \noalign{\smallskip}\hline\noalign{\smallskip}
     ANL 12FT    & ANL 12FT,12      & \cite{Day:1984nf} \\
     BNL 7FT     & BNL 7FT,8        &\cite{Kitagaki:1982dx} \\
     \noalign{\smallskip}\hline\hline
     \end{tabular}

    }
    \caption{Summary of data used for comparisons in Figs.~\ref{fig:incl_xsec_partialfit}, \ref{fig:excl_1pi}, \ref{fig:excl_2pi}, \ref{fig:inclusivePredictions}, \ref{fig:quasielasticPredictions}, \ref{fig:pPredictions}, \ref{fig:twoPredictions}, \ref{fig:nPredictions} and \ref{fig:ProfessorPred}. This table links the experiment and the tag used for the legend in each plot to the corresponding reference.}
    \label{tab:summary_data_legend}
\end{table*} 

\clearpage

\bibliography{sample.bib,database.bib}

\end{document}